\documentclass[prd,aps,floats,floatfix,superscriptaddress,preprintnumbers,
showpacs,eqsecnum,nofootinbib,twocolumn]{revtex4}
\usepackage{amsmath,amsthm,amssymb,mathptmx}
\usepackage[dvipdfmx]{graphicx}
\usepackage{color}
\input{colordvi.tex}
\usepackage{float}
\usepackage{booktabs}
\usepackage{bm}

\allowdisplaybreaks[1]

\newcommand{\bea}{\begin{eqnarray}}
\newcommand{\ena}{\end{eqnarray}}
\newcommand{\beann}{\begin{eqnarray*}}
\newcommand{\enann}{\end{eqnarray*}}

\newcommand{\Mp}{M_{\rm Pl}}

\begin{document}
\preprint{YITP-18-113}

\title{$\alpha$-attractor-type Double Inflation}

\author{Kei-ichi Maeda}
\affiliation{
Department of Physics, Waseda University,
3-4-1 Okubo, Shinjuku, Tokyo 169-8555, Japan
}

\author{Shuntaro Mizuno}
\affiliation{Center for Gravitational Physics, Yukawa Institute for Theoretical Physics, Kyoto University, 
	Kyoto 606-8502, Japan}

\author{Ryota Tozuka}
\affiliation{
Department of Physics, Waseda University,
3-4-1 Okubo, Shinjuku, Tokyo 169-8555, Japan
\footnote{This is the affiliation where R. T. had conducted this work. His current affiliation is
{\rm TIS Inc., 8-17-1, Nishishinjuku, Shinjuku, Tokyo, 160-0023, Japan.} }}

\date{\today}
\begin{abstract}
We study the background dynamics and primordial perturbation
in $\alpha$-attractor-type double inflation.
The model is composed of two minimally coupled scalar fields,
where each of the fields has a potential of the $\alpha$-attractor type. 
We find that  the background trajectory in this 
double inflation model is given by two almost straight lines connected with a turn in the field space.
With this simple background, we can classify the property of the perturbations generated 
in this double inflation model just depending on whether the turn occurs 
before the horizon exit for the mode corresponding to 
the scale of interest  or not. In the former case, the resultant primordial curvature perturbation
coincides with the one obtained by the single-field $\alpha$-attractor.
On the other hand, in the latter case,
it is affected by the multifield effects, like the mixing with the isocurvature perturbation and
the change of the Hubble expansion rate at the horizon exit.
We obtain the approximated analytic solutions for the background,
with which we can calculate the primordial curvature perturbation 
by the $\delta N$ formalism analytically, even when the multifield effects are significant.
We also impose observational constraints on the model parameters 
and initial values of the fields in this double inflation model based on the Planck result.

\end{abstract}

\maketitle
\section{Introduction}
\label{sec:Introduction}

Cosmic inflation is the simplest scenario for the origin of primordial perturbations
in our Universe, whose predictions are verified by the observations
of the cosmic microwave background (CMB) and large scale structure
of the Universe \cite{Guth:1982ec,Hawking:1982cz,Starobinsky:1982ee,Bardeen:1983qw} (see e.g. \cite{Kodama:1985bj,Mukhanov:1990me} for reviews). Actually, 
the recent Planck observations confirm that the primordial
curvature perturbations are almost scale invariant and Gaussian \cite{Ade:2015xua,Ade:2015ava}.
These observations are consistent with the predictions of
the simplest class of inflation models given by a single scalar field
with a canonical kinetic term and a sufficiently flat potential and couples minimally to gravity. 
Regardless of the phenomenological success, it is still nontrivial to embed 
such  a single-field slow-roll inflation   model
into a more fundamental theory (see \cite{Baumann:2014nda}, for a review). 
In this context, the so-called $\alpha$-attractor models
\cite{Kallosh:2013hoa,Kallosh:2013xya,Ferrara:2013rsa,Kallosh:2013yoa,Galante:2014ifa}
have been actively studied so far. Phenomenologically,  this class of models predict common values
of the spectral index $n_s$ and   the tensor-to-scalar ratio $r$,
well matching observational data. Theoretically, the attractive property of these models is
attributed to the conformal symmetry and
they are successfully embedded into supergravity through the hyperbolic geometry
\cite{Carrasco:2015uma,Carrasco:2015rva} (see also \cite{Cremonini:2010ua,Renaux-Petel:2015mga,
Achucarro:2016fby,Brown:2017osf,Mizuno:2017idt,Garcia-Saenz:2018ifx} for other interesting cosmological scenarios
based on the hyperbolic geometry). Furthermore, recently, the models with certain values of $\alpha$
were shown to be derived from string/M theory setups based on the seven-disk manifold
 \cite{Ferrara:2016fwe,Kallosh:2017ced}.

 However, since scalar fields such as moduli fields   are ubiquitous in the fundamental theories like supergravity 
and string/M theory, it is natural to consider the multifield extension of the $\alpha$-attractor models.
From the viewpoint of the primordial perturbations, the curvature perturbation can evolve even on
sufficiently large scales caused by the mixing with the isocurvature perturbation in multiple inflation,
which gives rich phenomenology \cite{Gordon:2000hv,GrootNibbelink:2001qt}.
Among the multiple inflation models, the so-called double inflation composed of two minimally coupled
massive scalar fields is a simple and concrete model including the multifield effects on the primordial perturbation
\cite{Polarski:1992dq,Polarski:1994rz,Langlois:1999dw}
and have been actively studied 
(see, e.g., \cite{Tsujikawa:2002qx,Vernizzi:2006ve,Byrnes:2006fr}).
The T-model, which is one of the simplest realizations of the $\alpha$-attractor
and has a plateau-like potential, is shown to appear out of a massive scalar field with a non-canonical
kinetic term with a pole \cite{Galante:2014ifa} with a field redefinition so that the new field has
a canonical kinetic term. Therefore, as a simple multi-field extension of the $\alpha$-attractor,
in this paper, we consider the $\alpha$-attractor-type double inflation, which is composed of 
two minimally coupled scalar fields and each of the fields has a potential of the $\alpha$-attractor-type.
\footnote{Although we concentrate on the phenomenology of the $\alpha$-attractor-type
double inflation and do not consider the link of the setup with fundamental theories in this paper,
it might be possible to derive this model based on the seven-disk manifold \cite{yamada:2018pc}.}
For other interesting works considering the multifield extension of the $\alpha$-attractor models
in different contexts, see \cite{Kallosh:2013daa,Kallosh:2017wku,Achucarro:2017ing,Krajewski:2018moi,Yamada:2018nsk,Linde:2018hmx,Christodoulidis:2018qdw,Dias:2018pgj,Iarygina:2018kee}.
\footnote{It is known that the Starobinsky inflation model  \cite{Starobinsky:1980te} is included in
the $\alpha$-attractor, as so-called the E-model \cite{Kallosh:2013xya}. 
The multifield extension of the Starobinsky inflation has been also actively studied recently, see, e.g., \cite{Ellis:2014gxa,Abe:2014vca,Ellis:2014opa,vandeBruck:2015xpa,Kaneda:2015jma,Wang:2017fuy,Mori:2017caa,Pi:2017gih,He:2018gyf}.}

The rest of this paper is organized as follows.  In Sec.~\ref{sec:model}, 
we present our $\alpha$-attractor-type multiple inflation model in general form.
We then analyze the background dynamics 
for the case of double inflation, which we focus in what follows, in Sec.~\ref{sec:background}, 
showing that the analytic solutions
based on the slow-roll approximation can approximate the numerical results very well.
In Sec.~\ref{sec:field_perturbations}, we calculate the primordial curvature perturbation
in this double inflation model that can be verified by the recent CMB observations.
We show that the  $\delta N$ formalism, where we can calculate the primordial curvature perturbation
analytically in this model, can reproduce the numerical results very well.
After imposing the constraints on the model parameters and initial values of the fields based on
the Planck result in Sec.~\ref{sec:constraints}, we summarize in Sec.~\ref{sec:conclusions}
with conclusions and discussions. Some technical issues in this model related with the background dynamics 
based on a potential without the symmetric property,
the excitation of the heavy field at the turn, and the primordial non-Gaussianity
are discussed in Appendix~\ref{sec:bd_asymmetriccase}, Appendix~\ref{sec:BD_TurningPhase}, 
Appendix~\ref{sec:PNG}, respectively.
In this paper, when we show plots, all quantities with dimensions are normalized by $\Mp$, unless otherwise mentioned.

\section{$\alpha$-attractor-type Multiple Inflation Model 
\label{sec:model}}

Before presenting  the $\alpha$-attractor-type multiple inflation model,
we briefly summarize the conventional (single-field) $\alpha$-attractor inflation model.
\cite{Kallosh:2013hoa,Kallosh:2013xya,Ferrara:2013rsa,Kallosh:2013yoa,Galante:2014ifa}.

\subsection{$\alpha$-attractor inflation \label{subsec:single}}

It is well known that inflation models of canonical scalar field are described by 
the following action:
\beann
S = \int d^4 x \sqrt{-g} \left(\frac{\Mp^2 R}{2}
 -\frac12 (\nabla\phi)^2- V (\phi)
\right)\,,
\label{single_field_inflation}
\enann
where $g$ is the determinant of  the spacetime metric $g_{\mu \nu}$, $R$ is the scalar curvature
and $\Mp$ is the Planck mass.
The $\alpha$-attractor models motivated by the conformal symmetry 
include various potentials whose asymptotic shape looks like
\bea
V (\phi) \simeq \Lambda^4 \left[1- \gamma \exp\left[\mp \sqrt{\frac{2}{3 \alpha}} \frac{\phi}{\Mp} \right] \right]\,,
\label{potential_single_field_alpha_attractor} 
\ena
for $|\phi| \gg \sqrt{\alpha} \Mp$, where $\mp$ in the parentheses above denotes 
$-$ for $\phi >0$ and $+$ for $\phi < 0$.
Here, $\Lambda$ is a parameter with mass dimension,
while $\gamma$ and $\alpha$ are dimensionless parameters. 
If  most of the last 60 $e$-folds of inflation is driven in the region, where the potential is given 
by (\ref{potential_single_field_alpha_attractor}), this class of models are shown to have the attractor property
on inflational observables independent of $\gamma$. Actually, the  $\alpha$-attractor models,
with $\alpha$ not too large, generically predict that the spectral index  $n_s$ and 
tenor-to-scalar ratio $r$  defined later  
are given by the  number of $e$-folding of inflation  $N$ at leading order in  $1/N$  as
\bea
1-n_s \simeq \frac{2}{N}\,,\quad\quad r \simeq   \frac{12 \alpha}{N^2}\,,\quad\quad
\label{rel_ns_r_alphaattractor_def_N}
\ena
with
$ N \equiv  \ln (a_{\rm end}/a)$, 
where $a_{\rm end}$ is the scale factor at the end of inflation.
In order to satisfy this condition, we assume that $\alpha$ is not significantly different from 
unity,
which includes the so-called conformal attractor of $\alpha=1$.
Although $\Lambda$ in Eq.~(\ref{potential_single_field_alpha_attractor})
does not appear in the expression of $n_s$ and $r$, it controls the amplitude of the curvature perturbation
$\mathcal{P}_{\mathcal{R}}$, which we will also define later.

\subsection{Multiple inflation model \label{subsec:model}}

In this paper, since we are interested in $\alpha$-attractor-type   multiple inflation,
we consider the following  action with $n$ scalar fields $\varphi^I (I=1,2,\cdots, n)$,
 \bea
S=\int d^4 x \sqrt{-g} \left(\frac{\Mp^2}{2} R -\frac12 \delta_{IJ} \nabla_\mu \varphi^I \nabla^\mu \varphi^J
-V (\varphi^I) \right)\,,
\label{action}
\ena
with
\beann
V (\varphi^I)=  \sum_{I=1} ^n V_I (\varphi^I)
\,. 
\label{potential}
\enann
We use Einstein's implicit summation rule 
for the scalar field indices $I, J, \cdots$ and all the field indices
can be lowered by the field metric $\delta_{IJ}$.
For the potential, we are interested in the case that each $V_I$ takes the form given 
by Eq.~(\ref{potential_single_field_alpha_attractor}) for $|\varphi^I| \gg \Mp$.
To be concrete, we may adopt  the simplest T-model potential as
 $V_I$  \cite{Kallosh:2013hoa},
\bea
V_I = \frac{\lambda_I M_I^4}{36} \tanh^2 \left[\frac{\varphi^I}{M_I}\right]\,,
\label{T_Model_potential}
\ena
which has two parameters $\lambda_I$ and $M_I$.

Before starting the analysis, let us consider the single-field T-model potential more 
and discuss the possible parameter regime for the  $\alpha$-attractor-type double inflation
based on Eq.~(\ref{T_Model_potential}) with $I=1$, $\varphi^1=\phi$, 
$\lambda_1 =\lambda$, $M_1 =M$.
For $ |\phi| \gg M$, it is approximated  by 
\bea
V (\phi) &\simeq& \frac{\lambda M^4}{36} 
 \left[1- 4 \exp \left[\mp 2  \frac{\phi}{M}  \right] \right]\,,
\label{potential_TModel_asym_large}
\ena
which shows the form of the $\alpha$-attractor given by Eq.~(\ref{potential_single_field_alpha_attractor}) 
with  $\Lambda = (\lambda^{1/4} M)/\sqrt{6}$, $\gamma=4$, $\alpha = M^2/(6 \Mp^2)$.
Then, the single-field T-model predicts $n_s$ and $r$ given by Eq.~(\ref{rel_ns_r_alphaattractor_def_N})
as long as $\phi$ stays at this region during most of the last 60 $e$-folds of inflation.
On the other hand, for $|\phi| \ll M$, the potential behaves just as the mass term driving 
the chaotic inflation  \cite{Linde:1983gd},
\bea
V (\phi) &\simeq& \frac{m^2}{2} \phi^2\,,\quad\quad{\rm with}\quad m \equiv \frac{\sqrt{2 \lambda}}{6} M\,.
\label{potential_TModel_asym_small}
\ena
With $M \gg \Mp$, since most of the last 60 $e$-folds of inflation are given by the potential (\ref{potential_TModel_asym_small}),
the  prediction on $n_s$ and $r$ in this model becomes
indistinguishable from the ones in the chaotic inflation.
Since the conventional double inflation based on two massive scalar fields have been actively studied,
we do not consider the case with $M \gg \Mp$ in this paper. Therefore, we assume that
one of $\varphi^I$ is larger than $M$ in most of the last 60 $e$-folds of inflation throughout this paper.

Notice that although we concentrate on the model based on the T-model,
as long as the parameters are chosen so that each $V_I$ takes the form of the $\alpha$-attractor-type 
potential given by (\ref{potential_single_field_alpha_attractor}) when the observable mode
is generated, most results shown in this paper
are also obtained by the  $\alpha$-attractor-type  multiple inflation based on different models,
like the E-model \cite{Kallosh:2013xya}.

In what follows, we focus on the case of two scalar fields (double inflation), 
which is simple but includes the interesting multifield effects on the primordial perturbations.
In concluding remarks, we mention briefly about some properties expected in
multiple inflation.

\section{Background Dynamics of $\alpha$-attractor-type Double Inflation
\label{sec:background}}
In what follows, we will analyze $\alpha$-attractor-type double inflation
with T-model potential (\ref{T_Model_potential}).
We describe two scalar fields as $\varphi^1=\phi$ and $\varphi^2=\chi$.
In the main text of this paper, we concentrate on the potential symmetric between 
$\phi$ and $\chi$  so that $M_1 = M_2 =M$ and $\lambda_1 = \lambda_2 = \lambda$.
For more general cases without imposing  $M_1 = M_2$ nor $\lambda_1 = \lambda_2$,
we discuss the background dynamics  in Appendix~\ref{sec:bd_asymmetriccase},
and we find that the phenomenologically essential aspects of the $\alpha$-attractor-type double inflation 
can be extracted sufficiently by this simple symmetric potential.

\begin{figure}[htbp]
\includegraphics[width=6.5cm]{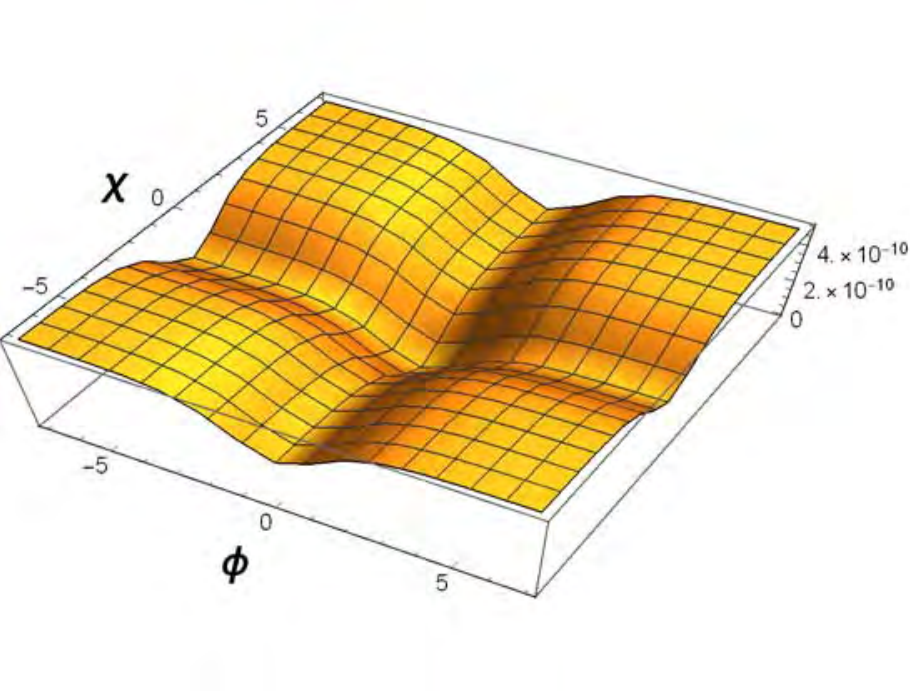}
\caption{The potential of the  $\alpha$-attractor-type double inflation based on two-field T-model 
given by Eqs.~(\ref{action}) and (\ref{T_Model_potential}) with $n=2$. 
We set   $M_1=M_2=M= \sqrt{3} \Mp$ and $\lambda_1=\lambda_2=\lambda = 10^{-9}$.  }
\label{TModelPotential1}
\end{figure}

\begin{figure}[htbp]
\includegraphics[width=4.5cm]{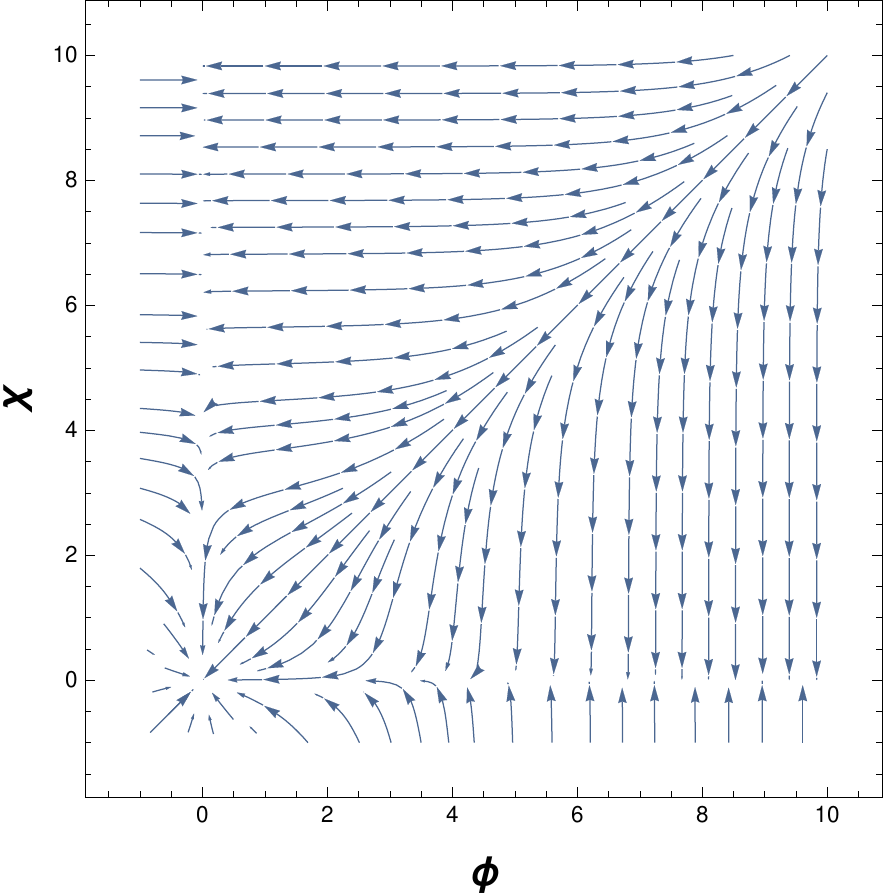}
\caption{Gradients of the inflationary potential
based on the two-field T-model, which suggests the direction of the background trajectories, 
with the same parameters as Fig.~\ref{TModelPotential1}.
Because of the symmetric property between $\phi$ and $\chi$,
the background trajectories towards $\phi=0$ and the ones towards $\chi=0$
in the early stage are separated by the line $\chi=\phi$.}
\label{TModelPotential2}
\end{figure}

In  Fig.~\ref{TModelPotential1}, we plot the shape of the potential of
the $\alpha$-attractor-type double inflation based on the symmetric two-field T-model.
From the symmetric property of the potential, 
we can restrict ourselves to the region $\chi \geq \phi \geq 0$ without loss of generality.
In  Fig.~\ref{TModelPotential2}, we depict the gradient of the potential.


In this section, we  analyze the background dynamics. 
We show that the analytic solutions based on the slow-roll
approximation can approximate the numerical results very well.

\subsection{Basic equations for background dynamics \label{subsec:background_dynamics}}

Suppose that the background Universe is homogeneous and isotropic with 
the flat Friedmann-Lema\^{i}tre-Robertson-Walker (FLRW) metric
\beann
ds^2 = g_{\mu\nu} dx^\mu dx^\nu = -dt^2+a^2(t) \delta_{ij} dx^i dx^j\,,
\label{FLRW}
\enann
where $a(t)$ is the scale factor whose evolution is governed by the Friedmann equations
\bea
H^2&=&\frac{1}{3 \Mp^2}  \left(\frac12 \delta_{IJ} \dot{\varphi}^I  \dot{\varphi}^J  +V \right)\,,
\label{BG_Friedmann}
\\
\dot{H} &=& -\frac{1}{2 \Mp^2} \delta_{IJ}  \dot{\varphi}^I   \dot{\varphi}^J\,,
\label{BG_Friedmann2}
\ena
where $H \equiv \dot{a}/a$ is the Hubble expansion rate and a dot denotes a derivative with respect to the cosmic time $t$. With  ${}_{,I}$  denoting a derivative with respect to $\varphi^I$,
the equations of motion for the homogeneous scalar fields $\varphi^I$ are
\bea
\ddot{\varphi}^I + 3H \dot{\varphi}^I + \delta^{IJ}V_{, J}=0\,.
\label{bg_field_eq_orig}
\ena

For later use, following Ref.~\cite{Gordon:2000hv}, 
we introduce the adiabatic and entropic directions based on the speed of the fields $\dot{\sigma}$ 
and the angle $\Theta$ characterizing the direction of the background trajectory, where
\bea
\dot{\sigma} \equiv \sqrt{\delta_{IJ} \dot{\varphi}^I \dot{\varphi}^J}\,,\quad\quad
\tan{\Theta} \equiv \frac{\dot{\chi}}{\dot{\phi}}\,.
\label{def_dotsigma_Theta}
\ena
The adiabatic direction is along  the background fields' evolution,
while the entropic direction is orthogonal to the adiabatic one.
The unit vectors corresponding to these directions $n^I$ and  $s^I$,
respectively are defined by
\bea
n^I &\equiv& \frac{\dot{\varphi}^I}{\dot{\sigma}}=(\cos \Theta,\; \sin \Theta)\,,
\label{def_unitvect_ad_en}
\\
 s^I &=& (-\sin \Theta,\; \cos \Theta)\,,
\label{def_unitvect_ad_en2}
\ena
which are related with each other through  $\dot{n}^I =  \dot{\Theta} s^I$,  $\dot{s}^I =  -\dot{\Theta} n^I$.
By projecting Eq.~(\ref{bg_field_eq_orig}) onto 
the adiabatic direction and entropic direction,
we can obtain the evolution equations for $\dot{\sigma}$ and $\Theta$, respectively as 
\beann
&&
\ddot{\sigma} + 3 H \dot{\sigma} + V_{,\sigma} =0\,,
\label{BG_EOM_ad_en}
\\
&&
\dot{\sigma} \dot{\Theta} + V_{,s} =0\,,
\label{BG_EOM_ad_en2}
\enann
with 
$
 V_{,\sigma} \equiv  V_{,I} n^I\,, \quad\quad V_{,s} \equiv V_{,I} s^I
$.

\subsection{Numerical results on background dynamics  \label{subsec:background_numerics}}

Before showing numerical results, 
 let us first give a brief overview of the background dynamics in the $\alpha$-attractor-type
 double inflation model, which can be easily expected from the potential form
 (Fig. \ref{TModelPotential1}).  
For the initial values of the fields $\varphi^I _{\rm ini}$ satisfying 
$\chi_{\rm ini} \geq \phi_{\rm ini} \gg M$, 
the slow-roll approximation becomes valid soon,
where the background trajectory starts following 
the gradient shown  in Fig.~\ref{TModelPotential2}.
In the first phase,  unless we fine-tune $\phi_{\rm ini} \simeq \chi_{\rm ini}$,
the motion towards $\chi=0$ soon  becomes 
negligible compared with that  towards $\phi=0$ 
and  the background trajectory becomes almost a  straight line towards $\phi=0$.
We expect a slow-roll inflation in this phase.
Then, when $\phi$ becomes smaller and approaches zero, 
the background fields are captured by the potential valley 
around $\phi = 0$ and changes the direction of the motion to  $\chi=0$.
 $\Theta$ defined by Eq.~(\ref{def_dotsigma_Theta}) changes from $\pi$ to $(3/2) \pi$.
After the direction has changed completely to $\chi=0$, 
the background trajectory becomes a straight line again towards $\chi = 0$.
This phase is characterized again as a single-field slow-roll inflation
driven by $\chi$ and it continues until $\chi$ becomes close to zero.
In this paper, for simplicity, we call these three evolutionary periods as the first inflationary phase,
the turning phase, and the second inflationary phase, respectively, 
although inflation may not be necessarily interrupted in the turning phase.

We perform numerical calculations
to show the typical behavior of the background dynamics in the $\alpha$-attractor-type double inflation.
We adopt $N$ defined by Eq.~(\ref{rel_ns_r_alphaattractor_def_N}) as the temporal variable.
As an example here, we set the model parameters  $M$ and $\lambda$ as
\bea
&&
M=\sqrt{3} \Mp\,, \quad\quad \lambda = 2.20 \times 10^{-10}\,,
\label{parameters_expample_BG}
\ena
while we choose the initial data as
  \bea
&&
\phi_{\rm ini} = 4.65 \Mp\,, \quad\quad \chi_{\rm ini}=5.50 \Mp\,,
\nonumber \\
&&
\dot{\varphi}^I_{\rm ini} = -{V_{\rm ini}}_{,I}/(3 H_{\rm ini})\,,
\label{initialvalues_expample_BG}
\ena
where 
$H_{\rm ini} = \sqrt{V_{\rm ini}}/(\sqrt{3} \Mp)$. 

Here we have assumed the slow-roll solution from the beginning.
Since it is an attractor, the numerical result does not depend on the choice of
$\dot{\varphi}^I _{\rm ini}$ significantly.

\begin{figure}[h]
\begin{center}
\includegraphics[width=4.5cm]{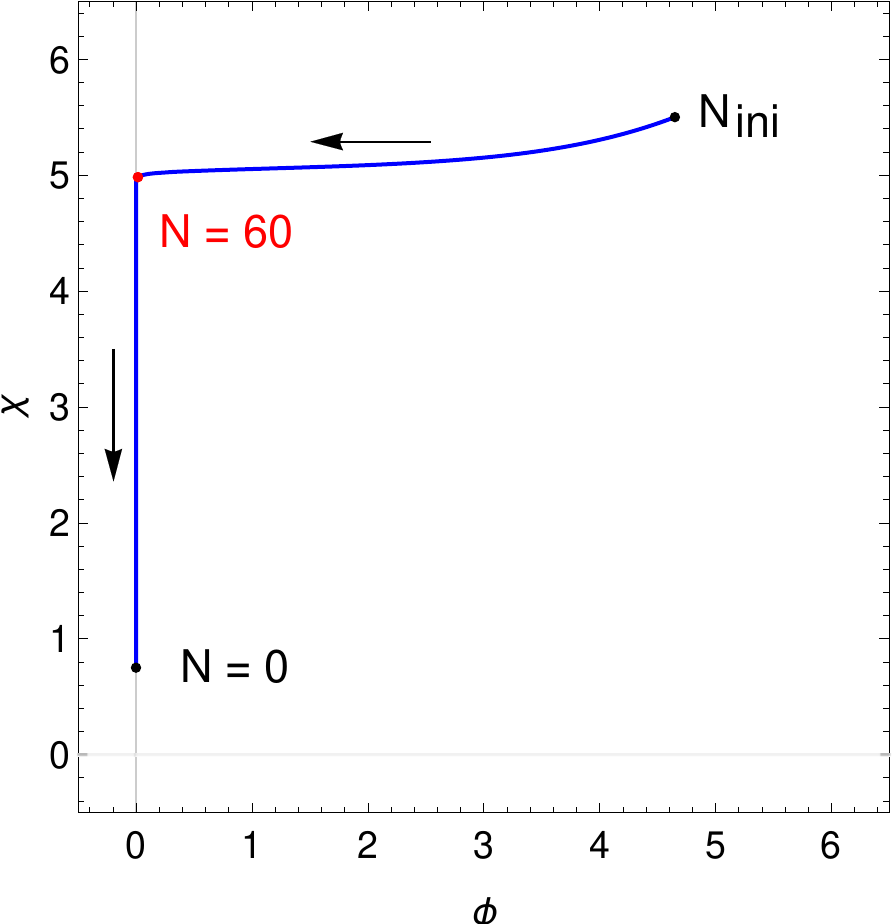}
\end{center}
\caption{ The background trajectory of field values $\varphi^I=(\phi,\chi)$  with $M=\sqrt{3}M_{\rm Pl}$, $\lambda=2.20\times 10^{-10}$. We choose the initial values as $\phi_{\rm ini}=4.65M_{\rm Pl}$
and $\chi_{\rm ini}=5.50M_{\rm Pl}$.
\label{trajectory_example}}
\end{figure}

In Fig.~\ref{trajectory_example}, we first show the background trajectory of two fields, which 
 confirms the above overview.
 This confirms our expectation mentioned at the end of Sec.~\ref{subsec:background_dynamics} 
that the motion of the background fields is towards $\phi=0$ with almost constant $\chi$ in the early stage,
and after $\phi$ reaches $0$, it changes the direction towards $\chi=0$ with keeping $\phi=0$.
Actually, until $N \sim 65$ the motion towards $\phi=0$ is more dominant compared with
that towards $\chi=0$, which we have called the first inflationary phase.
On the other hand, after then the motion is completely towards  $\chi=0$,
which we have called the second inflationary phase.

\begin{widetext}

\begin{figure}[htbp]
\begin{center}
\begin{tabular}{c}

\begin{minipage}{0.3\hsize}
\begin{center}
\includegraphics[width=5cm]{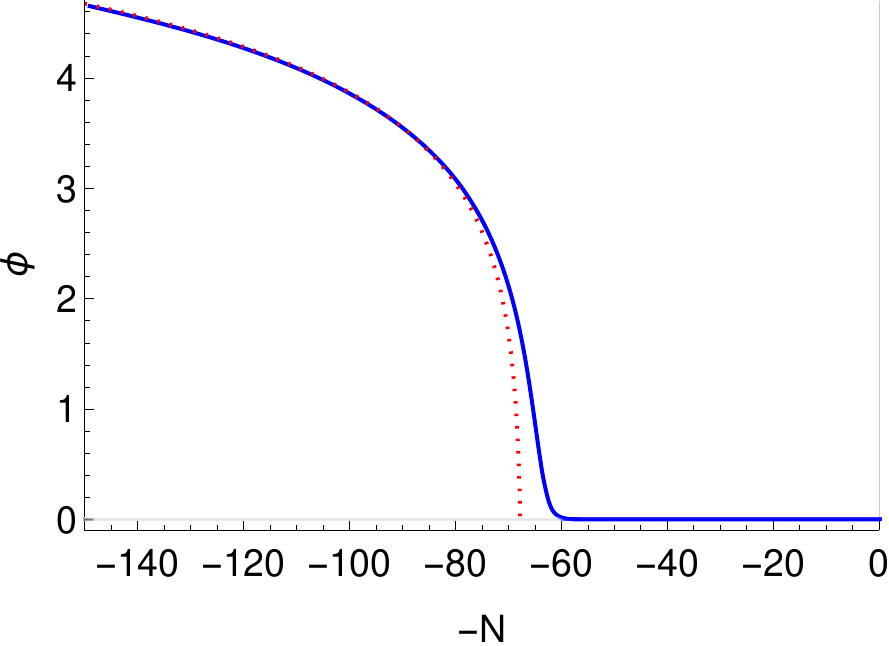}\\
\hspace{1.6cm}
\end{center}
\end{minipage}
\hspace{2cm}
\begin{minipage}{0.3\hsize}
\begin{center}
\includegraphics[width=5cm]{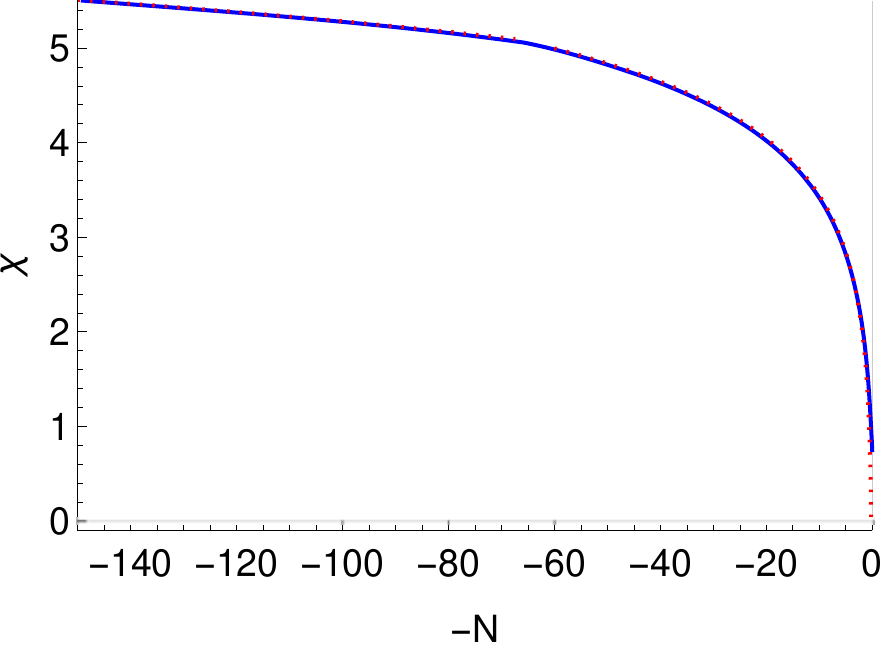}\\
\hspace{1.6cm}
\end{center}
\end{minipage}

\end{tabular}
\caption{Time evolution of field values $\varphi^I=(\phi,\chi)$ in terms of $-N$ with $M=\sqrt{3}M_{\rm Pl}$, $\lambda=2.20\times 10^{-10}$. We choose the initial values as $\phi_{\rm ini}=4.65M_{\rm Pl}$
and $\chi_{\rm ini}=5.50M_{\rm Pl}$.
In both panels,
the blue solid lines denote the numerical results, while the red
dotted lines denote the analytic results based on Eqs.~(\ref{phi_chi_slowroll_firstinflationaryphase_ito_N}) for the first inflationary phase
and Eq.~(\ref{secondinflationaryphase_chi_ito_N}) for the second inflationary phase.
In the left panel, we plot the analytic result only for the first inflationary phase.
For the analytic results, we choose the integration constants so that   
$N_{\rm ini}$ is given by Eq.~(\ref{concretevalue_N1_Nini}) and $N_2=0$.
}
\label{phi_chi_BG}
\end{center}
\end{figure}

In Fig.~\ref{phi_chi_BG}, we plot the time evolution of $\phi$ (left) and $\chi$ (right).
In the calculations, we regard that the inflation ends when 
the slow-roll parameter $\epsilon = 1$
and assign $N=0$ to that time. 
For the plots related with the time evolution, we use $-N$ as the temporal variable so that
the time evolves from left to right.
Later we will obtain analytic solutions which approximate these phases well.

In the left panel in Fig.~\ref{hubble_dottheta_BG}, we plot the time evolution of $H$,
where there is a gap between the first and second inflationary phases around $N \sim 65$,
This  is a unique characteristic of the $\alpha$-attractor-type double inflation, which we will discuss analytically later.
In the right panel in  Fig.~\ref{hubble_dottheta_BG},
we plot the time evolution of $\dot{\Theta}/H=d\Theta/d(-N)$ around the turn between the first and second
inflationary phases.
In this paper, formally, we define the turning phase as the period satisfying $\dot{\Theta}/H>0.05$.

\begin{figure}[htbp]
\begin{center}
\begin{tabular}{c}

\begin{minipage}{0.3\hsize}
\begin{center}
\includegraphics[width=5cm]{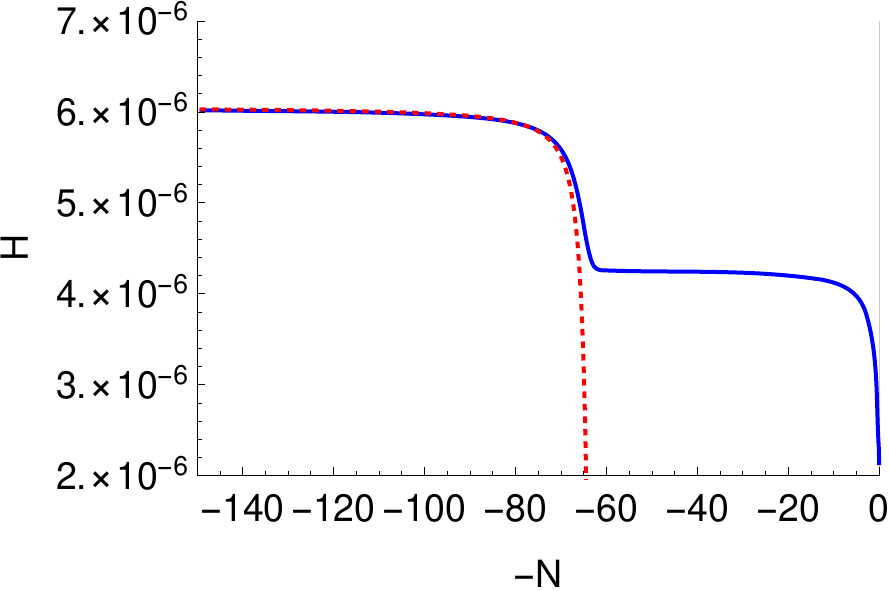}\\
\hspace{1.6cm}
\end{center}
\end{minipage}
\hspace{2cm}
\begin{minipage}{0.3\hsize}
\begin{center}
\includegraphics[width=5cm]{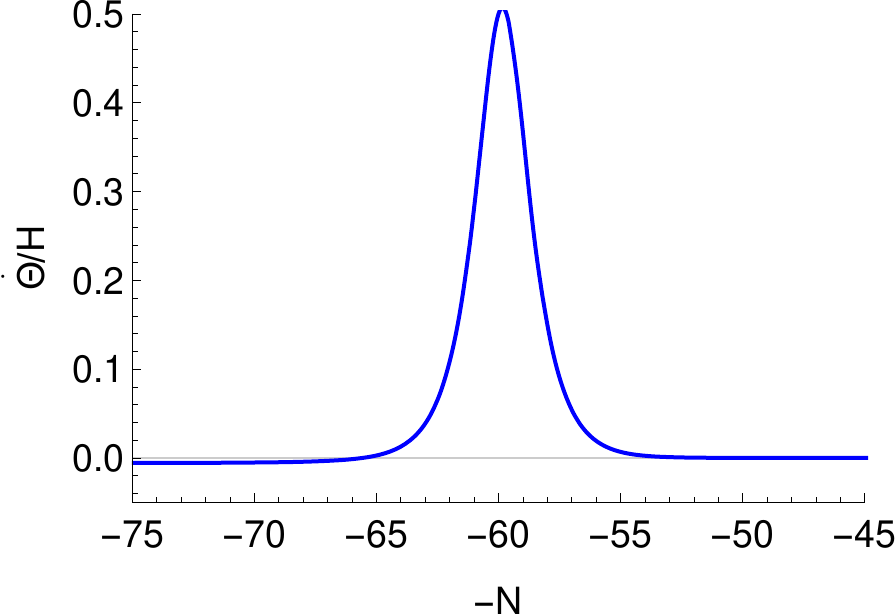}\\
\hspace{1.6cm}
\end{center}
\end{minipage}

\end{tabular}
\caption{ (Left) Time evolution of the Hubble expansion parameter $H$ in terms of $-N$  with the same parameters $M$, $\lambda$ and initial values $\varphi^I _{\rm ini}$ as in Fig. \ref{phi_chi_BG}.
(Right) Time evolution of  the angular velocity of the field variable $\dot \Theta/H = d\Theta/d(-N)$ 
around the turn between the first and second
inflationary phases.
In both panels, the blue solid lines denote the numerical results.
The red
dotted line in the left panel denotes  the analytic solution for the first inflationary phase
obtained by  the slow-roll approximation, 
$H = \sqrt{V}/(\sqrt{3} \Mp)$ with  Eqs.~(\ref{phi_chi_slowroll_firstinflationaryphase_ito_N})
and (\ref{phi_chi_slowroll_firstinflationaryphase_ito_N2}).
}
\label{hubble_dottheta_BG}
\end{center}
\end{figure}

\end{widetext}

\subsection{Analytic solutions based on slow-roll approximation  \label{subsec:background_app}}

Here, we present analytic solutions for the first and second inflationary phases 
in the $\alpha$-attractor-type double inflation, where the kinetic terms in Eq.  (\ref{BG_Friedmann})
and the second derivative in Eqs.~(\ref{bg_field_eq_orig}) are neglected. 
Then, we check the validity of the approximated solutions by comparing them with the numerical results.

\subsubsection{Second inflationary phase}

First we show the approximated solution for the second inflationary phase.
Although it has been already given as a single-field $\alpha$-attractor 
in Ref.~\cite{Kallosh:2013hoa},
it  gives hints to analyze the first inflationary phase.
Soon after the second inflationary phase starts,
where $\phi$ is trapped at zero, the potential can be approximated as
\bea
V(\phi=0, \chi) \simeq \frac{\lambda M^4}{36} \left[1-4 e^{-2 \frac{\chi}{M}} \right]\,,
\label{TModelPotential_approximated_secondinflationaryphase}
\ena
 for $\chi \gg M$\,.
Therefore, as long as $\chi \gg M$, the first term dominates
in Eq.~(\ref{TModelPotential_approximated_secondinflationaryphase})
and together with the slow-roll approximation, the Hubble expansion rate
in the second inflationary phase is given by 
$H/\Mp \simeq (\sqrt{\lambda} M^2)/(6 \sqrt{3} \Mp^2)$.
For the example we performed numerical calculations in Sec.~\ref{subsec:background_numerics},
with $M$ and $\lambda$ given by Eq.~(\ref{parameters_expample_BG}), we find
$H/\Mp \simeq 4.3 \times 10^{-6}$, which is consistent with the second plateau
shown in the left panel in Fig.~\ref{hubble_dottheta_BG}.
We can also obtain the analytic solution describing the evolution of $\chi$
based on the slow-roll approximation.
In the second inflationary phase,  the equation of motion for $\chi$ is approximately given by
\beann
\frac{d \chi}{d N} \simeq 8 \frac{\Mp^2}{M} e^{-2 \frac{\chi}{M}}\,,
\label{backgrounddynamics_chi_slowroll_secondinflationaryphase}
\enann
where $\simeq$ means that we have kept only up to the terms $\mathcal{O} ( e^{-2 \frac{\varphi^I}{M}})$.
Eq.~(\ref{backgrounddynamics_chi_slowroll_secondinflationaryphase})
is integrated with an integration constant $N_2$ as
\bea
\chi=\frac{M}{2} \ln \left[\frac{16 \Mp^2}{M^2} (N-N_2)\right]\,.
\label{secondinflationaryphase_chi_ito_N}
\ena
About $N_2$, it  was shown in Ref.~\cite{Kallosh:2013hoa} that
the integration constant corresponding to this is 
well approximated by the number of $e$-folding at the end of inflation, that is, zero,
in the single-field $\alpha$-attractor. Therefore, since the end of the second inflationary phase
coincides with the end of inflation in the $\alpha$-attractor-type double inflation, 
we can also set  $N_2=0$.
This is roughly explained by Eq.~(\ref{secondinflationaryphase_chi_ito_N}),
with which we can see that $\chi$ becomes zero slightly before $N=0$ if $N_2 = 0$.
In the right panel in Fig.~\ref{phi_chi_BG}, we confirm that the evolution of $\chi$
is well approximated by Eq.~(\ref{secondinflationaryphase_chi_ito_N}) with $N_2=0$.
This is mainly because the duration when $\chi$ is in the region 
$\chi > M$ is much longer than  that in the region $\chi<M$ in the second inflationary phase,
as we can see from the same plot.

For later use, we evaluate the time evolution of the slow-roll parameters in this phase here.
Under the slow-roll approximation,
\beann
&&
\frac{d \chi}{d N} \simeq \Mp^2 \frac{V_{,\chi}}{V}\,,\\
&&
\frac{d^2 \chi}{d N^2} \simeq \Mp^4 \frac{V_{,\chi}}{V} \left(\frac{V_{,\chi\chi}}{V} - \frac{V_{,\chi}^2}{V^2} \right)\,,
\enann
which gives  the following expression for the  slow-roll parameters 
defined by the derivatives of the potential
\bea
&&
\epsilon_\chi \equiv \frac{\Mp^2}{2} \left(\frac{V_{,\chi}}{V} \right)^2
 = \frac18 \frac{M^2}{\Mp^2} \frac{1}{N^2}\,,
 \label{epsilonchi_etachi_secondinflationaryphase}
\\
 &&
\eta_{\chi\chi} \equiv \Mp^2 \frac{V_{,\chi \chi}}{V}= -\frac{1}{N}+\frac14 \frac{M^2}{\Mp^2} \frac{1}{N^2}\,.
\label{epsilonchi_etachi_secondinflationaryphase2}
\ena
The slow-roll parameters obtained in Eqs.~(\ref{epsilonchi_etachi_secondinflationaryphase})
and (\ref{epsilonchi_etachi_secondinflationaryphase2})
are related with the ones defined in terms of the Hubble expansion rate as
\bea
&&
\epsilon = \epsilon_\chi\,,
\quad
\eta = 4 \epsilon_\chi -2 \eta_{\chi \chi}\,,
\label{rel_slowroll_hubble_chi}
\ena
with
\beann
\epsilon \equiv -\frac{\dot{H}}{H^2}\,,\quad\quad \eta \equiv \frac{\dot{\epsilon}}{H \epsilon}\,, 
\enann
when the time evolution of $\phi$ compared with that of $\chi$ is negligible.

\subsubsection{First inflationary phase}

Next, in a similar way as the previous subsubsection, 
we obtain the approximated solutions for the first inflationary phase.
With $\varphi^I _{\rm ini}$ mentioned previously, in the early stage,  we can approximate the potential as
\bea
V(\phi, \chi) \simeq \frac{\lambda M^4}{18} \left[1-2 e^{-2 \frac{\phi}{M}} -2e^{-2 \frac{\chi}{M}} \right]\,,
\label{TModelPotential_approximated}
\ena
for $ \phi \gg M\,,$ and $\chi \gg M$\,.
Therefore, by considering only the first term
in Eq.~(\ref{TModelPotential_approximated})
and making use of the slow-roll approximation, the Hubble expansion rate
in the first inflationary phase is given by 
$H/\Mp \simeq (\sqrt{\lambda} M^2)/(3 \sqrt{6} \Mp^2)$.
For the example we performed numerical calculations in Sec.~\ref{subsec:background_numerics},
with $M$ and $\lambda$  given by Eqs.~(\ref{parameters_expample_BG}) and (\ref{initialvalues_expample_BG}),
$H/\Mp \simeq 6.1 \times 10^{-6}$, which is consistent with the numerical results 
shown in the left panel in Fig.~\ref{hubble_dottheta_BG}.
The gap of the Hubble expansion rate between the first and second inflationary phases
is a typical feature in the $\alpha$-attractor-type double inflation.
We can also discuss the analytic solution describing the evolution of $\varphi^I$
in the first inflationary phase.
Together with the slow-roll approximation, the equations of motion for the scalar fields are given by
\bea
\frac{d \phi}{d N} \simeq 4 \frac{\Mp^2}{M} e^{-2 \frac{\phi}{M}}\,,\quad\quad
\frac{d \chi}{d N} \simeq 4 \frac{\Mp^2}{M} e^{-2 \frac{\chi}{M}}\,,
\label{eom_phi_chi_slowroll_firstinflationaryphase}
\ena
where $\simeq$ again  means that we have kept only up to the terms $\mathcal{O} ( e^{-2 \frac{\varphi^I}{M}})$.
Then,  Eqs.~(\ref{phi_chi_slowroll_firstinflationaryphase_ito_N})
and (\ref{phi_chi_slowroll_firstinflationaryphase_ito_N2}) can be integrated as
\bea
 &&
 \phi=\frac{M}{2} \ln \left[\frac{8 \Mp^2}{M^2} (N-N_{\rm ini}) + e^{2 \frac{\phi_{\rm ini}}{M} }\right]\,,
  \label{phi_chi_slowroll_firstinflationaryphase_ito_N}
\\
 &&
 \chi=\frac{M}{2} \ln \left[\frac{8 \Mp^2}{M^2} (N-N_{\rm ini}) + e^{2 \frac{\chi_{\rm ini}}{M}} \right]\,,
 \label{phi_chi_slowroll_firstinflationaryphase_ito_N2}
\ena
where $N_{\rm ini}$  is $N$ at initial time. 
The time evolution of $\phi$ in the first inflationary phase can be also written  as
\beann
\phi=\frac{M}{2} \ln \left[\frac{8 \Mp^2}{M^2} (N-N_1)\right]\,,
\enann
with
\bea
 N_1 \equiv N_{\rm ini} - \frac{M^2}{8 \Mp^2} e^{2 \frac{\phi_{\rm ini}}{M} }\,.
\label{def_N1}
\ena
Applying the same logic that fixed the integration constant $N_2$ that
appeared in Eq.~(\ref{secondinflationaryphase_chi_ito_N}) 
to $N_1$, we can regard
$N_1$ as the number of  $e$-folding at the end of the first inflationary phase.
In order to evaluate $N_1$ for given $\varphi^I _{\rm ini}$,
we assume that  the turning phase is instantaneous
and $N=N_1$ corresponds to not only the end of the first inflationary phase, 
but also the onset of the second inflationary phase.
Under this ``instantaneous turn'' assumption, 
$N_1$ becomes the number of $e$-folding of the second inflationary phase
and from Eq.~(\ref{secondinflationaryphase_chi_ito_N}), 
this depends on the value of $\chi$ when the second inflationary phase starts, which we denote $\chi_1$.
Since we can also regard $\chi_1$ as the value of $\chi$  
at the end of the first inflationary phase,
from Eqs.~(\ref{phi_chi_slowroll_firstinflationaryphase_ito_N})
and (\ref{phi_chi_slowroll_firstinflationaryphase_ito_N2}), $\chi_1$ can be estimated as
\beann
\chi_1= \frac{M}{2} \ln \left[1 + e^{2 \frac{\chi_{\rm ini}}{M}} - e^{2 \frac{\phi_{\rm ini}}{M}}\right]\,.
\label{chi1}
\enann
Here, we have assumed that the first inflationary phase ends at $\phi=0$ and
the expression (\ref{phi_chi_slowroll_firstinflationaryphase_ito_N})
and (\ref{phi_chi_slowroll_firstinflationaryphase_ito_N2}) are valid until that time.
Making use of Eq.~(\ref{secondinflationaryphase_chi_ito_N}) with $N=N_1$ and $N_2=0$,
and Eq.~(\ref{def_N1}) we can express 
$N_1$ and $N_{\rm ini}$ in terms of $\varphi^I _{\rm ini}$
as
\bea
&&
N_1 = \frac{M^2}{16 \Mp^2} \left(1 + e^{2 \frac{\chi_{\rm ini}}{M}} - e^{2 \frac{\phi_{\rm ini}}{M}} \right)\,,
\label{concretevalue_N1_Nini}
\\
&&
N_{\rm ini}= \frac{M^2}{16 \Mp^2} \left(1 + e^{2 \frac{\chi_{\rm ini}}{M}} + e^{2 \frac{\phi_{\rm ini}}{M}} \right)\,.
\label{concretevalue_N1_Nini2}
\ena
Regardless of the fact that Eqs.~(\ref{concretevalue_N1_Nini}) and
(\ref{concretevalue_N1_Nini2}) are based on a couple of assumptions,
in Fig.~\ref{phi_chi_BG}, we confirm that the evolution of $\varphi^I$
is well approximated by Eqs.~(\ref{phi_chi_slowroll_firstinflationaryphase_ito_N})
and (\ref{phi_chi_slowroll_firstinflationaryphase_ito_N2}) with 
$N_{\rm ini}$ given by Eqs.~(\ref{concretevalue_N1_Nini2}).
This is again explained by that the duration when $\phi$ is in the region 
$\phi > M$ is much longer than  that in the region $\phi<M$ in the first inflationary phase,
as we can see in the left panel in  Fig.~\ref{phi_chi_BG}. Now that we have obtained the time evolution
of $\varphi^I$ in the first inflationary phase, we can express the relation between $\phi$ and $\chi$
along the background trajectory originating from $\varphi^I _{\rm ini}$ and $\Theta$, introduced in Eq.~(\ref{def_dotsigma_Theta}) as
\bea
&&
\chi = \frac{M}{2} \ln \left[e^{2 \frac{\phi}{M} } + e^{2 \frac{\chi_{\rm ini}}{M}} - e^{2 \frac{\phi_{\rm ini}}{M}}\right]\,,
\label{trajectory_Theta_firstinflationaryphase}
\\
&&
\Theta = \tan^{-1} \left[\frac{\frac{8 \Mp^2}{M^2} (N-N_{\rm ini}) + e^{2 \frac{\phi_{\rm ini}}{M} }}
{\frac{8 \Mp^2}{M^2} (N-N_{\rm ini}) + e^{2 \frac{\chi_{\rm ini}}{M} }} \right]\,.
\label{trajectory_Theta_firstinflationaryphase2}
\ena
The information given by Eqs.~(\ref{trajectory_Theta_firstinflationaryphase}) and (\ref{trajectory_Theta_firstinflationaryphase2}), confirmed numerically, explains
our expectation mentioned at the end of Sec.~\ref{subsec:background_dynamics} 
that the motion of the background fields is towards $\phi=0$ with almost constant $\chi$ in the early stage
in a more quantitative way.

In the first inflationary phase, after $\phi$ becomes smaller than $M$, the potential becomes of the form
\bea
V(\phi, \chi) \simeq \frac12 m^2 \phi^2 + \frac{\lambda M^4}{36} \left[1 -4 e^{-2 \frac{\chi}{M}} \right]\,,
\label{potential_turning}
\ena
for
$ 0 < \phi \ll M\,,\quad\chi \gg M$, 
where $m$ is defined by Eq.~(\ref{potential_TModel_asym_small}).
With this potential, the behavior of the background trajectory around $\phi=0$,
which we have called the turning phase, when $\Theta$ roughly changes from $\pi$ to $(3/2) \pi$,
is classified into two cases.
In one case, $\phi$ asymptotes smoothly to $0$ during the turn, 
while the motion towards $\chi=0$ becomes gradually important, which was shown in the right panel 
in Fig.~\ref{hubble_dottheta_BG}.
In the other case, the turn is accompanied by the oscillations in the $\phi$ direction
caused by the excitation of the heavy field.
As we have mentioned, since the duration  when $\phi$ is in the region 
$\phi > M$ is much longer than  the one in the region $\phi<M$, the existence of the turning phase
or whether the excitation of the heavy field occurs or not does not affect the expression 
(\ref{eom_phi_chi_slowroll_firstinflationaryphase}) describing the background dynamics
in the first inflationary phase. Regardless of this, as we will discuss in detail
in Appendix~\ref{sec:BD_TurningPhase}, the perturbations generated in inflation before the turn 
are affected if there is heavy field excitation. After this turning phase, 
the second inflationary phase, about which we have already discussed in the previous subsubsection,
starts.

\subsubsection{Correspondence between the initial field values and number of e-folding}

\begin{widetext}

\begin{figure}[htbp]
\begin{center}

\begin{tabular}{c}
\begin{minipage}{0.45\hsize}
\begin{center}
\includegraphics[width=4.5cm]{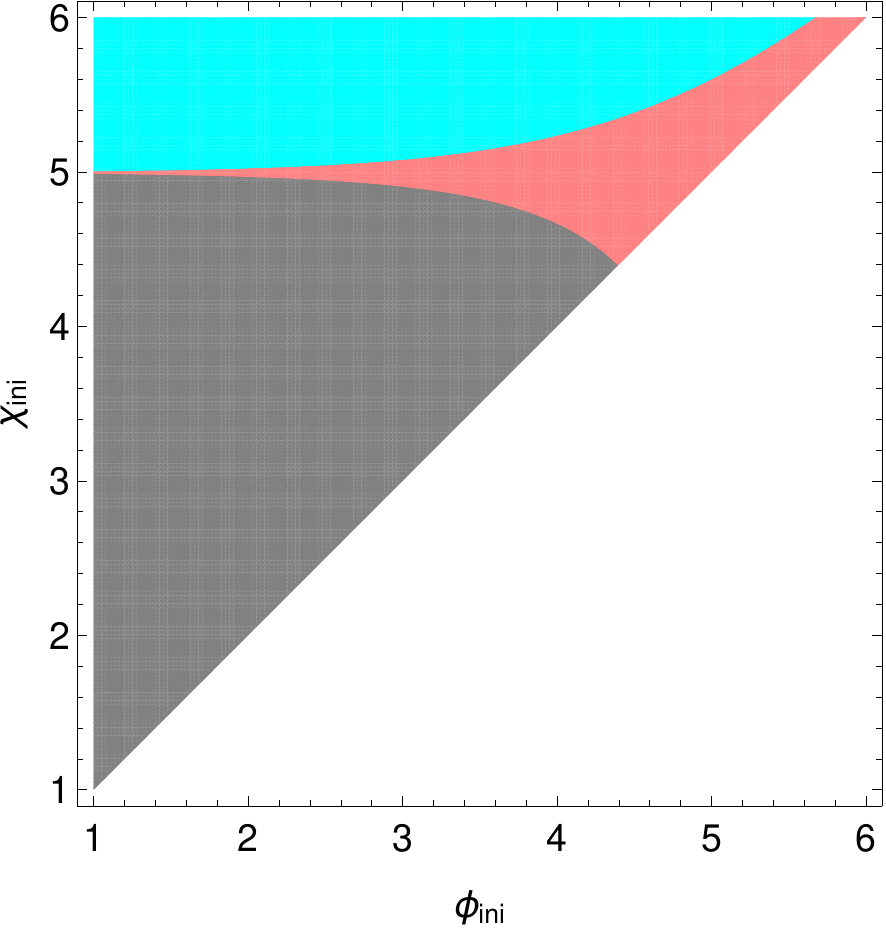}\\
\hspace{1.6cm}
\end{center}
\end{minipage}
\hskip 1cm
\begin{minipage}{0.45\hsize}
\begin{center}
\includegraphics[width=4.5cm]{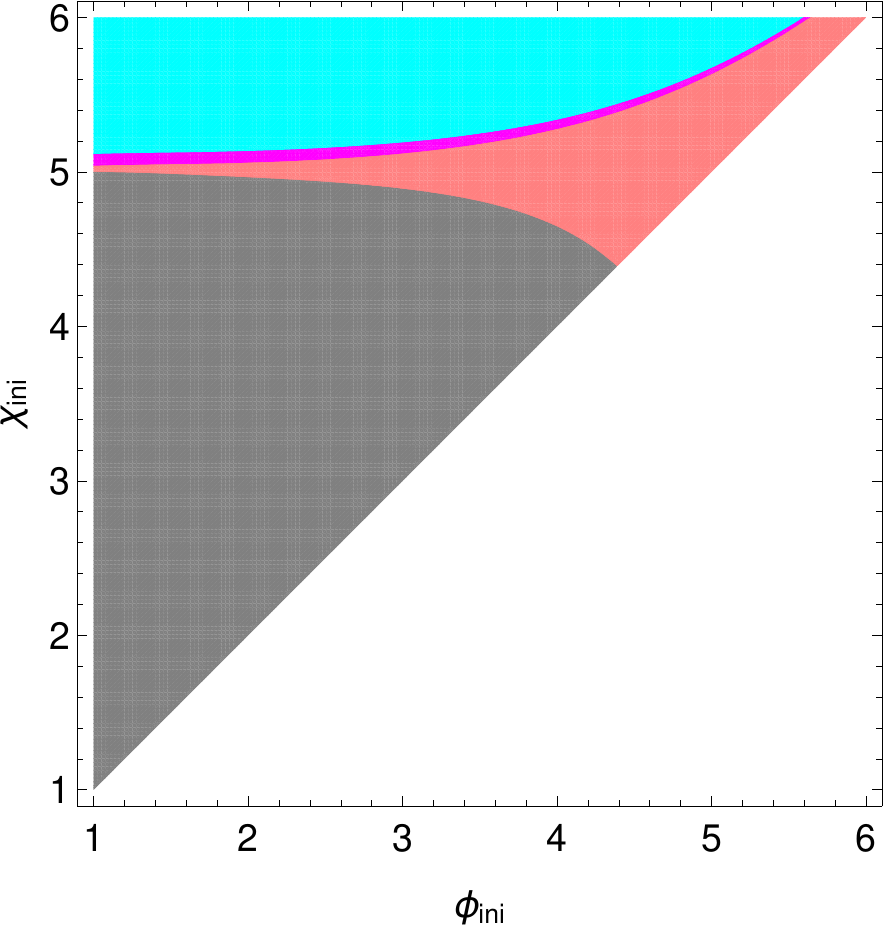}\\
\hspace{1.6cm}
\end{center}
\end{minipage}
\end{tabular}
\caption{
Classification of the region in $\varphi^I_{\rm ini}$ space based on 
in which phase the pivot scale of the recent CMB observations exits the horizon scale
with  $M=\sqrt{3} \Mp$.
In the gray part, $N_{\rm ini} < 60$, where the total number of $e$-folding during inflation
is insufficient. 
Among the other region in the light blue part, 
the scale exits the horizon scale in the second inflationary phase,
while in the light red part, it occurs in the first inflationary phase.
(Left)  The classification is based on $N_1$ defined by Eqs.~(\ref{concretevalue_N1_Nini}) and
(\ref{concretevalue_N1_Nini2}).
(Right) The classification is based on numerical calculations. 
The part in $\varphi^I_{\rm ini}$ space,
where the scale exits the horizon scale in the turning phase, is colored with purple.
}
\label{fatemsquare3_BG}

\end{center}

\end{figure}

\end{widetext}

Before moving to the part of  perturbations,
since we have obtained intuitive understanding of the integration constants
$N_1$, $N_{\rm ini}$ given by Eqs.~(\ref{concretevalue_N1_Nini}) and
(\ref{concretevalue_N1_Nini2}), 
it is helpful to discuss more on the correspondence between
the initial values $\varphi^I_{\rm ini}$ and two $e$-folds $N_1$, $N_{\rm ini}$.
Later in this paper, when we discuss the perturbations,
we assume that the pivot scale of the recent
CMB observations exits the horizon scale at $N=N_*=60$, 
where $*$ denotes that the quantity is evaluated 
when the scale exits the horizon scale, that is,
$60$ $e$-folds before the inflation ends.
This means that for given $M$ and $\lambda$, the region in $\varphi^I_{\rm ini}$ space
with $N_{\rm ini} < 60$ is ruled out (the gray region in Fig. \ref{fatemsquare3_BG}).
For the remaining region in $\varphi^I_{\rm ini}$ space that satisfies $N_{\rm ini} > 60$, 
since the property of the perturbations exiting the horizon scale
in the first inflationary phase and that in the second inflationary phase are completely different,
it is useful to classify the region
based on in which phase the scale of interest exits the horizon scale.
 In the left panel in Fig.~\ref{fatemsquare3_BG},
the classification is based on $N_1$ defined by Eq.~(\ref{concretevalue_N1_Nini}).
From the discussion in the previous subsection,
we expect that if $N_1 < 60$, $N=N_*=60$ is in the first inflationary phase (the light red region), 
while if $N_1 > 60$, $N=N_*= 60$ is in the second inflationary phase (the light blue region).
Since $N_1$ is  obtained analytically with the instantaneous turn assumption,
in order to be more precise,  in the right panel of  Fig.~\ref{fatemsquare3_BG}
we classify the region by specifying  in which phase the mode exits the horizon scale 
with given $\varphi^I_{\rm ini}$
based on the numerical calculations.
This shows that even though some small purple region corresponding to near $N_1 = 60$
 in the left panel is not clearly seen,
  in most part the classification based on $N_1$ is correct.
As we have already explained, we think that  the minor disagreement
comes from the existence of the finite turning phase
and the period when $\phi$ is in the region $\phi < M$ in the first inflationary phase.
We have adopted $M= \sqrt{3} \Mp$ here 
and it is worth mentioning that these plots are independent of $\lambda$
as $N_{\rm ini}$ and $N_{1}$ do not depend on $\lambda$.

\section{Dynamics of the primordial curvature perturbation
\label{sec:field_perturbations}}

In this section,  we calculate the primordial curvature perturbation
in the $\alpha$-attractor-type double inflation model that is verified by the recent CMB observations.
We also show that the  $\delta N$ formalism, where we can calculate the primordial curvature perturbation
analytically, can reproduce the numerical results very well.

\subsection{Perturbations of fields}

We decompose the scalar fields $\varphi^I$ into the background values  $\bar{\varphi}^I$
plus the perturbations $\delta \varphi^I$ as
\begin{eqnarray*}
\varphi^I = \bar{\varphi}^I + \delta \varphi^I\,.
\label{2of2}
\end{eqnarray*}
In the following, we will simply write the background homogeneous values as $\varphi^I$
as long as this does not cause confusion. Here, we restrict ourselves to the linear perturbation
and work with the spatially-flat gauge given by
\begin{eqnarray}
ds^2 = - \left( 1+2A \right) dt^2 +2a \partial_i B dx^i dt +a^2 \delta _{ij} dx^i dx^j\,,
\label{1of2}
\end{eqnarray}
where the scalar-type degrees of freedom are completely encoded in $\delta \varphi^I$
owing to the Hamiltonian and momentum constraints
\beann
&&
3 H^2 A+ \frac{H}{a}\partial^2 B=- \frac{1}{2 \Mp^2} 
 \left[\dot{\varphi}_I\left(\delta \dot{\varphi}^I - \dot{\varphi}^I A \right) + V_{,I} \delta \varphi^I \right]\,,
\label{constraints_flat}
\\
&&
H A = \frac{1}{2 \Mp^2} \dot{\varphi}_I \delta \varphi^I\,.
\label{constraints_flat2}
\enann

The quadratic action for $\delta \varphi^I$ can be obtained by expanding the action (\ref{action})
to quadratic order in perturbations. In terms of the
 Mukhanov-Sasaki variables 
 $v^I   = (v_\phi, v_\chi) \equiv a \delta \varphi^I$ \cite{Sasaki:1986hm,Mukhanov:1988jd}
 and  the conformal time  $\tau \equiv \int dt/a(t)$, 
it becomes 
\beann
S^{(2) }&=&\frac12 \int d \tau d^3 x \Big[
\delta_{IJ} (v^I)' (v^J)'- \delta_{IJ} \delta^{ij} \partial_i v^I \partial_j v^J
\nonumber \\
&&
~~~~~~~~~~~~
-a^2 \left(\mu_{IJ} -H^2 (2-\epsilon) \delta_{IJ} \right) v^I v^J  \Big]\,,
\label{second_order_action}
\enann 
where a prime denotes a derivative with respect to $\tau$,
the matrix $\mu_{IJ}$ corresponds to the (squared) mass matrix given by
\beann
\mu_{IJ} \equiv V_{,IJ} + \frac{1}{\Mp^2 a^3} \Biggl(\frac{a^3}{H} \dot{\varphi}_I \dot{\varphi}_J \Biggr)^{\cdot}\,.
\label{massmatrix}
\enann
Varying the action with respect to $v^I$ and moving to the Fourier space, we obtain the linear equations of motion,
\begin{eqnarray}
({v^I})'' + \left( k^2 - \frac{a^{''}}{a} \right) v^I + a^2 \mu_{IJ}   v^J = 0\,,
\label{8of2}
\end{eqnarray}
where $k$ is the comoving wave number. 
If we choose the initial time $\tau_0$ when the corresponding mode is on sufficiently small scales
satisfying $|a''/a|, |\mu_{IJ}| \ll k^2$ and thus Eqs.~(\ref{8of2}) can be approximated as 
\begin{eqnarray*}
({v^I})'' +k^2 v^I = 0\,,
\label{9of2}
\end{eqnarray*}
it is legitimate to choose the Bunch-Davies vacuum \cite{Bunch:1978yq} as an initial condition of our calculation:
\begin{eqnarray}
v^I (\tau_0, k) = \frac{1}{\sqrt{2 k}} e^{-i k \tau_0} e^I (k)\,.
\label{10of2}
\end{eqnarray}
Here $e^I (k)$ are independent unit Gaussian random variables and with $\langle \rangle$
to be the ensemble average they
satisfy
\beann
\langle  e^I (k) \rangle = 0\,,\quad\quad \langle  e^I (k) {e^J (k')}^* \rangle =\delta^{IJ} \delta (k-k')\,.
\enann

\subsection{Curvature perturbation and isocurvature perturbation\label{subsec:curvature_isocurvature}}

In order to compare the prediction of  linear perturbation in the $\alpha$-attractor-type double inflation 
with the CMB observations, it is necessary to relate $\delta \varphi^I$ to other perturbed quantities
describing the metric perturbations, which are also the physical degrees of freedom.
One of the useful quantities is $\mathcal{R}$ that is gauge-invariant and coincides with
curvature perturbation in the comoving slice  \cite{Lyth:1984gv}.
It can be shown that $\mathcal{R}$ is related to  the perturbation of the fields in the spatially-flat gauge
$\delta \varphi^I$ as\footnote{In Ref.~\cite{Gordon:2000hv}, the sign of $\mathcal{R}$ is chosen so that
$\mathcal{R} \equiv \psi - H/(\rho + P) \delta q$, where $\psi$ is the diagonal part of the spatial metric perturbation
defined by $a^2[(1-2 \psi) \delta_{ij}]dx^i dx^j$ and $\delta q$ is the energy flux for
the scalar fields $\delta q \equiv -\dot{\varphi}_I \delta \varphi^I$. In this paper, for the consistency
with the latter part, where we consider the $\delta N$ formalism, 
we choose the sign opposite to Ref.~\cite{Gordon:2000hv}.
Although this choice does not change the result as long as we consider the quantities related with
$\mathcal{P}_{\mathcal{R}}$, when we consider the primordial non-Gaussianity in Appendix~\ref{sec:PNG},
it affects the sign of $f_{\rm NL}$.}
\bea
\mathcal{R} = -\frac{H}{\dot{\sigma}^2} \dot{\varphi}_I \delta \varphi^I = -\frac{H}{\dot{\sigma}} \delta \sigma\,,
\label{comoving_cp}
\ena
where $\delta \sigma$ is the adiabatic perturbation defined based on the unit adiabatic vector $n^I$
given the first equation of  (\ref{def_unitvect_ad_en}) as
\bea
\delta \sigma \equiv n_I \delta \varphi^I\,.
\ena
The time derivative of Eq.~(\ref{comoving_cp}) is written as
\bea
\dot{\mathcal{R}} = -\frac{H}{\dot{H}} \frac{k^2}{a^2} \Psi -\frac{2 H}{\dot{\sigma}} \dot{\Theta} \delta s\,, 
\label{time_evol_comoving_cp}
 \ena 
 where $\Psi$ is  the Bardeen potential \cite{Bardeen:1980kt} that is related to $B$
 in the spatially-flat gauge given by Eq.~(\ref{1of2}) as
 \bea
 \Psi=-aHB\,,
 \ena 
 and $\delta s$ is the entropy perturbation defined in terms of the unit entropic vector
 $s^I$ given by Eq. (\ref{def_unitvect_ad_en2}) as
 \bea
 \delta s \equiv s_I \delta \varphi^I\,.
 \ena
 On sufficiently large scales, in the absence of the anisotropic stress, which holds 
 in the current model, the first term in Eq.~(\ref{time_evol_comoving_cp}) 
 is negligible. However, if the entropy perturbation is not suppressed and if the background trajectory
 changes the direction, i.e., $\dot{\Theta} \neq 0$, the curvature perturbation evolves
 even on sufficiently large scales.
In this paper,  assuming that the curvature perturbation at the end of inflation is 
 connected to the adiabatic perturbation at late time, we will follow the evolution of curvature perturbation
in the inflation and estimate the amplitude at the end of inflation
 through the power spectrum defined by
 \bea
 \mathcal{P}_{\mathcal{R}} = \frac{k^3}{2 \pi^2} |\mathcal{R}|^2\,.
 \label{def_Pzeta}
 \ena 
From the Planck result \cite{Ade:2015xua}, 
$\mathcal{P}_{\mathcal{R}}$ at the pivot scale 
$k_{*} = 0.05 {\rm Mpc}^{-1}$ is constrained as 
$(2.198 ^{+0.076}_{-0.085}) \times 10^{-9}$ with $1 \sigma$ errors. 
In order to constrain model parameters more, 
we also calculate the spectral index of $\mathcal{P}_{\mathcal{R}}$ 
at the same pivot scale defined by
\bea
n_s-1 \equiv \frac{d \ln  \mathcal{P}_{\mathcal{R}}}{d \ln k}\,,
\label{def_ns}
\ena
which is constrained as $(0.9655 \pm 0.0062)$  with $1 \sigma$ errors \cite{Ade:2015xua}. 
In single-field slow-roll inflation, $\mathcal{P}_{\mathcal{R}}$ and $n_s$ are expressed by
\bea
\mathcal{P}_{\mathcal{R}}  = \frac{H^2 _*}{8 \pi^2 \Mp^2 \epsilon_*}\,,\quad\quad
n_s = 1-2 \epsilon_* - \eta_*\,,
\label{observables_singlefield_slowroll}
\ena
where the quantities with $*$ are evaluated when the pivot scale exits the horizon scale.
As we will see later, for some cases, the estimation based on Eqs.~(\ref{observables_singlefield_slowroll})
is valid, while for the other cases, the multifield effects 
during inflation give modifications from the estimation based on Eqs.~(\ref{observables_singlefield_slowroll}).
We also calculate the tensor-to-scalar ratio $r$ defined by
\bea
r \equiv \mathcal{P}_{h}/ \mathcal{P}_{\mathcal{R}}\,,
\label{def_r}
\ena
where $\mathcal{P}_{h}=(2/\pi^2) (H_*^2/\Mp^2)$ is the power spectrum of the  primordial gravitational waves at the same scale 
\cite{Starobinsky:1979ty}.
The upper bound of $r$ from the Planck result is  $r < 0.10$ with $2 \sigma$ and
in single-field slow-roll inflation, $r$ is expressed by $r = 16 \epsilon_*$.

In this paper, in order to see how the entropy perturbation affects the dynamics of curvature perturbation
during inflation, we will also follow the evolution of entropy perturbation
during the inflation based on the power spectrum defined by
 \beann
 \mathcal{P}_{\mathcal{S}} = \frac{k^3}{2 \pi^2} |\mathcal{S}|^2\,,\quad\quad
  \mathcal{S} \equiv  \frac{H}{\dot{\sigma}}\delta s\,,
\label{def_isocurvature}
 \enann
 where $ \mathcal{S}$ is the isocurvature perturbation defined 
 so that it becomes dimensionless and has a similar normalization factor to the  curvature perturbation.
 One technical thing for calculating $\mathcal{P}_{\mathcal{R}}$, 
$\mathcal{P}_{\mathcal{S}}$ in double inflation is that we must treat 
$v_\phi$ and $v_\chi$
as two independent stochastic variables for the modes well inside the horizon as Eqs.~(\ref{10of2}).
To reflect this independent property, as pointed out by Ref.~\cite{Langlois:1999dw},
we must perform two numerical  computations. One computation with an initial condition that
$v_\phi$ is in the Bunch Davies vacuum and $v_\chi=0$ giving the solutions
$\mathcal{R}=\mathcal{R}_1$ and $\mathcal{S} = \mathcal{S}_1$ corresponds to $e^1 (k)$.
Another computation with an initial condition that $v_\phi = 0$ and $v_\chi$ is in the Bunch-Davies vacuum
giving the solutions $\mathcal{R}=\mathcal{R}_2$ and $\mathcal{S} = \mathcal{S}_2$
corresponds to $e^2 (k)$.
Then, we can obtain the correct  power spectra  in terms of $\mathcal{R}_1$, $\mathcal{R}_2$,
 $ \mathcal{S}_1$, and $\mathcal{S}_2$, as
 \beann
 &&
 \mathcal{P}_{\mathcal{R}} = \frac{k^3}{2 \pi^2} (|\mathcal{R}_1|^2 + |\mathcal{R}_2|^2)\,,\\
 &&
  \mathcal{P}_{\mathcal{S}} = \frac{k^3}{2 \pi^2} (|\mathcal{S}_1|^2 + |\mathcal{S}_2|^2)\,.
 \enann

\subsection{Typical Examples}

In double inflation models, the background trajectories depends on $\varphi^I _{\rm ini}$,
from which the resultant inflationary observables like 
$\mathcal{P}_{\mathcal{R}}$, $n_s$, $r$ depend on  $\varphi^I _{\rm ini}$.
Here, with fixed $M$ and $\lambda$ given by Eqs.~(\ref{parameters_expample_BG}) and (\ref{initialvalues_expample_BG}),
we follow the dynamics of the comoving curvature perturbation  numerically  and 
obtain $\mathcal{P}_{\mathcal{R}}$, $n_s$ and $r$ at the end of inflation 
for the pivot scale of the recent CMB observations, 
with some representative $\varphi^I _{\rm ini}$. 
In this paper, as we mentioned, we assume that the scale exits the horizon scale 
at $N=N_*=60$.

\begin{widetext}

\begin{figure}[h]
\begin{center}
\begin{tabular}{c}

\begin{minipage}{0.45\hsize}
\begin{center}
\includegraphics[width=4.5cm]{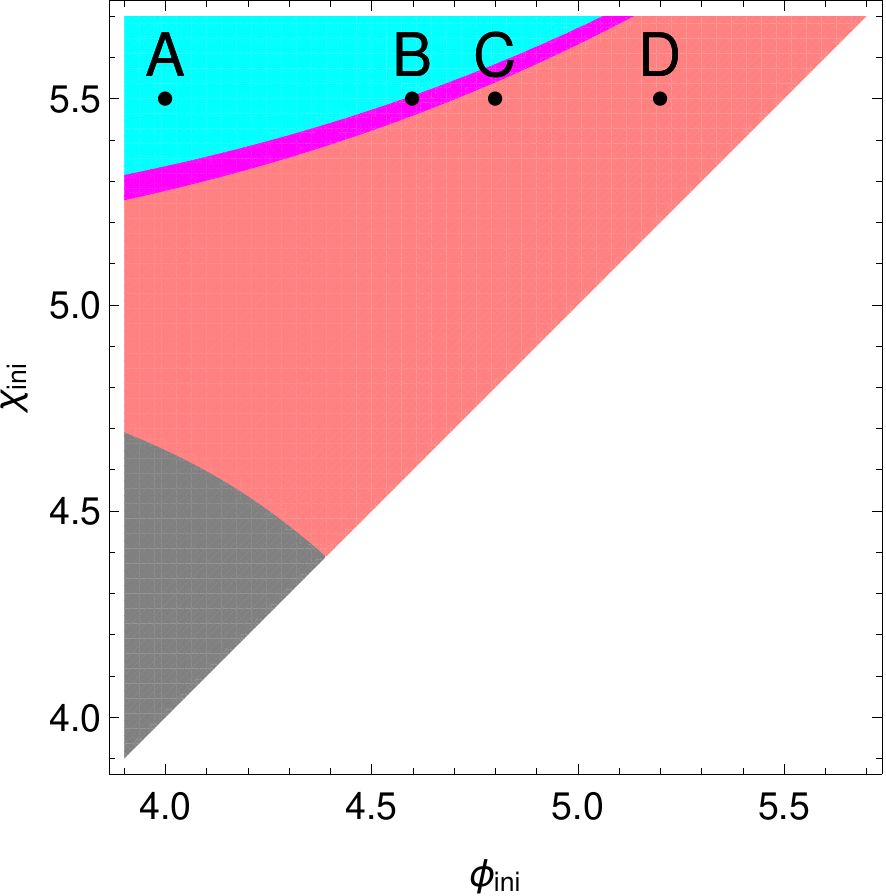}\\
\hspace{1.6cm}
\end{center}
\end{minipage}

\begin{minipage}{0.45\hsize}
\begin{center}
\includegraphics[width=4.5cm]{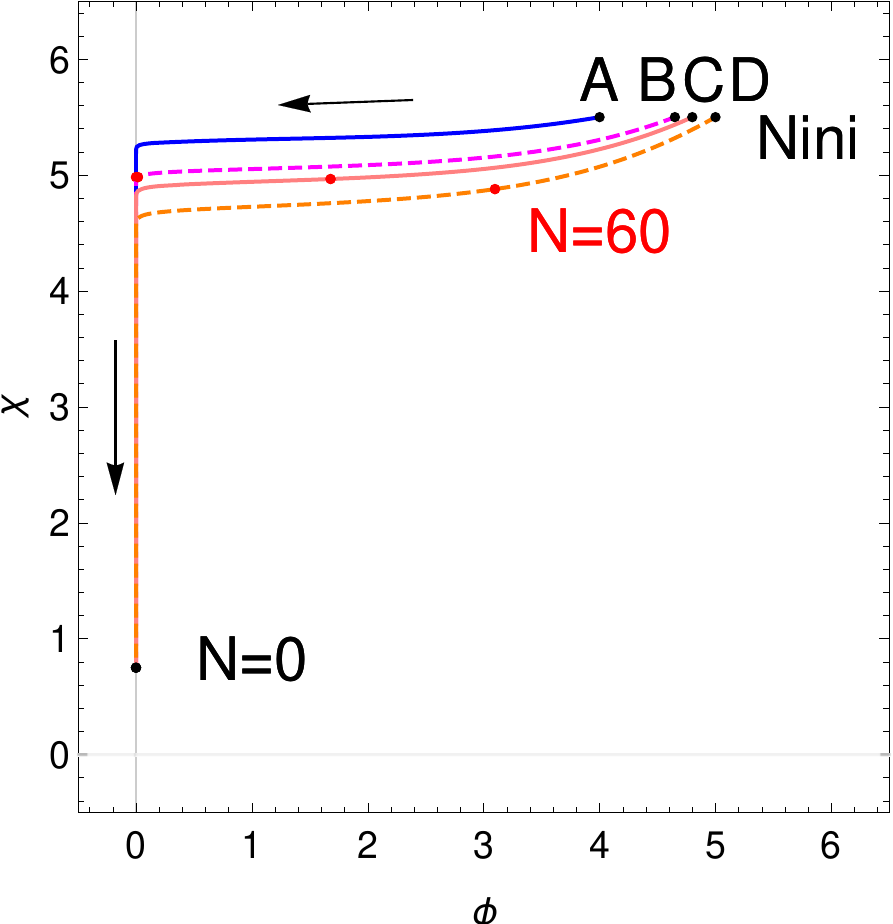}\\
\hspace{1.6cm}
\end{center}
\end{minipage}

\end{tabular}
\caption{ (Left) Four initial values $\varphi^I _{\rm ini}$ shown in the magnified
version of the right panel of Fig.~\ref{fatemsquare3_BG}, where the region in $\varphi^I _{\rm ini}$ space
is classified based on in which phase the pivot scale of the recent CMB observations exits the horizon scale, i.e., the light red and light blue regions correspond to
the horizon exits in the first and second inflation phases, respectively. The purple region is the marginal case.
(Right) The background trajectories with the four $\varphi^I _{\rm ini}$ shown in the left panel. 
The correspondences are as follows: (A) the solid-blue line, 
(B) the dashed-purple line, (C)  the solid pink line,  and  (D)
the dashed-orange line.
We also show the points $N=N_{*}=60$ on the trajectories by red circles.
\label{fig_summary_example}}
\end{center}
\end{figure}
\end{widetext}

To show the typical examples in double inflation, we 
consider the following
four initial values: \\
\beann
(\phi_{\rm ini}/\Mp, \chi_{\rm ini}/\Mp) =\left\{
\begin{array}{cccc}
 (4.00, &5.50)&~{\rm case}&{\rm A}\\
 (4.65, &5.50) &~{\rm case}&{\rm B}\\
  (4.80, &5.50)&~{\rm case}& {\rm C}\\
  (5.00, &5.50)&~{\rm case}& {\rm D}\\
\end{array}
\right.
\enann
which are plotted in Fig.~\ref{fig_summary_example}.
We also depict their background evolutions in the right panel.
In  case A, 
the horizon exit occurs in the first inflation phase, while 
it happens in the second inflation phase for cases C and D. 
Case B is the marginal.

The left panel in Fig.~\ref{pert_example_second_turn} shows the time evolution of 
$\mathcal{P}_{\mathcal{R}}$ and  $\mathcal{P}_{\mathcal{S}}$
in case A. 
In this example, from the left panel in Fig.~\ref{fig_summary_example},
the scale exits the horizon scale in the second inflationary phase
and since the turn occurs when the scale stays at deep inside the horizon scale, 
the effect of the turn is suppressed compared with $k^2$
from Eqs.~(\ref{8of2}). Therefore,  the resultant $\mathcal{P}_{\mathcal{R}}$, $n_s$ and $r$
are expected to coincide with the ones in the $\alpha$-attractor based on the single-field T-model 
given by $\chi$ with $\phi=0$. 
Indeed, with this $\varphi^I _{\rm ini}$, we obtain $\mathcal{P}_{\mathcal{R}} \simeq 2.213 \times 10^{-9}$,
$n_s \simeq 0.9667 $, $r \simeq 0.0017$. From Eqs.~(\ref{observables_singlefield_slowroll}) and 
making use of the relations $H^2 \simeq (\lambda M^4)/(108 \Mp^2)$,
Eqs.~(\ref{epsilonchi_etachi_secondinflationaryphase})
and (\ref{epsilonchi_etachi_secondinflationaryphase2}) with $N_*=60$ 
and Eqs.~(\ref{rel_slowroll_hubble_chi}), we can see that these really coincide with the ones  in single-field model.
These are consistent with the Planck result as we have chosen $M$ and $\lambda$
so that the resultant  $\mathcal{P}_{\mathcal{R}}$, $n_s$ and $r$ 
are so in the single-field model.
The right panel in Fig.~\ref{pert_example_second_turn} shows the time evolution of 
$\mathcal{P}_{\mathcal{R}}$ and  $\mathcal{P}_{\mathcal{S}}$ in case B.
In this example,  from the left panel in Fig.~\ref{fig_summary_example}, 
the scale exits the horizon scale in the turning phase
and around that time, the behavior of $\mathcal{R}_{\mathcal{R}}$ is slightly  modified
by the mixing with $\mathcal{S}$. However, since the mixing occurs
during just one or two $e$-foldings around the tern, the modification on 
 $\mathcal{P}_{\mathcal{R}}$, $n_s$ and $r$ from those in single-field model is small.
Actually, with this $\varphi^I _{\rm ini}$,  we obtain $\mathcal{P}_{\mathcal{R}} \simeq 2.213 \times 10^{-9}$,
$n_s \simeq 0.9665 $, $r \simeq 0.0017$, which are still consistent with the Planck result.

\begin{widetext}

\begin{figure}[h]
\begin{center}
\begin{tabular}{c}

\begin{minipage}{0.3\hsize}
\begin{center}
\includegraphics[width=5cm]{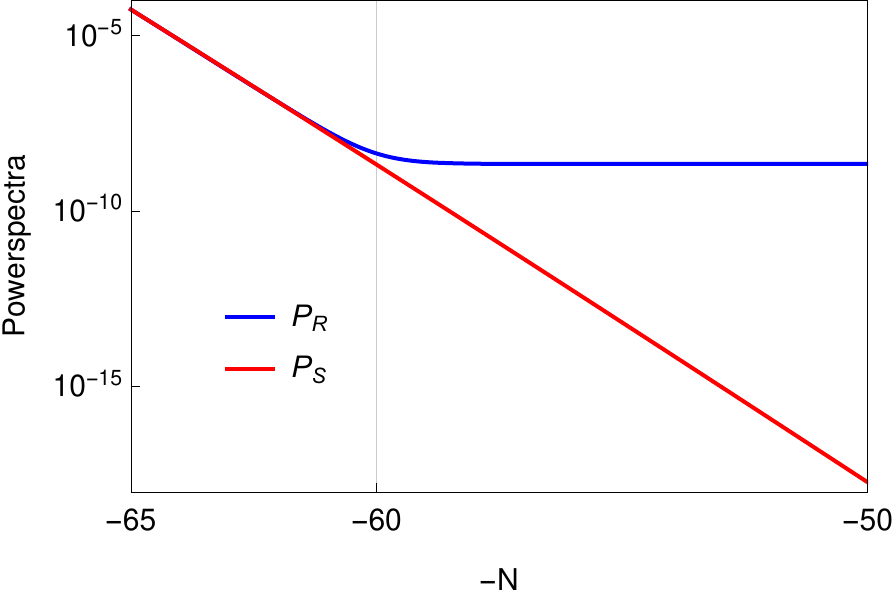}\\
\hspace{1.6cm}
\end{center}
\end{minipage}
\hskip 2cm
\begin{minipage}{0.3\hsize}
\begin{center}
\includegraphics[width=5cm]{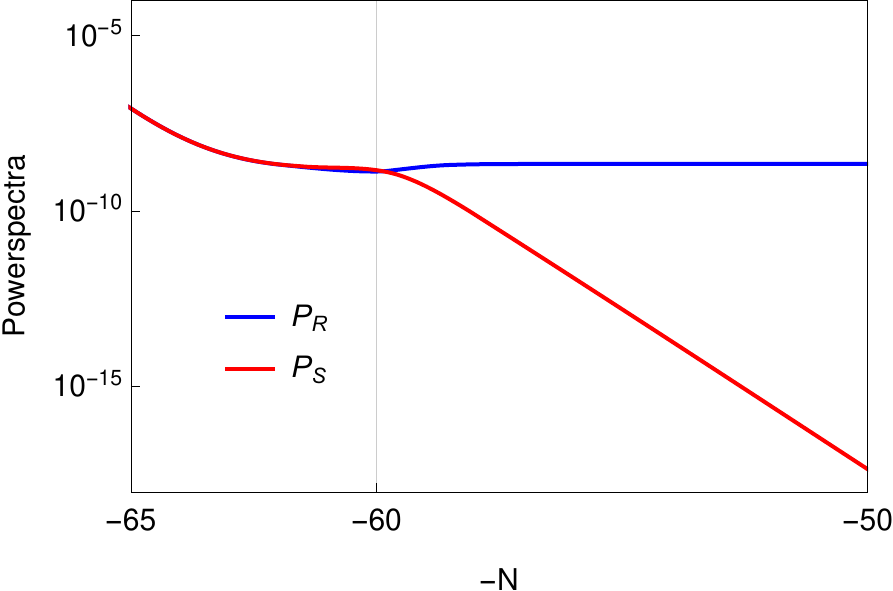}\\
\hspace{1.6cm}
\end{center}
\end{minipage}

\end{tabular}
\caption{Time evolution of $\mathcal{P}_{\mathcal{R}}$ and  $\mathcal{P}_{\mathcal{S}}$  in terms of $-N$,
where the blue and red lines correspond to $\mathcal{P}_{\mathcal{R}}$ and 
$\mathcal{P}_{\mathcal{S}}$,  
respectively.
(Left) Case A,
where the scale stays at deep inside the horizon scale at the turn.
The curvature perturbation conserves after the horizon exit  as in the single-field model. 
(Right) 
Case B, 
where the scale is about to exit the horizon scale at the turn.
Because of the mixing with the isocurvature perturbation around the horizon exit,
the curvature perturbation is slightly enhanced.\label{pert_example_second_turn}}
\end{center}
\end{figure}


\begin{figure}[htbp]
\begin{center}
\begin{tabular}{c}

\begin{minipage}{0.3\hsize}
\begin{center}
\includegraphics[width=5cm]{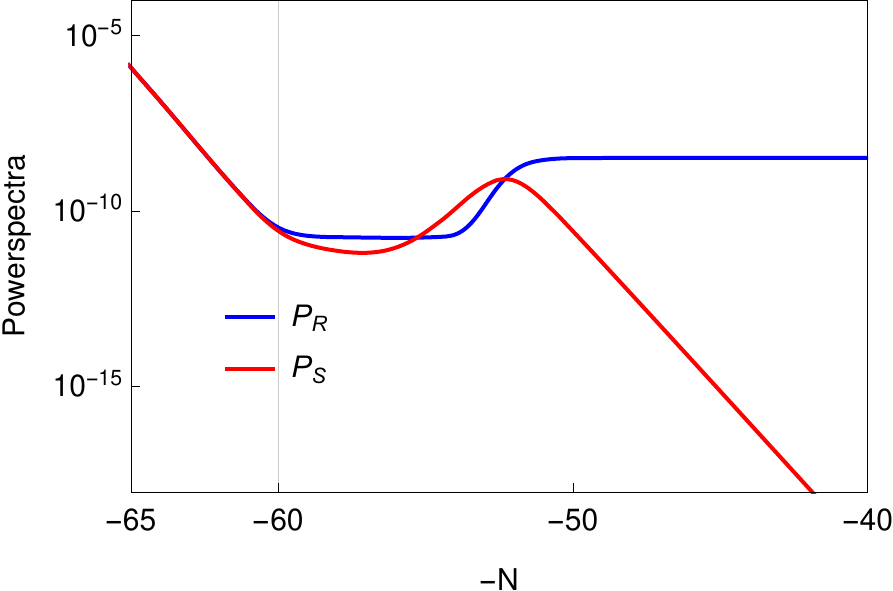}\\
\hspace{1.6cm}
\end{center}
\end{minipage}
\hskip 2cm
\begin{minipage}{0.3\hsize}
\begin{center}
\includegraphics[width=5cm]{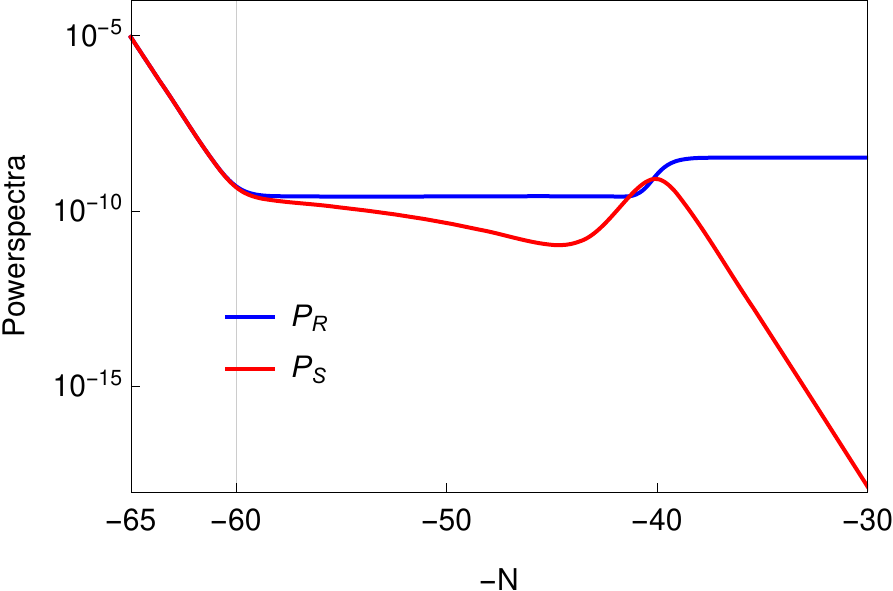}\\
\hspace{1.6cm}
\end{center}
\end{minipage}

\end{tabular}
\caption{Time evolution of $\mathcal{P}_{\mathcal{R}}$ and  $\mathcal{P}_{\mathcal{S}}$  in terms of $-N$,
where the blue and red lines correspond to $\mathcal{P}_{\mathcal{R}}$ and 
$\mathcal{P}_{\mathcal{S}}$,  
respectively. In both cases,  the scale 
exits the horizon scale before the turn and because of the mixing with the isocurvature perturbation
after the horizon exit, the curvature perturbation is enhanced.
(Left) Case C. 
(Right)  Case D.
Comparing these two, since the interval between the horizon exit and the turn is longer, 
the enhancement of the curvature perturbation is smaller in case D.}
\label{pert_example_first}
\end{center}

\end{figure}
\end{widetext}

Fig.~\ref{pert_example_first} shows the time evolution of 
$\mathcal{P}_{\mathcal{R}}$ and  $\mathcal{P}_{\mathcal{S}}$
in case C (left) and case D (right).
In these examples,  from the left panel in Fig.~\ref{fig_summary_example}, 
the scale exits the horizon scale in the first inflationary phase
and during the turn when the scale is on superhorizon scales, 
the curvature perturbation is sourced by the isocurvature perturbation.
At the end of inflation, in case C, 
we obtain $\mathcal{P}_{\mathcal{R}} \simeq 3.286 \times 10^{-9}$, 
$n_s \simeq 0.917$,  $r \simeq 0.0017$, while in case D, 
we obtain $\mathcal{P}_{\mathcal{R}} \simeq 3.352 \times 10^{-9}$, 
$n_s \simeq 0.974$,  $r \simeq 0.0021$.
Comparing these two cases,
since the isocurvature perturbation decays on superhorizon scales in the first inflationary phase,
the effect of the isocurvature perturbation is more significant for case C, 
where the interval between the horizon exit and the turn is shorter.
This explains the fact that the deviation of $n_s$ from one becomes larger for case C,
where the single-field model predicts $n_s \simeq 0.967$.
On the other hand,  $\mathcal{P}_{\mathcal{R}}$ is larger for case D, which can be explained
as follows. Since the first inflationary phase can be regarded as the single-field inflation
with $\phi$, Eq.~(\ref{observables_singlefield_slowroll}) is valid when we estimate
$\mathcal{P}_{\mathcal{R} *}$.
Comparing the two  $\varphi^I _{\rm ini}$, from Fig.~\ref{fig_summary_example},  
since the values of $\phi_*$ and $H_*$ are larger for  case D,
$\mathcal{P}_{\mathcal{R} *}$ is also larger for  case D.
Although $\mathcal{P}_{\mathcal{R}}$ experiences the enhancement by the end of inflation caused by the effect
of isocurvature perturbation,
which is  larger for  case C, the difference at $N=N_*=60$
can not be compensated.
Regardless of this discussion on the amplitude, as a result,
for  cases C and 
D, $\mathcal{P}_{\mathcal{R}}$ is too large to be consistent with the Planck result.

Next, in Fig.~\ref{pzeta_ns_pert_example}, we  show the numerical results
of $\mathcal{P}_{\mathcal{R}}$ and $n_s$ by changing $\phi_{\rm ini}$ 
but fixing the parameters $M$, $\lambda$
and the initial value $\chi_{\rm ini}$ as before. 
Here, for the constraint on  $\mathcal{P}_{\mathcal{R}}$ we adopt the $1 \sigma$ shown below 
Eq.~(\ref{def_Pzeta}), while for the constraints on $n_s$ we adopt the
$2 \sigma$ constraints shown in Ref.~\cite{Ade:2015xua}.
For $0 \leq \phi_{\rm ini}/\Mp \leq 4.58$, 
the scale exits the horizon scale in the second inflationary phase,
where $\mathcal{P}_{\mathcal{R}}$ and $n_s$ are same as the one in the single-field model. 
Since both $\mathcal{P}_{\mathcal{R}}$ and $n_s$ are consistent with the Planck result
with $\varphi^I _{\rm ini}$ in this region,
we omit to show the result for $\phi_{\rm ini}/\Mp < 4.00$ in Fig.~\ref{pzeta_ns_pert_example}.
For $4.58 < \phi_{\rm ini}/\Mp < 4.71$, the scale exits
the horizon scale near the turn, where the influence of the isocurvature perturbation
on the curvature perturbation is expected to be small. Actually, the observational constraints
of $\mathcal{P}_{\mathcal{R}}$ and $n_s$ give $\phi_{\rm ini}/\Mp < 4.71$
and $\phi_{\rm ini}/\Mp < 4.70$, respectively, which shows that
for $\phi _{\rm ini}$ in the most part of this region, 
the resultant $\mathcal{P}_{\mathcal{R}}$ and $n_s$ are consistent with Planck result.
For $4.71 \leq \phi_{\rm ini}/\Mp \leq 5.50$, 
 the scale exits the horizon scale in the first inflationary phase,
where because of the influence of the isocurvature perturbation and the change of the Hubble expansion rate
 at $N=N_*=60$, the modification of  $\mathcal{P}_{\mathcal{R}}$ and $n_s$ 
 from those in the single-field model is significant.
 For $\phi_{\rm ini}$ in this region, 
 the observational constraints on  $\mathcal{P}_{\mathcal{R}}$ and $n_s$ give
 $4.84 \leq \phi_{\rm ini}/\Mp$ and  $  \phi_{\rm ini}/\Mp\geq 5.42$, respectively.

In the left panel of Fig.~\ref{r_pert_example}, 
similarly, we show the numerical results of $r$  by changing $\phi_{\rm ini}$.  
For $\phi_{\rm ini}/\Mp \leq 4.7$, where $\mathcal{P}_\mathcal{R}$ and $n_s$
are the ones in the single-field inflation, $r$ is almost $0.00167$,
consistent with Eq.~(\ref{rel_ns_r_alphaattractor_def_N}) 
with our parameter set, $N=N_*=60$ and $\alpha=M^2/(6 \Mp^2)=1/2$, which is valid in the single-field inflation.
On the other hand, for $\phi_{\rm ini}/\Mp \geq 4.7$, $r$ is different from the one in the single-field case
caused by the multifield effects.
Although we find that $r$ is always sufficiently small to be consistent with the Planck result
and this result itself cannot give additional constraints, the information of $r$
is also important observationally. As we have explained previously, in the case of single-field $\alpha$-attractor
model, from Eq.~(\ref{rel_ns_r_alphaattractor_def_N}), there is a universal relation between $n_s$ and $r$,
that is, $(1-n_s)/r = N/(6 \alpha)$. In the right panel, we show the numerical result of $(1-n_s)/r$,
 by changing $\phi_{\rm ini}$.  With our parameter set, 
 the quantity is about $20$ in the single-field model and this result shows that the
universal relation for $\alpha$-attractors between $n_s$ and $r$ in the single-field model
no longer valid when the multifield effects is important. Making use of this fact,
in principle, we can distinguish between the single-field model and this double inflation model.

 Although it seems that multifield effects are so complicated that we must rely on numerical calculations
 for the quantitative understanding, as we will show in the next subsection,
 we can reproduce the results analytically with a very high accuracy.

\begin{widetext}

\begin{figure}[htbp]
\begin{center}
\begin{tabular}{c}

\begin{minipage}{0.3\hsize}
\begin{center}
\includegraphics[width=5cm]{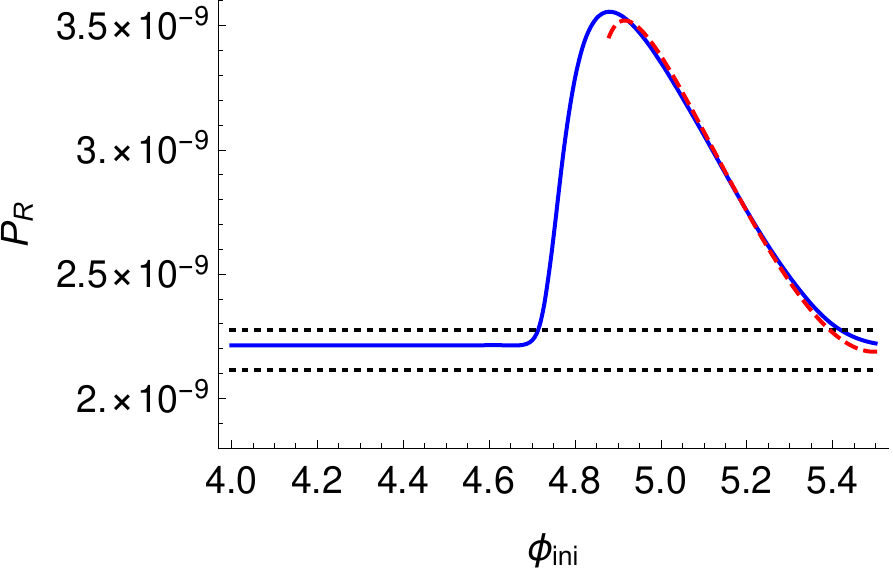}\\
\hspace{1.6cm}
\end{center}
\end{minipage}
\hskip 2cm
\begin{minipage}{0.3\hsize}
\begin{center}
\includegraphics[width=5cm]{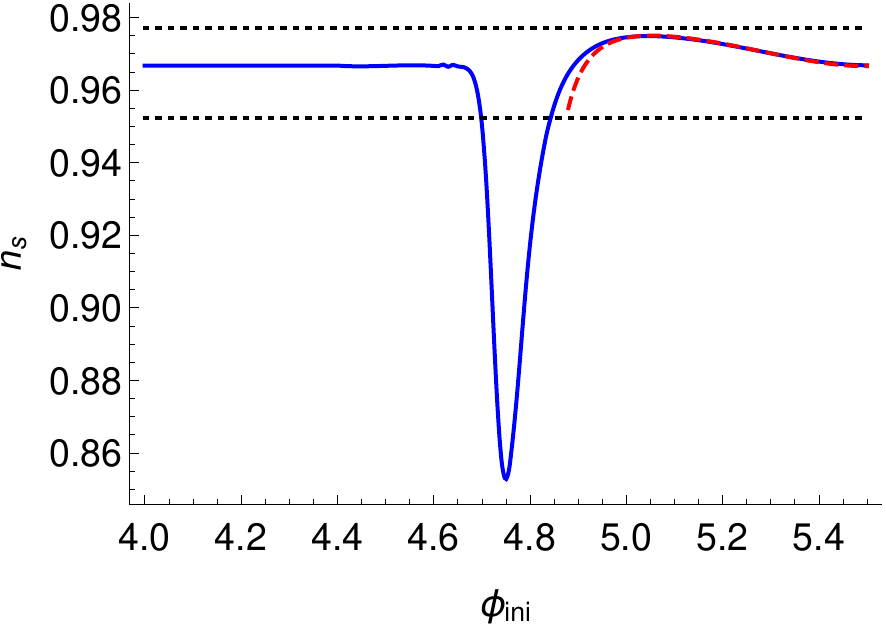}\\
\hspace{1.6cm}
\end{center}
\end{minipage}

\end{tabular}
\caption{$\phi_{\rm ini}$ dependence of $\mathcal{P}_{\mathcal{R}}$ (left) and $n_s$ (right) 
with $M=\sqrt{3} \Mp\,,  \lambda = 2.20 \times 10^{-10}$\,,
and  $\chi_{\rm ini}/\Mp = 5.50$.
With this $M$, $\lambda$ and $\chi_{\rm ini}$  the scale exits the horizon scale
in the second inflationary phase
for $0 \leq  \phi_{\rm ini}/\Mp \leq 4.58$,  near the turn for $4.58 < \phi_{\rm ini}/\Mp  < 4.71$
and the first inflationary phase for $4.71 \leq \phi_{\rm ini}/\Mp \leq 5.50$. In both panels,
the blue and red dashed lines correspond to the numerical calculation
and analytic $\delta N$ formalism, respectively.  We also show the current upper and lower bounds 
based on the Planck result with black dotted lines.}
\label{pzeta_ns_pert_example}
\end{center}

\end{figure}

\begin{figure}[htbp]
\begin{center}
\begin{tabular}{c}

\begin{minipage}{0.3\hsize}
\begin{center}
\includegraphics[width=5cm]{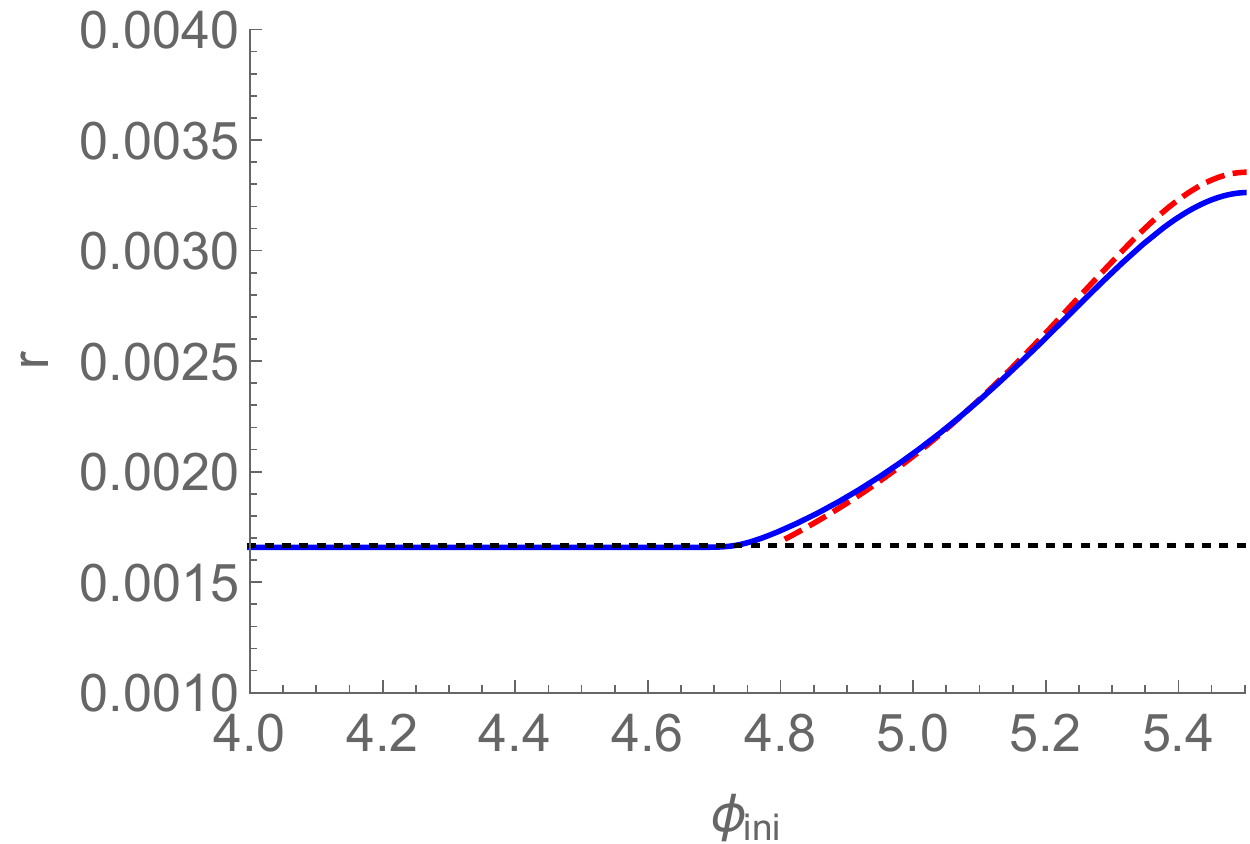}\\
\hspace{1.6cm}
\end{center}
\end{minipage}
\hskip 2cm
\begin{minipage}{0.3\hsize}
\begin{center}
\includegraphics[width=5cm]{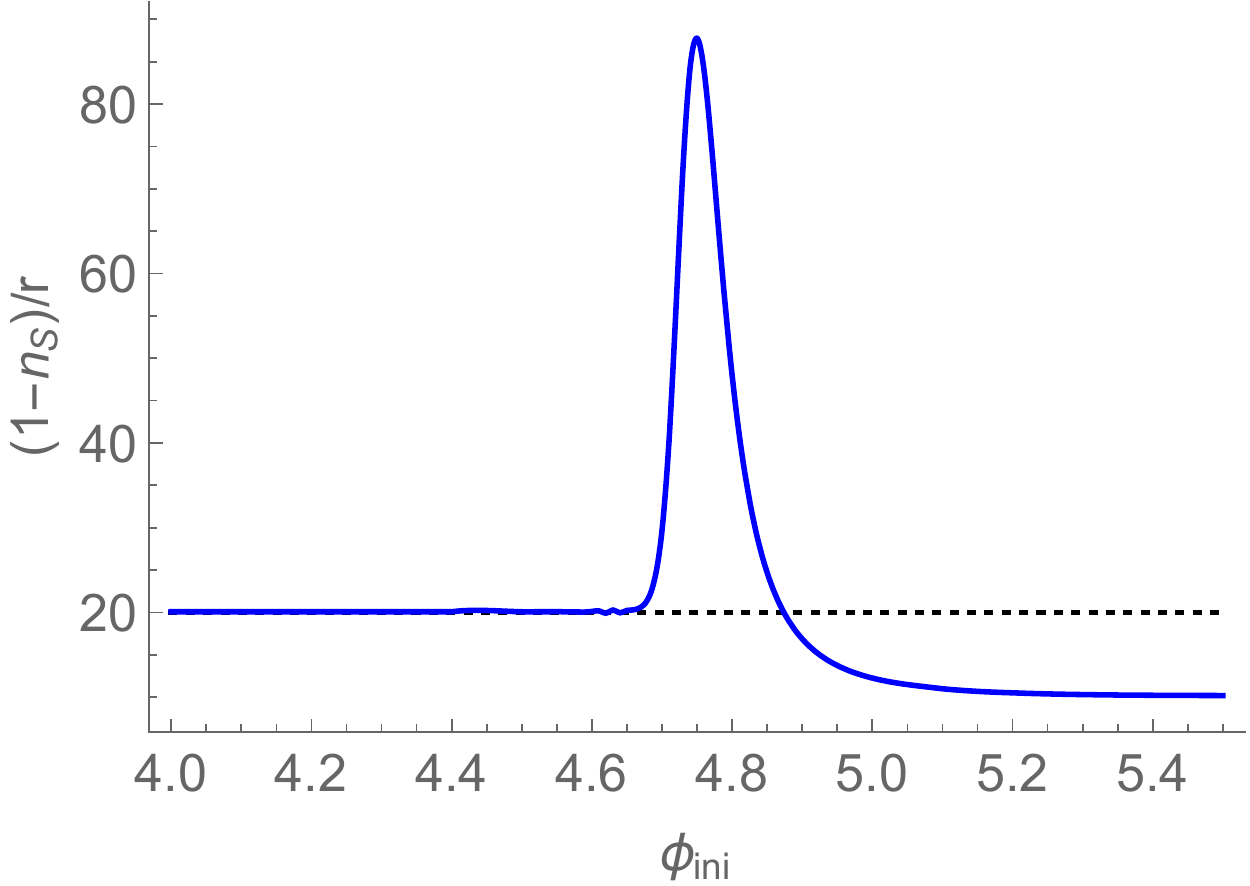}\\
\hspace{1.6cm}
\end{center}
\end{minipage}

\end{tabular}
\caption{(Left) $\phi_{\rm ini}$ dependence of $r$ 
with $M=\sqrt{3} \Mp\,,  \lambda = 2.20 \times 10^{-10}$\,,
and  $\chi_{\rm ini}/\Mp = 5.50$.
With this $M$, $\lambda$ and $\chi_{\rm ini}$  the scale exits the horizon scale
in the second inflationary phase
for $0 \leq  \phi_{\rm ini}/\Mp \leq 4.58$,  near the turn for $4.58 < \phi_{\rm ini}/\Mp  < 4.71$
and the first inflationary phase for $4.71 \leq \phi_{\rm ini}/\Mp \leq 5.50$. 
The blue and red dashed lines correspond to the numerical calculation
and analytic $\delta N$ formalism, respectively. (Right)  $\phi_{\rm ini}$ dependence of $(1-n_s)/r$,
where the blue line corresponds to the numerical calculation. In both panels, we also show
the values in the single-field model 
from Eqs.~(\ref{rel_ns_r_alphaattractor_def_N}) with our parameter set
$N=60$, $\alpha=1/2$ by black dotted line. }
\label{r_pert_example}
\end{center}

\end{figure}
\end{widetext}

\subsection{$\delta N$ formalism}

In  the previous subsection, we have found
two regions in the initial values $\phi _{\rm ini}$
 with fixed $M\,,\lambda$, and $\chi_{\rm ini}/\Mp$ 
in order to
obtain a consistent result with the CMB data.
In one part,  we can evaluate $\mathcal{P}_{\mathcal{R}}$ and $n_s$ based on the 
single-field model consisting of only the second inflationary phase.
In the other part, $\mathcal{P}_{\mathcal{R}}$ is enhanced compared with that in
the single-field model caused by the mixing with isocurvature perturbation on super horizon scales
as well as the change of $H$ at the horizon exit.
We also numerically calculated $\mathcal{P}_{\mathcal{R}}$, $n_s$  and $r$
to estimate these multifield effects there.
Here, we provide a simple way to understand these multifield effects
based on the $\delta N$ formalism \cite{Starobinsky:1986fxa,Sasaki:1995aw,Sasaki:1998ug}, which turns out to be very powerful in the $\alpha$-attractor-type double inflation model. 
According to the $\delta N$ formalism, on super horizon scales,
$\mathcal{R}$ evaluated at some time $t=t_{\rm f}$ coincides with the perturbation of the
number of $e$-foldings from an initial spatially flat hypersurface at $t=t_{\rm i}$ to a final comoving
hypersurface at $t=t_{\rm f}$, i.e.,
\beann
&&\mathcal{R} (t_f, {\bf x}) \simeq \delta N (t_{\rm f}, t_{\rm i}, {\bf x})
\equiv \mathcal{N} (t_{\rm f}, t_{\rm i}, {\bf x}) - N (t_{\rm f}, t_{\rm i})\,,
\enann
with
\beann
&&
\mathcal{N} (t_{\rm f}, t_{\rm i}, {\bf x}) \equiv \int_{t_{\rm i}} ^{t_{\rm f}}  \mathcal{H} (t, {\bf x}) dt\,,\\\
&&
N  (t_{\rm f}, t_{\rm i}, {\bf x}) \equiv \int_{t_{\rm i}} ^{t_{\rm f}} H (t) dt\,,
\enann
where $ \mathcal{H} (t, {\bf x})$ is the inhomogeneous Hubble expansion rate.
Since the local expansion on sufficiently large scales behaves like a locally homogeneous 
and isotropic Universe  (separate universe approach) \cite{Salopek:1990jq,Wands:2000dp}, 
we can calculate $\delta N$ on large scales using the homogeneous equation of motion. 
Suppose that we consider inflation with multiple scalar fields $\varphi^I$ and
choose the initial time $t_{\rm i}$ to be during inflation, when some scale exits the Hubble horizon,
$k_i = a H |_{t_{\rm i}}$, and $t_{\rm f}$ as some time 
($t_{\rm f} > t_{\rm i}$) during or after inflation when $\mathcal{R}$ has become constant.
Then,  we can regard $\mathcal{N}$ as a function of the configuration of the fields
on the spatially flat hypersurface at $t=t_{\rm i}$, $\varphi^I (t_{\rm i}, {\bf x})$ and $t_{\rm f}$,
\beann
\mathcal{N} (t_{\rm f}, t_{\rm i}, {\bf x}) = \mathcal{N} (t_{\rm f}, \varphi^I (t_{\rm i}, {\bf x}))\,.
\enann
Expanding the above expression in terms of the fields' fluctuations on the spatially flat hypersurface
$\delta \varphi^I$ at $t=t_{\rm i}$, we obtain
\beann
\mathcal{R} (t_{\rm f}, {\bf x}) \simeq \delta N (t_{\rm f}, t_{\rm i}, {\bf x}) \simeq N_{,I} 
\delta \varphi^I (t_{\rm i}, {\bf x})\,,
\enann
with
\bea
 N_{,I} \equiv \frac{\partial N}{\partial \varphi^I}
\biggr|_{t=t_{\rm i}}\,.
\label{rel_deltan}
\ena
Here, $N_{,I}$ is the derivative of the unperturbed number of $e$-foldings
$N(t_{\rm f}, t_{\rm i})$, with respect to the unperturbed values of the fields at $t_i$.
Moving to the Fourier space and if we take $\varphi^I$ as uncorrelated stochastic variables
with scale invariant spectrum of massless scalar fields in de-Sitter spacetime at $t=t_{\rm i}$ 
much before the turn so that 
$\mathcal{P}_{\varphi^I} (t_{\rm i}) \equiv \mathcal{P} (t_{\rm i}) = H^2 / (4 \pi^2) |_{t=t_{\rm i}}$,
we can obtain
\bea
\mathcal{P}_{\mathcal{R}} (t_{\rm f}) = N^{,I} N_{,I}\mathcal{P} (t_i) \simeq N^{,I} N_{,I} \frac{V (t_i)}{12 \pi^2 \Mp^2}\,,
\label{pzeta_deltan_gen}
\ena
where for the last equality, we have used the slow-roll approximation.
 On the other hand, since the tensor perturbations are conserved on super Hubble scales,
the power spectrum of the tensor perturbations is given by 
$\mathcal{P}_h (t_f)=\mathcal{P}_h (t_i)=(2/\pi^2) (H^2 /\Mp^2)|_{t=t_{\rm i}}=(8\mathcal{P} (t_i))/\Mp^2$, which gives
\bea
r(t_f) = \frac{8}{\Mp^2 N^{,K} N_{,K}}\,.
\label{r_deltan}
 \ena 
Applying Eq.~(\ref{def_ns}) to Eq.~(\ref{pzeta_deltan_gen}), we can also obtain the spectral index $n_s$
by the $\delta N$ formalism, and when the slow-roll approximation is valid, it is given by \cite{Sasaki:1995aw}
\bea
n_s (t_f) & =& 1-2 \epsilon (t_i) -\frac{2}{\Mp^2 N^{,K} N_{,K}} 
\nonumber \\
&&
+ \frac{2 \Mp^2 N^{,I} N^{,J} V (t_i) _{,IJ}}{V (t_i) N^{,K} N_{,K}}\,.
\label{ns_deltan}
 \ena 

Next, we apply the $\delta N$ formalism to the $\alpha$-attractor-type double inflation model, 
especially for the cases with
significant multifield effects, where the scale exits the horizon scale
during the first inflationary phase. 
From the discussion in Sec.~\ref{sec:background} C, 
if we choose $t_i$ as $t_{\rm ini}$ and $t_f$ as the end of inflation,
$N_{\rm ini}$ given Eqs.~(\ref{concretevalue_N1_Nini}) and
(\ref{concretevalue_N1_Nini2}) serves as $N$
and $\varphi^I _{\rm ini}$ serves as $\varphi^I$ in the $\delta N$ formalism.
Regardless of this, since the scale exits the horizon scale
not at $t_{\rm ini}$ but at $t_*$, 
the correct choice to calculate $\mathcal{P}_{\mathcal{R}}$ and $n_s$ 
by the $\delta N$ formalism is $t_{\rm i} = t_{*}$.
Therefore, for this purpose,
it is necessary to express $N_{*}$ in terms of 
$\varphi^I _{*}$ and in general multifield inflation models
we need to integrate the background equations numerically.
However, in this model, for a given set of $\varphi^I _{\rm ini}$, 
we have already specified the background trajectory in the first inflationary phase as 
Eqs.~(\ref{phi_chi_slowroll_firstinflationaryphase_ito_N})
and (\ref{phi_chi_slowroll_firstinflationaryphase_ito_N2}), with which we can obtain
\bea
&&
\phi_* = \frac{M}{2} \ln \biggl[\frac{8 \Mp^2}{M^2} N_* -\frac{1}{2} +\frac12 e^{2 \frac{\phi_{\rm ini}}{M}}
-\frac12 e^{2 \frac{\chi_{\rm ini}}{M}}\biggr]\,,
\label{phiexit_chiexit_deltan}
\\
&&
\chi_* = \frac{M}{2} \ln \biggl[\frac{8 \Mp^2}{M^2} N_* -\frac{1}{2} +\frac12 e^{2 \frac{\chi_{\rm ini}}{M}}
-\frac12 e^{2 \frac{\phi_{\rm ini}}{M}}\biggr]\,,
\label{phiexit_chiexit_deltan2}
\ena
where we have used Eqs.~(\ref{concretevalue_N1_Nini}) and
(\ref{concretevalue_N1_Nini2}) to express $N_{\rm ini}$ in terms of
$\varphi^I _{\rm ini}$ and we 
can substitute $N_* =60$. Just repeating the discussion in
Sec.~\ref{sec:background} C with the replacements $N_{\rm ini} \to N_*$ and 
$\varphi^I _{\rm ini} \to \varphi^I _*$, we can express $N_*$ in terms of $\varphi^I_*$ as
\bea
N_* = \frac{M^2}{16 \Mp^2} \left(1 + e^{2 \frac{\chi_*}{M}} + e^{2 \frac{\phi_*}{M}} \right)\,,~~
\label{nexit_deltan}
\ena
with $\varphi^I _*$ given by Eqs.~(\ref{phiexit_chiexit_deltan}) and
 (\ref{phiexit_chiexit_deltan2}).
Then, by just regarding $N_*$ as $N$ and $\varphi^I _*$ as $\varphi^I$ 
in  Eqs.~(\ref{pzeta_deltan_gen}) and (\ref{ns_deltan}), 
we can calculate
$\mathcal{P}_{\mathcal{R}}$ and $n_s$ by the $\delta N$ formalism analytically.
Actually, $\mathcal{P}_{\mathcal{R}}$ at the end of inflation is given by
\beann
\mathcal{P}_{\mathcal{R}}  \simeq \frac{1}{12  \pi^2 \Mp^2}
\left[({N_*}_{,\phi_*})^2 + ({N_*}_{,\chi_*})^2\right] V_*\,,
\enann
 with
 \bea
V_* \simeq \frac{\lambda M^4}{18}\left[1-\frac{M}{4 \Mp^2} ({N_*}_{,\phi_*})^{-1}
-\frac{M}{4 \Mp^2} ({N_*}_{,\chi_*})^{-1}\right]\,,
\label{pzeta_deltan_example}
\ena
where ${N_*}_{,\phi_*}$ and ${N_*}_{,\chi_*}$ are given by
\bea
{N_*}_{,\phi_*} = \frac{M}{8 \Mp^2} \left(\frac{8 \Mp^2}{M^2} N_{*} -\frac12 + \frac12 e^{2 \frac{\phi_{\rm ini}}{M}}
-\frac12e^{2 \frac{\chi_{\rm ini}}{M}}\right)\,,
\label{delNoverdelphi_delNoverdelchi}
\\
{N_*}_{,\chi_*} = \frac{M}{8 \Mp^2} \left(\frac{8 \Mp^2}{M^2} N_{*} - \frac12+ \frac12 e^{2 \frac{\chi_{\rm ini}}{M}}
-\frac12e^{2 \frac{\phi_{\rm ini}}{M}}\right)\,.
\label{delNoverdelphi_delNoverdelchi2}
\ena
Similarly,  $r$ and
$n_s$ at the end of inflation is given by
\bea
r&=& \frac{8}{\Mp^2 (({N_*}_{,\phi_*})^2 + ({N_*}_{,\chi_*})^2)}\,,
\label{r_deltan_example}\\
n_s  &=& 1 - 2 \epsilon_* -\frac{2}{\Mp^2 (({N_*}_{,\phi_*})^2 + ({N_*}_{,\chi_*})^2)} 
\nonumber \\
&+&
\frac{2 \Mp^2 (({N_*}_{,\phi_*})^2 {V_*}_{,\phi_* \phi_*}
+ ({N_*}_{,\chi_*})^2  {V_*}_{,\chi_* \chi_*}    ) }{V_*  (({N_*}_{,\phi_*})^2 + ({N_*}_{,\chi_*})^2)}\,,
\label{ns_deltan_example}
\ena
where under the slow-roll approximation, the quantities $\epsilon_*$, ${V_*}_{,\phi_* \phi_*}$
and ${V_*}_{,\chi_* \chi_*}$ can be expressed as
\beann
 &&\epsilon_* =\frac12 \Mp^2 \frac{({V_{*}}_{,\phi_*})^2+({V_{*}}_{,\chi_*})^2 }{V_{*}^2}\,,\\
 &&
 {V_{*}}_{,\phi_*} = \frac{\lambda M^4}{36 \Mp^2} ({N_*}_{,\phi_*})^{-1}\,,\quad\quad
  {V_{*}}_{,\chi_*} = \frac{\lambda M^4}{36 \Mp^2} ({N_*}_{,\chi_*})^{-1}\,,\\
  && {V_{*}}_{,\phi_* \phi_*} = -\frac{\lambda M^3}{18 \Mp^2} ({N_*}_{,\phi_*})^{-1}\,,
~\,
  {V_{*}}_{,\chi_* \chi_*} = -\frac{\lambda M^3}{18 \Mp^2} ({N_*}_{,\chi_*})^{-1}\,.
\enann
From these, as long as the mode exits the horizon scale in the first inflationary phase,
based on the $\delta N$ formalism, the resultant $\mathcal{P}_{\mathcal{R}}$ and $n_s$
can be expressed analytically in terms of $M$, $\lambda$ and $\varphi^I _{\rm ini}$. 
The plots in Fig.~\ref{pzeta_ns_pert_example} show that  this method can reproduce
the numerical results very well 
for $\varphi^I _{\rm ini}$ in the region giving $n_s$ consistent with the Planck result.

\section{Observational constraints
\label{sec:constraints}}

With the tendency explained in Sec.~\ref{sec:field_perturbations} C in mind, 
we impose observational constraints  on the $\alpha$-attractor-type double inflation model based on the Planck result.
For this, we take into account the constraints on
$\mathcal{P}_{\mathcal{R}}$ and $n_s$  only, as that on $r$ does not give an additional constraint.
As in the previous sections, if we assume that the initial velocity of the fields
obey the slow-roll approximation, the predictions on $\mathcal{P}_{\mathcal{R}}$ and $n_s$
in this model depend on $M$, $\lambda$, and $\varphi^I _{\rm ini}$.
Here, in order to constrain them, first we fix $M$
and specify possible $\lambda$ for given $\varphi^I_{\rm ini}$.
First we set $M=\sqrt{3} \Mp$ so that
the analysis in the previous sections are included.
Then we repeat similar procedures for different $M$ to see 
``$M$ dependence'' of the observational constraints,
where we consider the cases with $M=\Mp$ as an example of smaller $M$
and $M=\sqrt{6} \Mp$  as that of larger $M$.

\subsection{Case with $M=\sqrt{3} \Mp$}

\begin{widetext}

\begin{figure}[htbp]
\begin{center}
\begin{tabular}{c}

\begin{minipage}{0.45\hsize}
\begin{center}
\includegraphics[width=4.5cm]{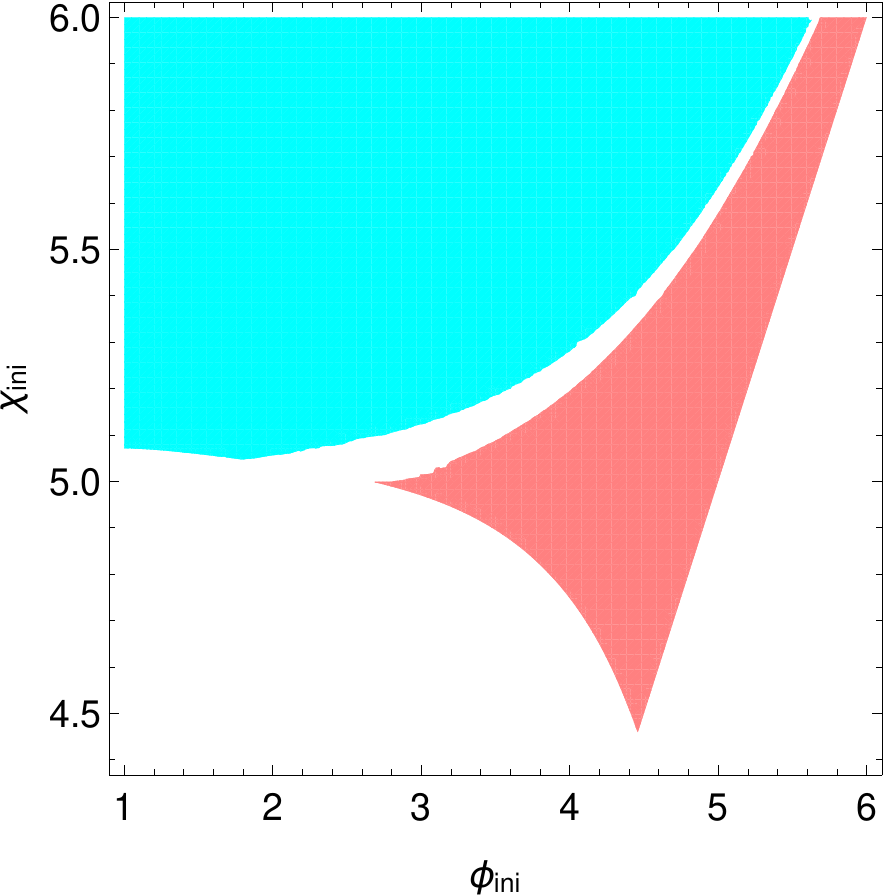}\\
\hspace{1.6cm}
\end{center}
\end{minipage}

\begin{minipage}{0.45\hsize}
\begin{center}
\includegraphics[width=5.5cm]{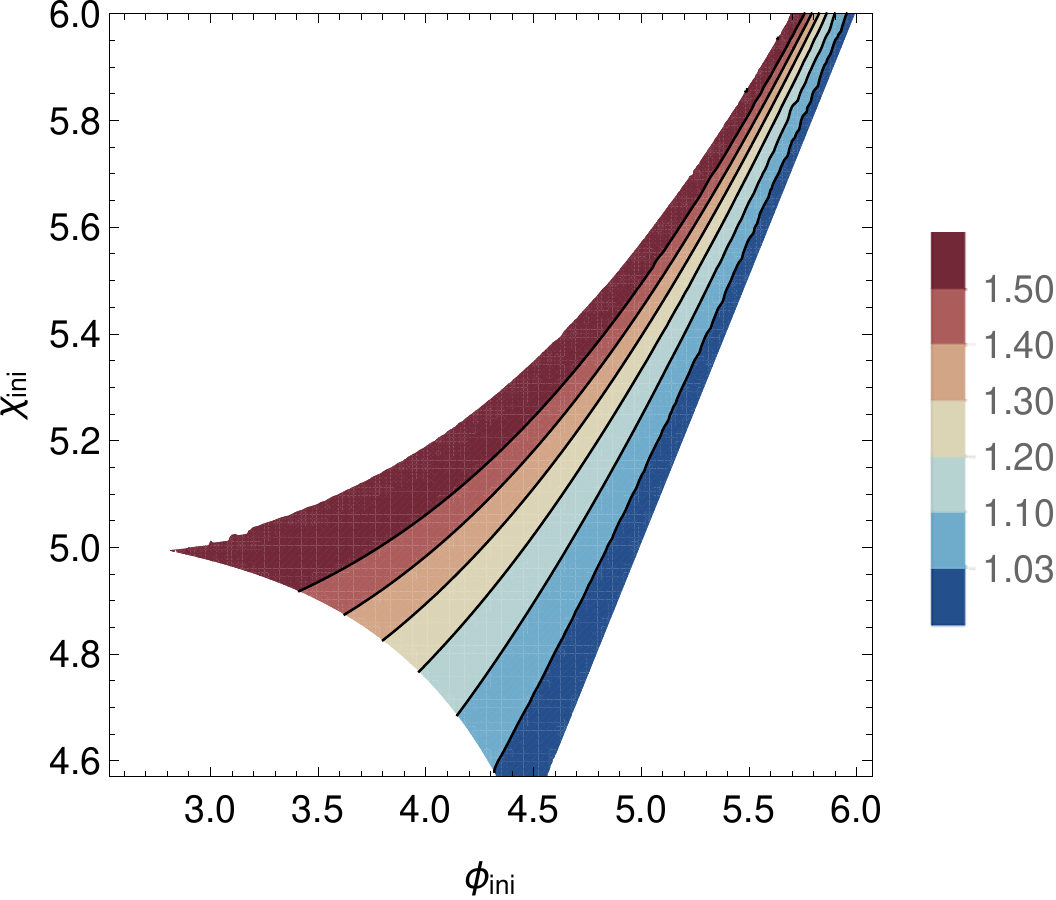}\\
\hspace{1.6cm}
\end{center}
\end{minipage}

\end{tabular}
\caption{The constraints on $\varphi^I _{\rm ini}$ from $n_s$ and $\mathcal{P}_{\mathcal{R}}$
obtained by the Planck result
with $M=\sqrt{3} \Mp$ and $\lambda=2.20 \times 10^{-10}$.  (Left) The region of $\varphi^I _{\rm ini}$ giving 
$n_s$  consistent with the Planck result. In the light blue region 
the resultant $\mathcal{P}_{\mathcal{R}}$
can be calculated based on the single-field model and is consistent with the Planck result, 
while in the light red part the resultant $\mathcal{P}_{\mathcal{R}}$ is
modified from that in the single-field model by  $\varphi^I _{\rm ini}$ dependent multifield effects.
(Right) Contour plot of $\mathcal{P}_{\mathcal{R}}$ in the light red region in the left panel.
We normalize the resultant $\mathcal{P}_{\mathcal{R}}$ by that in the single-field model
and only the dark blue region gives $\mathcal{P}_{\mathcal{R}}$ consistent with the Planck result.
\label{obsconst_Msqrt3_num}}
\end{center}
\end{figure}

\end{widetext}

Here, we impose the observational constraints for the case with $M=\sqrt{3} \Mp$
and first  we perform numerical calculations with various $\varphi^I _{\rm ini}$
and fixed  $\lambda$ given by  Eqs.~(\ref{parameters_expample_BG}) and (\ref{initialvalues_expample_BG}).
The left panel in Fig.~\ref{obsconst_Msqrt3_num} shows 
the region of $\varphi^I _{\rm ini}$
giving 
$n_s$  consistent with the Planck result.\footnote{Here, we also require that the 
total number of $e$-folding during inflation is sufficiently large.
Usually, its lower bound is 60, where inflation
can explain the horizon problem. However, in the numerical calculations,
in order to pick up the information of the mode exits the horizon scale 60 $e$-foldings before
the end of inflation without instability, we need more period of inflation. Because of this technical reason,
here,  instead of 60 we adopt 65 as the lower bound.}
The region is composed of the two parts, that is, the light blue and light red ones. 
In the light blue region, $\mathcal{P}_{\mathcal{R}}$ can be calculated based on the single-field model,
while in the light red part, the resultant $\mathcal{P}_{\mathcal{R}}$ is
modified from that in the single-field model by $\varphi^I _{\rm ini}$ dependent multifield effects. 
For  fixed $\chi_{\rm ini}$, the dependence of the resultant $n_s$ on $\phi_{\rm ini}$
becomes similar to the right panel in Fig.~\ref{pzeta_ns_pert_example}.
Among the examples in Sec.~\ref{sec:field_perturbations} C,
(A) and (B),  where the scale 
exits the horizon scale during the second inflationary phase or  in the turning phase
belong to the light blue region.
On the other hand, about (C) and (D),
where  the  scale exits the horizon scale during the first inflationary phase,
(D) belongs to the light red part, while (C)
belongs to the white region between the light blue and light red parts.
This is because, as we explained in  Sec.~\ref{sec:field_perturbations} C,  
the interval between the horizon exit and the turn is longer for $\varphi^I _{\rm ini}$
in the light red part, and the multifield effects on $n_s$ is smaller. 

Since we have only considered the constraint on $n_s$ so far, in order to specify the allowed region
for this set of $M$ and $\lambda$, we further consider the constraint on
$\mathcal{P}_{\mathcal{R}}$. As we mentioned before,
when $\mathcal{P}_{\mathcal{R}}$ can be calculated based on the single-field model,
 $\lambda=2.20 \times 10^{-10}$ gives $\mathcal{P}_{\mathcal{R}}=2.213 \times 10^{-9}$, 
which is consistent with the Planck result.
Therefore, the light blue region in the left panel in Fig.~\ref{obsconst_Msqrt3_num}
is allowed. 
On the other hand, when the multifield effects are important,
the resultant $\mathcal{P}_{\mathcal{R}}$ depends on $\varphi^I _{\rm ini}$.
The right panel in Fig.~\ref{obsconst_Msqrt3_num} shows
the resultant $\mathcal{P}_{\mathcal{R}}$ normalized by that in the single-field model
for given $\varphi^I _{\rm ini}$  in the light red region
in the left panel in Fig.~\ref{obsconst_Msqrt3_num} as a contour plot,
which briefly shows $\varphi^I _{\rm ini}$ dependence of 
$\mathcal{P}_{\mathcal{R}}$. We find that the maximum value is about $1.6$.
 For  fixed $\chi _{\rm ini}$, the behavior of $\mathcal{P}_{\mathcal{R}}$ becomes
similar to the left panel in Fig.~\ref{pzeta_ns_pert_example}.
Since the Planck result admits $\mathcal{P}_{\mathcal{R}}$ up to about $3 \%$
greater than $\mathcal{P}_{\mathcal{R}}=2.213 \times 10^{-9}$,  only the dark blue region in the contour plot is allowed.

\begin{widetext}
 
\begin{figure}[htbp]
\begin{center}
\begin{tabular}{c}

\begin{minipage}{0.45\hsize}
\begin{center}
\includegraphics[width=4.5cm]{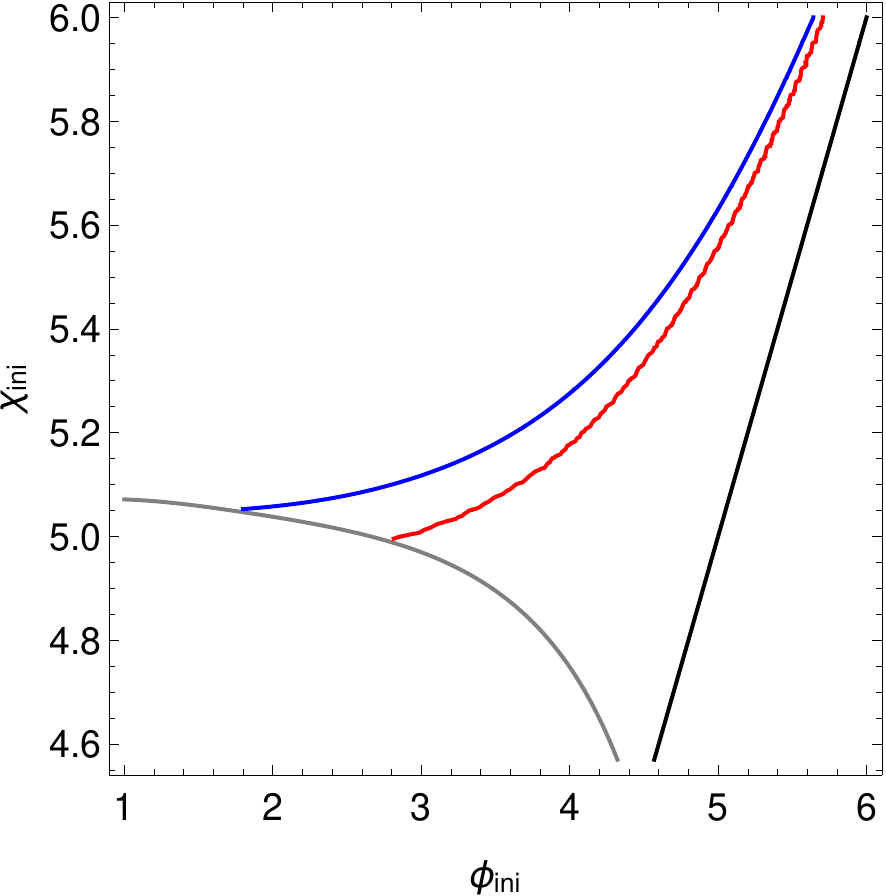}\\
\hspace{1.6cm}
\end{center}
\end{minipage}

\begin{minipage}{0.45\hsize}
\begin{center}
\includegraphics[width=5.5cm]{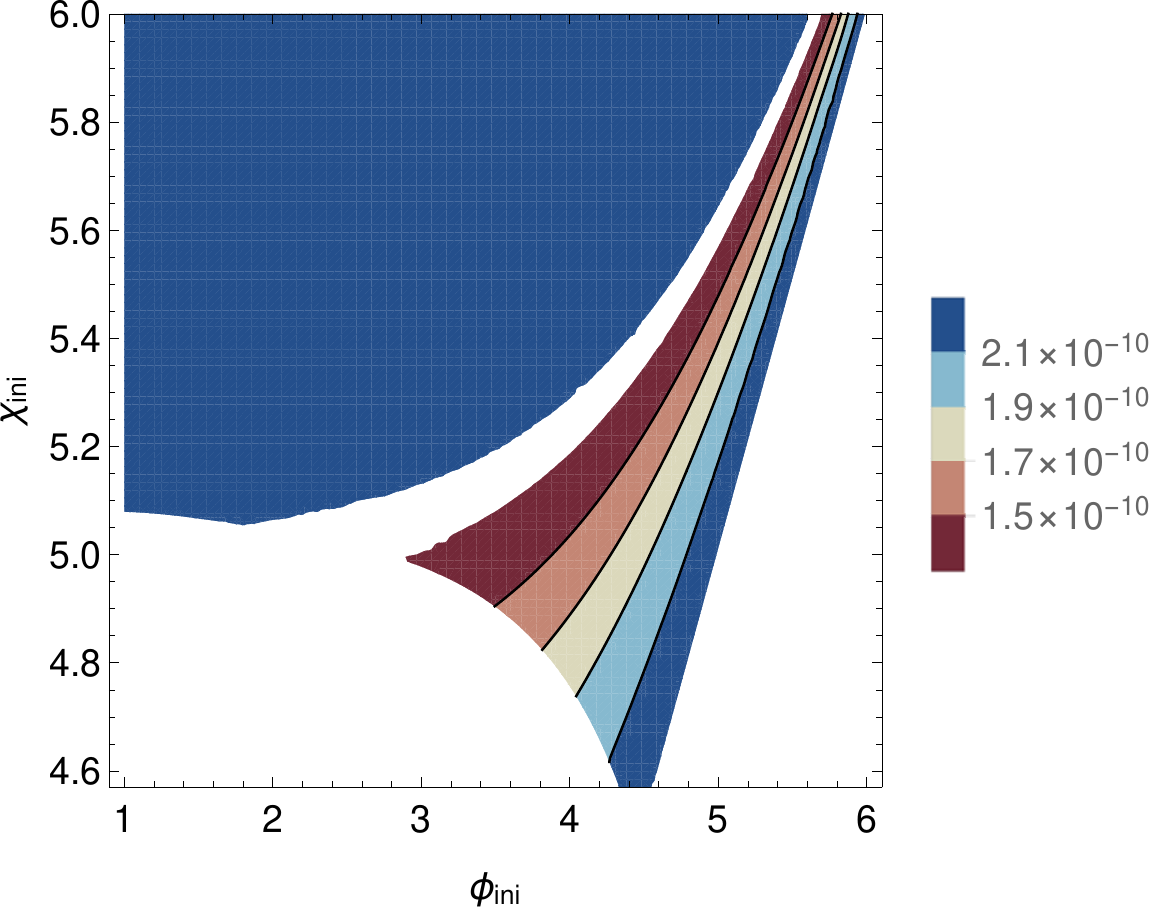}\\
\hspace{1.6cm}
\end{center}
\end{minipage}

\end{tabular}
\caption{
 (Left) Summary of the boundaries of the parts shown in the left panel in Fig~\ref{obsconst_Msqrt3_num}.
 The black line denotes $\chi_{\rm ini} = \phi_{\rm ini}$.
 The gray line denotes about  $N_{\rm ini}=65$.
 The blue and red lines denote $n_s=0.9521$.
 In this plot, for the last two lines, we specify the blue line as about $N_1=60$ and 
 the red line is obtained based on the analytic $\delta N$ formalism,
 which approximate the numerical results shown  in the left panel in Fig~\ref{obsconst_Msqrt3_num} very well.
 (Right) Contour plot of $\lambda$ giving $\mathcal{P}_{\mathcal{R}}$ consistent with the Planck result
for $\varphi^I _{\rm ini}$ in the region giving  $n_s$ consistent with the Planck result.
To obtain this plot, we set the fiducial value of  $\lambda$ giving $\mathcal{P}_{\mathcal{R}}$ 
consistent with the Planck result in the single-field model to be $2.20 \times 10^{-10}$.
The fiducial value as well as $\lambda$ in the plot can be up to $4.52 \%$ smaller  and   $2.76 \%$ greater 
within the regime consistent with the Planck result.
\label{obsconst_Msqrt3_deltan}}
\end{center}
\end{figure}

\end{widetext}

So far, we have imposed observational constraints on $\varphi^I _{\rm ini}$
with fixed $M$ and $\lambda$, while our original  purpose here is to impose 
the observational constraints
on $\varphi^I _{\rm ini}$ and $\lambda$ with fixed $M$.
One straightforward way  is simply repeating numerical calculations
with varying $\lambda$, which takes a considerable amount of time.
Contrary to this, we make use of the fact that some ingredients of the plots 
in Fig.~\ref{obsconst_Msqrt3_num} are $\lambda$ independent.
We begin with considering the boundary 
corresponding to the lower bound for the sufficient total number of $e$-folding during inflation, 
which is written as the gray line in the left panel in Fig.~\ref{obsconst_Msqrt3_deltan}.
Although we have obtained this line by solving the background equation of motion numerically
in order to be precise,
as in the discussion in Sec.~\ref{sec:background} D, 
the total number of $e$-folding is well approximated by $N_{\rm ini}$ 
defined by Eq.~(\ref{concretevalue_N1_Nini}) and
(\ref{concretevalue_N1_Nini2}).
Since the correspondence between $\varphi^I _{\rm ini}$ and
$N_{\rm ini}$ for given $M$  is independent of $\lambda$,
we expect that the location of this boundary does not depend on $\lambda$.

Next, we consider the boundary
between the part where  the resultant $\mathcal{P}_{\mathcal{R}}$ 
coincides with that in the single-field model and the part where that is modified by the multifield effects,
which is written as the blue line in the left panel in Fig.~\ref{obsconst_Msqrt3_deltan}.
In the left panel in Fig.~\ref{obsconst_Msqrt3_num}, we have obtained this
by specifying the line giving $n_s = 0.952$ based on the numerical calculation on the perturbations
with fixed $\lambda$.
However, in  Sec.~\ref{sec:background} D,
we have already shown that this boundary is qualitatively understood as $N_1 = 60$,
where $N_1$ is defined by Eq.~(\ref{concretevalue_N1_Nini}).
Since  $N_1$ is also independent of $\lambda$ for given $M$ and $\varphi^I _{\rm ini}$,
we expect that the location of this boundary does not depend on $\lambda$, either.
From the discussion in Sec.~\ref{sec:field_perturbations} C, if the pivot scale of the recent CMB observations
exits the horizon scale in the turning phase like the example (B) there, the multifield effects
are still sufficiently small, where the resultant $n_s$ is consistent with the Planck result. 
Therefore, in order to be more precise, instead of $N_1=60$, we regard the boundary as  that between 
the regions where the scale exits the horizon scale in the turning phase
and in the first inflationary phase specified by the numerical calculation on the background, 
which is shown as the blue line in the left panel in Fig.~\ref{obsconst_Msqrt3_deltan}.
Compared with the corresponding boundary shown in the left panel in Fig.~\ref{obsconst_Msqrt3_num}, 
we see that this method can reproduce the result from the numerical calculations
on the perturbations with a very high accuracy.
Combining the above two discussions, the locations of the boundaries of the light blue region in the left panel
in Fig.~\ref{obsconst_Msqrt3_num} are $\lambda$ independent.
Therefore, the remaining thing for $\varphi^I _{\rm ini}$ in the light blue region,
 where the resultant $n_s$ is consistent with the Planck result,
is to assign $\lambda$ 
so that the resultant  $\mathcal{P}_{\mathcal{R}}$ that coincides with the one in
the single-field model is consistent with the Planck result.
Within the regime consistent with the Planck result,
$\lambda$ must be between  $4.52 \%$ smaller  than the ``fiducial value'', $2.20 \times 10^{-10}$,  and   
$2.76 \%$ greater than that. 

Let us move on to the region where the resultant $\mathcal{P}_{\mathcal{R}}$ is modified from that in
the single-field model 
and first think the left boundary of the light red region in the left panel in Fig.~\ref{obsconst_Msqrt3_num}.
This line corresponds to $n_s=0.952$, the lower bound of $n_s$ according to the Planck result.
In the left panel in Fig.~\ref{obsconst_Msqrt3_num}
we have obtained it by specifying the line giving $n_s = 0.952$ from the numerical calculation on the perturbations
with fixed $\lambda$.
For this, we make use of the fact that 
the analytic $\delta N$ formalism works well
when the mode exits the horizon scale in the first inflationary phase sufficiently before the turn,
as shown in Sec.~\ref{sec:field_perturbations} D. 
Actually, this boundary in the left panel in Fig.~\ref{obsconst_Msqrt3_deltan} written as a red line
is obtained by specifying the line giving $n_s = 0.952$  
based on the analytic $\delta N$ formalism.  
Compared with the plots in Fig.~\ref{obsconst_Msqrt3_num}, we see that
the result from the numerical calculations on the perturbations can be reproduced  with very high accuracy.
Although we do not show explicitly the whole plot corresponding to the right panel
in  Fig.~\ref{obsconst_Msqrt3_num}, we can also reproduce the numerical result on 
$\mathcal{P}_{\mathcal{R}}$ by the analytic $\delta N$ formalism very well. 
In the part giving $n_s$ consistent with Plnck result,
we show that the two power spectra $\mathcal{P}_{\mathcal{R}}$ obtained by two different methods
differ at most  $3 \%$.

Once we confirm the validity of the analytic $\delta N$ formalism in the light red part
in the left panel in Fig.~\ref{obsconst_Msqrt3_num},
we can discuss $\lambda$ dependence on $\mathcal{P}_{\mathcal{R}}$ and $n_s$ 
with $\varphi^I _{\rm ini}$ in this part quantitatively.
Based on the analytic $\delta N$ formalism, $n_s$ is given by Eq.~(\ref{ns_deltan_example})
and $\lambda$ appears only through $V_*$ and $V_{*,IJ}$, which are eventually canceled.
Therefore, although we have specified the line giving $n_s = 0.952$,
which is  the left boundary of the 
light red region in the left panel in Fig.~\ref{obsconst_Msqrt3_num}, or equivalently
the red line in the left panel in Fig.~\ref{obsconst_Msqrt3_deltan}
with fixed $\lambda$, 
the location of this line does not depend on $\lambda$.
To complete the discussion, 
we assign $\lambda$ for given $\varphi^I_{\rm ini}$ in the light red part
in the left panel in Fig.~\ref{obsconst_Msqrt3_num}, which was shown to give $n_s$
consistent with the Planck result independent of $\lambda$
so that the resultant $\mathcal{P}_{\mathcal{R}}$ is  also consistent with the Planck result.
For this, we can use the fact that
$\varphi^I _{\rm ini}$ dependent  enhancement factor in this part 
obtained in the right panel in Fig.~\ref{obsconst_Msqrt3_num} is independent of $\lambda$.
This can be seen  by that
based on the  $\delta N$ formalism, $\mathcal{P}_{\mathcal{R}}$ is given by 
Eq.~(\ref{pzeta_deltan_example}) and  $\lambda$ appears only through $V_* \propto \lambda$,
while $\mathcal{P}_{\mathcal{R}}$ in the single-field model also depends on $\lambda$
only through $V_* \propto \lambda$. 
Then, the desirable  $\lambda$ for given $\varphi^I _{\rm ini}$ is just the ``fiducial value'' of
$\lambda$ giving $\mathcal{P}_{\mathcal{R}}$  consistent with the Planck result in the single-field model,
divided by $\varphi^I_{\rm ini}$ dependent enhancement factor obtained in the right panel in  Fig.~\ref{obsconst_Msqrt3_num}.

Combining all the discussions in this subsection, in the right panel in Fig.~\ref{obsconst_Msqrt3_deltan},
we summarize the observational constraints with $M=\sqrt{3} \Mp$.
We have specified the region giving $n_s$  consistent with the Planck result in $\varphi^I_{\rm ini}$ space,
and  assigned there $\lambda$ so that the resultant
$\mathcal{P}_{\mathcal{R}}$ is also consistent with the Planck result.
In the plot,  we set the ``fiducial value'' of  $\lambda$
giving consistent  $\mathcal{P}_{\mathcal{R}}$ in the single-field model  to be $2.20 \times 10^{-10}$.
Since the ``fiducial value''  can be $4.52 \%$ smaller  and   $2.76 \%$ greater
in the single-field model,  $\lambda$ shown in the contour plot 
can also vary within this range.
 Although we have imposed the constraint on $\lambda$ for each $\varphi^I _{\rm ini}$ with $M=\sqrt{3}\Mp$,
practically, there is no way for us to know the initial values of the fields.
Then, this result shows that the observational constraint on $\lambda$,
where it is $2.10 \times 10^{-10} < \lambda < 2.26 \times 10^{-10}$ in the single field inflation,
becomes weaker so that $1.31 \times 10^{-10} < \lambda < 2.26 \times 10^{-10}$ in this double inflation
 with $M=\sqrt{3} \Mp$.

\subsection{Cases with $M=\Mp$ and $M=\sqrt{6}\Mp$}

Here  we impose the observational constraints on $\lambda$ and $\varphi^I _{\rm ini}$ with different $M$
and to see how the constraints change. As an example with $M$ smaller than $M=\sqrt{3} \Mp$ considered before,
we consider the case with  $M=\Mp$, while as an example with $M$ larger than $M=\sqrt{3} \Mp$,
we consider the case with $M=\sqrt{6}\Mp$.
When we change $M$, we choose the fiducial value of $\lambda$ that gives $\mathcal{P}_{\mathcal{R}}$
consistent with the Planck result in the single-field model so that it does not change 
 $\mathcal{P}_{\mathcal{R}}$ given by Eqs.~(\ref{observables_singlefield_slowroll}) from the case with
 $M=\sqrt{3} \Mp$. To be more concrete, since $H ^2 \simeq V/(3 \Mp^2) \simeq (\lambda M^4)/(108 \Mp^2)$
 and $\epsilon \simeq \epsilon_{\chi} \simeq M^2/(8 N_*^2 \Mp^2)$, we change $M$
 and the fiducial value of $\lambda$ so that $(\lambda M^2)/\Mp^2$ is fixed.
 It is worth mentioning that  with this choice, we eventually fix $m$ given by Eq.~(\ref{potential_TModel_asym_small})
 to be constant given by  $m = (\sqrt{2 \lambda} M)/6  \simeq  6.1 \times 10^{-4} \Mp$, while $H$ changes as
 $H \propto M$. Then, the combination $m/H$, which turn out to be important to
 discuss the effect of the heavy field excitation, as we will see later, changes as
 $m/H \propto \Mp/M$.

Before starting the analysis,
it is worth summarizing  the  boundaries of the parts
in $\varphi^I _{\rm ini}$ space  shown in the left panel in Fig.~\ref{obsconst_Msqrt3_deltan}.
The black line denotes just $\chi_{\rm ini} = \phi_{\rm ini}$ above which we only consider from the symmetric
property of the potential. The gray line denotes the lower bound for the region with sufficient total number
of $e$-folding during inflation, which is approximately given by $N_{\rm ini} =65$.
The red and blue lines denote $n_s = 0.951$, and we first specified them based on numerical calculations
on perturbations. However, as we mentioned above, the blue line is approximately given by 
$N_1 =60$ and the red line is obtained based on the analytic $\delta N$ formalism.
Although the above is not sufficient for the accurate specification of the gray and blue lines, 
it turns out that we can reproduce the results
with a very high accuracy by solving the background equations of motion numerically.
Here, we obtain the plots corresponding to the one in the right panel in Fig.~\ref{obsconst_Msqrt3_deltan}
with different $M$ in a similar form as the previous subsection, but based on the quicker way without relying 
on the numerical calculations on perturbations. 
Although we have not checked the full region in $\varphi^I _{\rm ini}$ giving $n_s$ consistent with Planck result
 presented in the following plots,
along a line with some fixed $\chi_{\rm ini}$, 
we see that the two power spectra $\mathcal{P}_{\mathcal{R}}$ obtained by 
the numerical calculations on the perturbations and the analytic $\delta N$ formalism 
differ at most  $1 \%$ for the case with $M=\Mp$ and $6 \%$ for the case with $M=\sqrt{6} \Mp$.

First, in order to see $M$ dependence of the background dynamics, we show the plots corresponding
to the one in the left panel in Fig.~\ref{fatemsquare3_BG}   in  Fig.~\ref{fatemsquare1_fatemsquare6_BG}
with $M=\Mp$ (left) and $M=\sqrt{6} \Mp$ (right).
From Eqs.~(\ref{concretevalue_N1_Nini}), for given $\phi_{\rm ini}$,  
$\chi_{\rm ini}$ giving $N_1=60$ is always larger than the one giving $N_{\rm ini} =60$.
On the other hand, in this paper, we avoid the situation with $M  \gg \Mp$
so that the approximated background solution is valid,
where we will discuss later in this section about the estimation of this condition.
We find that as long as this condition is satisfied, both $\chi_{\rm ini}$ giving $N_1=60$
and the one giving $N_{\rm ini} =60$ with fixed $\phi_{\rm ini}$ increase as $M$ increases.
Then, even if some  $\varphi^I _{\rm ini}$ gives  $N_1 > 60$ and is in the light blue region
for small $M$, if we increase $M$, above some $M$, this  $\varphi^I _{\rm ini}$ no longer gives  $N_1 >60$.
But since $\chi_{\rm ini}$ giving $N_1=60$ is always larger than the one giving $N_{\rm ini} = 60$
for given $\phi_{\rm ini}$, there is some regime $M$,
where $\varphi^I _{\rm ini}$ does not give $N_1 > 60$, but still gives $N_{\rm ini} > 60$
and  is in the light red part.
If we further increase $M$, this $\varphi^I _{\rm ini}$ eventually cannot give $N_{\rm ini}>60$, either
and is in the gray part.

The left panel in Fig.~\ref{obsconst_M1_Msqrt6} shows the observational constraints with
$M=\Mp$ in a similar form as the one in the right panel in Fig.~\ref{obsconst_Msqrt3_deltan}
with $M=\sqrt{3} \Mp$ and 
we set the fiducial value of $\lambda$
giving  consistent  $\mathcal{P}_{\mathcal{R}}$ in the single-field model to be
 $6.60 \times 10^{-10}$.
 Since $\lambda$ can be $3.99 \%$ smaller  and   $3.32 \%$ greater
in the single-field model,  $\lambda$ shown in the contourplot 
can also vary within this range.
Compared with the case with $M=\sqrt{3} \Mp$, the most evident difference is that
in the part where the multifield effects modify $\mathcal{P}_{\mathcal{R}}$ from that in the single-field model,
there is region not consistent with the Planck result around $\lambda \sim 4.0 \times 10^{-10}$.
This is because in this region, the resultant $n_s$ is larger than the upper bound of the
$2 \sigma$ constraints in the Planck result, $0.9771$.
We also see that the maximum value of $\varphi^I _{\rm ini}$ dependent enhancement factor of $\mathcal{P}_{\mathcal{R}}$ from that
in the single-field model  (about $1.8$)  is  larger than the case with $M=\sqrt{3} \Mp$ (about $1.6$) as well as the width of the region in $\varphi^I _{\rm ini}$ space
giving $n_s$ smaller than the lower bound of the $2 \sigma$ constraints in the Planck result, $0.9521$
is narrower than  the case with  $M=\sqrt{3} \Mp$. 
 Regardless of detailed discussion of $\varphi^I _{\rm ini}$ dependence,
as in the discussion in the last part of Subsec.~V A, practically,
there is no way for us to know the initial values of the fields.
Then, this result shows that the observational constraint on $\lambda$,
where it is $6.34 \times 10^{-10} < \lambda < 6.82 \times 10^{-10}$ in the single field inflation,
becomes weaker so that $3.52 \times 10^{-10} < \lambda < 6.82 \times 10^{-10}$ in this double inflation
 with $M=\Mp$.

 The difference of the features of the left panel in Fig.~\ref{obsconst_M1_Msqrt6} with 
those of the right panel in Fig.~\ref{obsconst_Msqrt3_deltan}
can be explained by the excitation of the heavy field during the turn in multifield inflation
(see e.g. \cite{Tolley:2009fg,Achucarro:2010da,Shiu:2011qw,Pi:2012gf,Gao:2012uq,Noumi:2012vr}).
As we discuss in Appendix~\ref{sec:BD_TurningPhase},  
by analyzing the background dynamics of the turning phase, when $H < (2/3) m$, or equivalently
$M < (2 \sqrt{6} /3) \Mp$, we find that the background trajectory experiences oscillations
during the turn.
Actually, in the left panel in Fig.~\ref{heavy_field_excitation}, we show the time evolution of $\phi$
around the turn with $M=\Mp$. In this model, the efficiency of the heavy field excitation
depends on the dynamics of $\phi$ at $0 < \phi < M$ in the first inflationary phase, 
where the potential can not be approximated
by neither Eq.~(\ref{TModelPotential_approximated}) nor (\ref{potential_turning}).
As in the right panel in Fig.~\ref{heavy_field_excitation}, we find that
the efficiency of the heavy field excitation monotonously increases if $M$ decreases.
Then, if we decrease $M$ further, we expect that the wider region is excluded by the constraint on $n_s$
and the maximum value of $\varphi^I _{\rm ini}$ dependent enhancement factor of $\mathcal{P}_{\mathcal{R}}$ from 
that in the single-field model is larger.
 Notice that in the left panel in Fig.~\ref{obsconst_M1_Msqrt6},
in the part, where the multifield effects modify the resultant $\mathcal{P}_{\mathcal{R}}$ and $n_s$ from that
in the single-field model, we change $\lambda$ from the fiducial value, which gives different 
$m$ and $H$ at the turn. However, since both $H$ and $m$ scale as $\propto \sqrt{\lambda}$ with fixed $M$,
the ratio $m/H$, parametrizing the efficiency of the heavy field excitation does not change. 
Therefore, we expect that the region excluded by too large $n_s$ does not change with $\lambda$ 
giving $\mathcal{P}_{\mathcal{R}}$ consistent with the Planck result.
On the other hand, the larger $\varphi^I _{\rm ini}$ dependent enhancement factor of $\mathcal{P}_{\mathcal{R}}$ 
from that in the single-field model and narrower region in $\varphi^I _{\rm ini}$ space
giving too small $n_s$ is the typical behavior of the sharp turn accompanied with 
the heavy field excitation (see,  \cite{Gao:2012uq}).

\begin{widetext}

\begin{figure}[htbp]
\begin{center}
\begin{tabular}{c}

\begin{minipage}{0.45\hsize}
\begin{center}
\includegraphics[width=4.5cm]{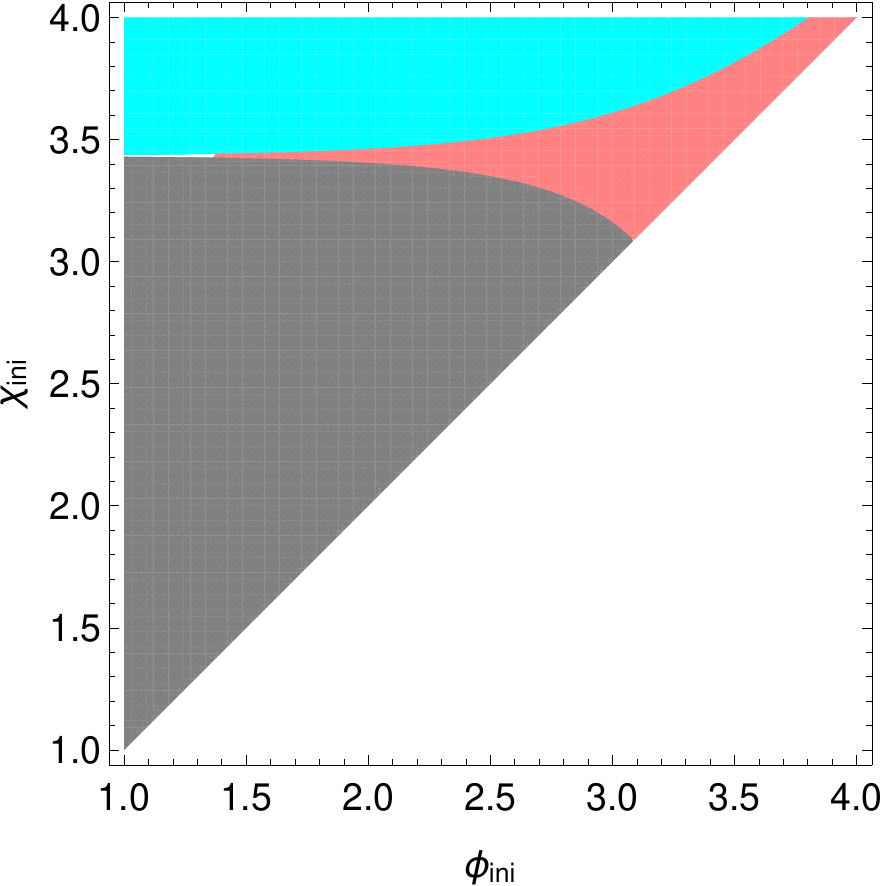}\\
\hspace{1.6cm}
\end{center}
\end{minipage}

\begin{minipage}{0.45\hsize}
\begin{center}
\includegraphics[width=4.5cm]{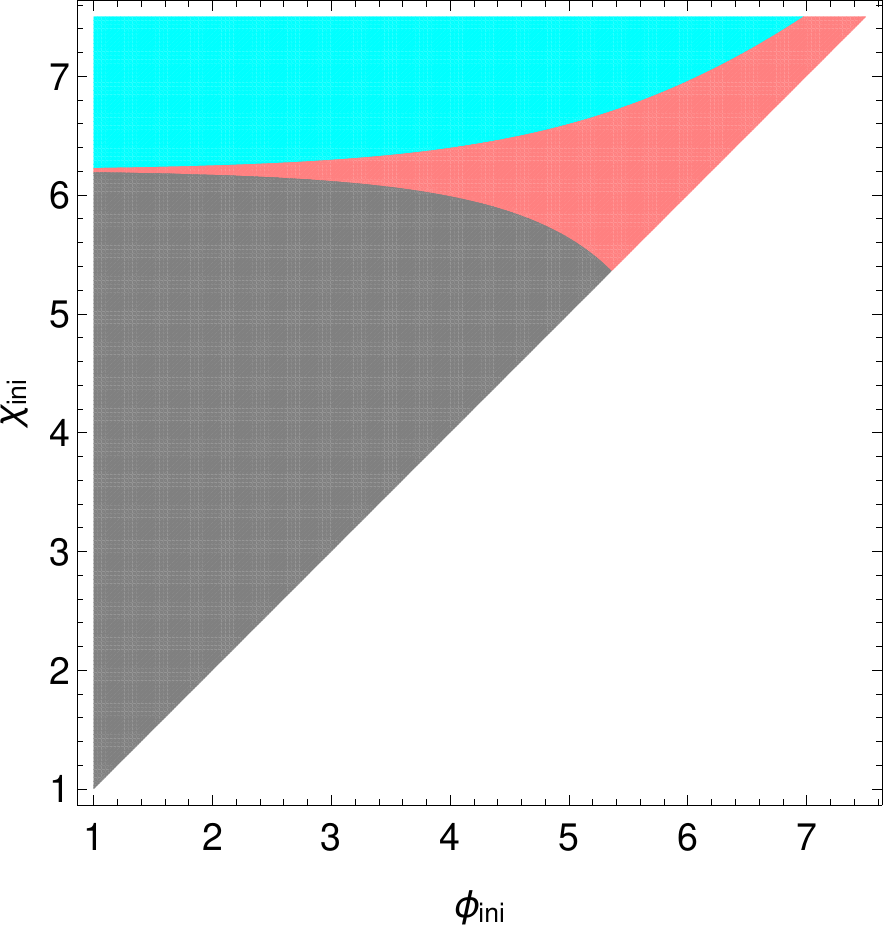}\\
\hspace{1.6cm}
\end{center}
\end{minipage}

\end{tabular}
\caption{Classification of the region in $\varphi^I _{\rm ini}$ space based on in which phase 
the pivot scale of the recent CMB observations exits the horizon scale
based on $N_1$ defined by Eqs.~(\ref{concretevalue_N1_Nini}).
In the gray part, $N_{\rm ini} < 60$, where the total number of $e$-folding during inflation is insufficient.
Among the other region in the light blue part, the scale exits the horizon scale in the second inflationary phase,
while in the light red part, it exits the horizon scale in the first inflationary phase. 
We choose $M=\Mp$ (left)
and $M=\sqrt{6} \Mp$ (right), respectively.}
\label{fatemsquare1_fatemsquare6_BG}
\end{center}
\end{figure}


\begin{figure}[htbp]
\begin{center}
\begin{tabular}{c}

\begin{minipage}{0.45\hsize}
\begin{center}
\includegraphics[width=5.5cm]{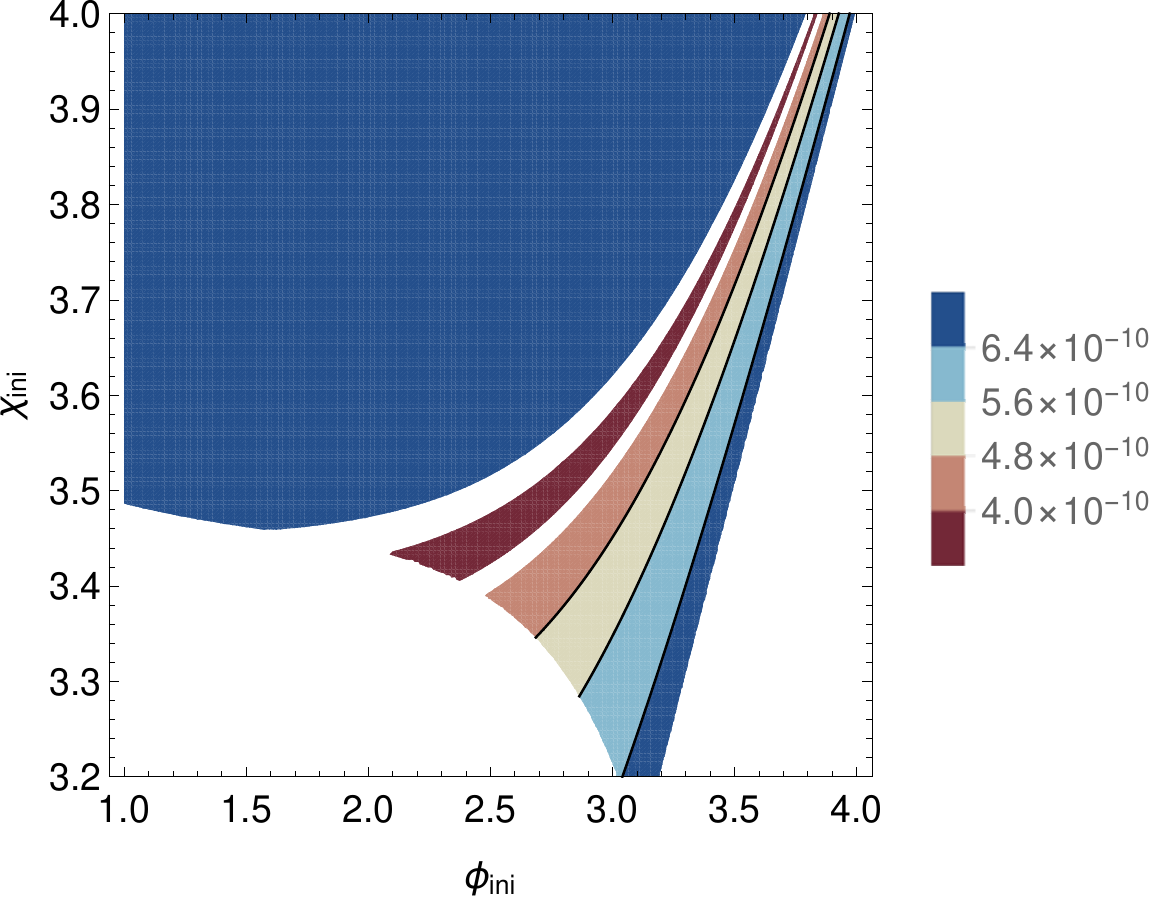}\\
\hspace{1.6cm}
\end{center}
\end{minipage}

\begin{minipage}{0.45\hsize}
\begin{center}
\includegraphics[width=5.5cm]{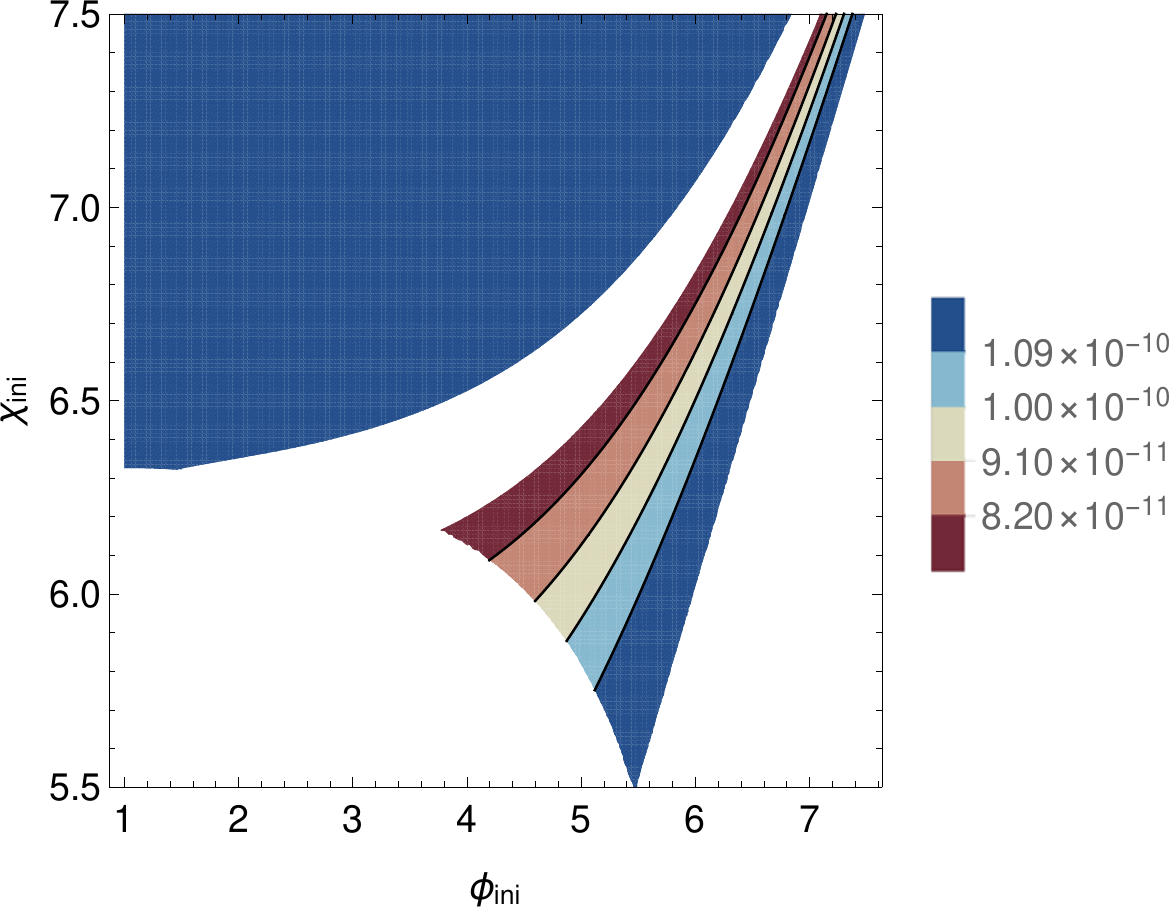}\\
\hspace{1.6cm}
\end{center}
\end{minipage}

\end{tabular}
\caption{Contour plot of $\lambda$ giving $\mathcal{P}_{\mathcal{R}}$ consistent with the Planck result
for $\varphi^I _{\rm ini}$ in the region giving $n_s$  consistent with the Planck result.
  To obtain these plots, we set the fiducial value of  $\lambda$ giving $\mathcal{P}_{\mathcal{R}}$ 
consistent with the Planck result in the single-field model to be 
$6.60 \times 10^{-10} (\Mp^2/M^2)$.
(Left) The case with $M=\Mp$, where
the fiducial value as well as $\lambda$ in the plot can be up to $3.99 \%$ smaller  and   $3.32 \%$ greater 
within the regime consistent with the Planck result.
 (Right) 
The case with $M=\sqrt{6} \Mp$, where the
fiducial value as well as $\lambda$ in the plot can be up to $4.86 \%$ smaller  and   $2.39 \%$ greater 
within the regime consistent with the Planck result.
\label{obsconst_M1_Msqrt6}}
\end{center}
\end{figure}

\end{widetext}

The right panel in Fig.~\ref{obsconst_M1_Msqrt6} shows the observational constraints with
$M=\sqrt{6} \Mp$ in a similar form as the one in the right panel in Fig.~\ref{obsconst_Msqrt3_deltan}
with $M= \sqrt{3} \Mp$ and 
we set the fiducial value of $\lambda$
giving  consistent  $\mathcal{P}_{\mathcal{R}}$ in the single-field model to be
 $1.10 \times 10^{-10}$.
 Since the fiducial value  can be $4.86 \%$ smaller  and   $2.39 \%$ greater
in the single-field model,  $\lambda$ shown in the contour plot 
can also vary within this range.
In this case, compared with the case with $M=\sqrt{3} \Mp$, 
the maximum value of $\varphi^I _{\rm ini}$ dependent enhancement factor of 
$\mathcal{P}_{\mathcal{R}}$ from the one in the single-field model is smaller 
(about $1.4$)
as well as the width of the region in $\varphi^I _{\rm ini}$ space
giving $n_s$ smaller than the lower bound of the $2 \sigma$ constraints in the Planck result, $0.9521$
is wider than  the case with  $M=\sqrt{3} \Mp$, 
which can be explained by the lower efficiency of the heavy field excitation.
As in the case with $M=\sqrt{3} \Mp$ there is no region giving $n_s$
 larger than the upper bound of the $2 \sigma$ constraints in the Planck result.
 Similar to the cases with $M=\sqrt{3} \Mp$ and $M=\Mp$,
we can consider the constraints on $\lambda$ taking into account the fact that
there is no way for us to know the initial values of the fields, practically.
Then, this result shows that the observational constraint on $\lambda$,
where it is $1.05 \times 10^{-10} < \lambda < 1.13 \times 10^{-10}$ in the single field inflation,
becomes weaker so that $7.48 \times 10^{-11} < \lambda < 1.13 \times 10^{-10}$ in this double inflation
 with $M=\Mp$.

 If we increase $M$ further, although there is no heavy field excitation,
 we must care other effects. As we mentioned, in this paper, in order for the $\alpha$-attractor-type
 double inflation to occur, we must avoid the situation with $M \gg \Mp$. 
 Here, we discuss the upper bound of $M$ based on $\varphi^I _{\rm ini}=(0, \chi_{\rm ini})$
 giving $N=N_*=60$. 
 When the approximated background solution of the $\alpha$-attractor-type double inflaton is valid, 
 this is given by Eqs.~(\ref{concretevalue_N1_Nini}) and
(\ref{concretevalue_N1_Nini2}) 
 with $\phi_{\rm ini} =0$ and $N_{\rm ini} = N_* =60$,  which we denote $\chi_{\rm ini, *} ^{\rm anal}$,
 an increasing function of $M$.
 The concrete form of $\chi_{\rm ini, *} ^{\rm anal} (M)$ and $M_{\chi_{\rm ini, *}}$ defined by the relation
 $M_{\chi_{\rm ini, *}} \equiv \chi_{\rm ini, *} ^{\rm anal} (M_{\chi_{\rm ini, *}} )$ are given by
 \bea
 &&
 \chi_{\rm ini, *} ^{\rm anal}=\frac{M}{2} \log \left[\frac{960 \Mp^2}{M^2}  -2 \right]\,,
 \label{chi_ini_*_anal}
\\
 &&
 \frac{M_{\chi_{\rm ini, *}}}{\Mp} = 8 \sqrt{\frac{15}{2 + e^2}} \simeq 10.1\,.
 \label{chi_ini_*_anal2}
 \ena
 As a very rough order estimation, if $M$ becomes as large as $M= M_{\chi_{\rm ini, *}}$,
 the condition $\chi \gg M$ is no longer valid during the last 60 $e$-foldings in the inflation, 
 which means that the inflationary background is not described by this double inflation.
 Furthermore, since $\chi_{\rm ini, *} ^{\rm anal}$  is obtained by assuming that $\chi \gg M$ holds
 during the last  60 $e$-foldings in the inflation, the actual upper bound of $M$ 
 should be smaller than $M_{\chi_{\rm ini, *}}$.
On the other hand, we can obtain actual $\varphi^I _{\rm ini}=(0, \chi_{\rm ini})$ giving $N=N_*=60$
 by solving the background equations numerically, which we denote $\chi_{\rm ini, *} ^{\rm num}$.
 We find that $\Delta \chi_{\rm ini, *} \equiv \chi_{\rm ini, *} ^{\rm num} - \chi_{\rm ini, *} ^{\rm anal}$
 increases as we increase $M$.
 
 We finalize this section by briefly explaining what happens 
 if we further increase $M$ based on $\Delta \chi_{\rm ini, *}/\chi_{\rm ini, *} ^{\rm num}$.
 At $M=3 \Mp$, we find that $\Delta \chi_{\rm ini, *}/\chi_{\rm ini, *} ^{\rm num} \simeq 0.0042$.
 Although this difference seems not so significant, 
the two power spectra $\mathcal{P}_{\mathcal{R}}$ obtained by 
the numerical calculations on the perturbations and the analytic $\delta N$ formalism 
differ at most as much as  $10 \%$ in the part corresponding to the light red part
in the left panel in Fig.~\ref{obsconst_Msqrt3_num}.
This is because in the analytic $\delta N$ formalism, the accurate specification of 
$N_*$ given by Eq.~(\ref{nexit_deltan}) is indispensable. Regardless of this, with this $M$,
the main property of the $\alpha$-attractor-type double inflation that the background trajectory is composed of
two almost straight lines and can be specified once we fix $\varphi^I _{\rm ini}$ still holds. 
Then, with any $\varphi^I _{\rm ini}$ giving the horizon exit of the mode
in the second inflationary phase, it occurs at  $\chi_{\rm ini, *} ^{\rm num}$ and because of this 
attractor property the resultant  $\mathcal{P}_{\mathcal{R}}$ and $n_s$ are constant. 
How about $\varphi^I _{\rm ini}$ giving the horizon exit of the mode in the first inflationary phase, where
the multifield effects modify the resultant  $\mathcal{P}_{\mathcal{R}}$ and $n_s$  from
that in the single-field model?  With such  $\varphi^I _{\rm ini}$,
the fact that the modification depends on the interval between the horizon exit and the turn,
and if the interval is shorter, the modification is larger, will not change.
Then if we calculate $\mathcal{P}_{\mathcal{R}}$ and $n_s$ with such $\varphi^I _{\rm ini}$
with fixed $M$ and $\lambda$, we expect that similar results as shown in the plots in Fig.~\ref{obsconst_Msqrt3_num} 
are obtained, although the analytic $\delta N$ formalism no longer works efficiently.
If we further increase $M$, at $M= 6 \Mp$, we find that
$\Delta \chi_{\rm ini, *}/\chi_{\rm ini, *} ^{\rm num} \simeq 0.041$.
With this $M$, the direction of the background trajectory starts changing  much before $\phi=0$
and it is no longer regarded as two almost straight lines.
With $M$ above this, the two-field model no longer possesses the good property
of the $\alpha$-attractor-type double inflation 
 discussed in this paper and the calculation and understanding of $\mathcal{P}_{\mathcal{R}}$ and $n_s$ 
 become much more complicated, as in the conventional double inflation \cite{Polarski:1992dq,Polarski:1994rz,Langlois:1999dw,Tsujikawa:2002qx,Vernizzi:2006ve,Byrnes:2006fr}.

\section{Conclusions and Discussions
\label{sec:conclusions}}

Recently, the $\alpha$-attractor models have been very actively studied.
Not only this class of models phenomenologically explain the observational results
$1-n_s \sim 1/N$ and $r \sim 1/N^2$, they can be derived from fundamental theories
like supergravity or string theory. Since scalar fields are ubiquitous
in these theories, it is natural to consider the multifield extension of
the $\alpha$-attractor models. Among the multifield inflation, 
the so-called double inflation composed of two minimally coupled massive scalar fields is the simplest
toy model including sufficient ingredients giving rich phenomenology with multifield effects 
in the primordial perturbation.
On the other hand, the common property of the potential in the $\alpha$-attractor
is that it has a plateau for the large field values and especially in the T-model,
the potential is obtained by stretching the massive potential.
Therefore, as a simple multifield extension of the $\alpha$-attractor,
in this paper, 
we have considered the $\alpha$-attractor-type double inflation,
which is composed of two minimally coupled scalar fields $\varphi^I = (\phi, \chi)$ 
and each of the fields has a potential
of the $\alpha$-attractor-type.
Although our analysis is based on the symmetric two-field T-model,
where the potential is given by Eq.~(\ref{T_Model_potential}),
we expect that most results shown in this paper are also obtained by
other models as long as the asymptotic forms of the potentials are $\alpha$-attractor type.

About the background dynamics, 
first we showed the numerical results
and explained the brief overview. If the initial values of the fields $\varphi^I_{\rm ini}$
satisfies $\phi_{\rm ini} < \chi_{\rm ini}$,
in the early stage, the motion of the fields is towards $\phi=0$
with almost a straight trajectory. Then, when $\phi$ approaches zero,
the background fields are captured by the potential valley around $\phi=0$
and changes the direction of the motion to $\chi=0$.
After the direction has changed completely to $\chi=0$, the background trajectory
again becomes a straight line towards $\chi=0$.
In this paper, briefly, we called these three phases 
as the first inflationary phase, the turning phase, and the second inflationary phase, respectively.
Then, we obtained the analytic solutions describing the first and second
inflationary phases based on the slow-roll approximation.
Since the second inflationary phase can be regarded as single-field $\alpha$-attractor-type inflation
driven solely by $\chi$, the approximated solutions for this phase are just the conventional
$\alpha$-attractor-type ones. On the other hand, the approximated solutions
for the first inflationary phase, given by Eqs.~(\ref{phi_chi_slowroll_firstinflationaryphase_ito_N})
and (\ref{phi_chi_slowroll_firstinflationaryphase_ito_N2})
are new. Especially, the integration constant $N_{\rm ini}$ appeared there, or equivalently
$N_1$ defined by Eq.~(\ref{def_N1}) are shown to be regarded as the total number of $e$-folding
during inflation and the number of $e$-folding at the end of the first inflationary phase.
As we showed in Fig~\ref{fatemsquare3_BG}, in terms of $N_{\rm ini}$ and $N_1$,
we can roughly classify the region in $\varphi^I _{\rm ini}$ space based on whether
the total number of $e$-folding during inflation is sufficient and if so,
in which phase the pivot scale of the recent CMB observations exits the horizon scale. 
This classification gives intuitive understanding of the property of the
the perturbations generated by this double inflation model.

Then,
we have considered the primordial perturbation in the  $\alpha$-attractor-type double inflation. 
Since we have assumed that the curvature perturbation at the end of inflation is connected to 
the adiabatic perturbation at late time, the power spectrum 
$\mathcal{P}_{\mathcal{R}}$, the spectral index $n_s$  and the tensor-to-scalar ratio $r$
constrained by the Planck result are given by
Eqs.~(\ref{def_Pzeta}), (\ref{def_ns}), and (\ref{def_r}), respectively.
In multiple inflation, the curvature perturbation can evolve even on sufficiently large scales 
caused by the mixing with the isocurvature perturbation 
if the background trajectory changes the direction.
 In  the  $\alpha$-attractor-type double inflation, whether the mixing with the isocurvature perturbation
 affects the curvature perturbation or not depends on whether 
the mode corresponding to the scale of interest exits the horizon scale before the turn or not.
In this model, since we obtained the analytic expression for the background dynamics,
this can be seen by the similar figure as  Fig~\ref{fatemsquare3_BG} with appropriate $M$.
With fixed $M$, $\lambda$ and some sets of $\varphi^I_{\rm ini}$, we performed the numerical calculations
on perturbations to obtain $\mathcal{P}_{\mathcal{R}}$, $n_s$ and $r$.
We find that $r$ is always sufficiently small to be consistent
with the Planck result, which is the basic property of $\alpha$-attractor,
also holds in this double inflation model. However, it is worth mentioning that
the universal relation between $n_s$ and $r$ in the single-field $\alpha$-attractor model
$(1-n_s)/r=N/(6 \alpha)$ obtained from Eqs.~(\ref{rel_ns_r_alphaattractor_def_N}) no longer holds
when the multifield effects are significant. This fact is, in principle, helpful to distinguish  between 
single-field $\alpha$-attractor models and this  $\alpha$-attractor-type double inflation.
On the other hand, since we obtained the approximated analytic solutions for the background,
we can calculate $\mathcal{P}_{\mathcal{R}}$ and $n_s$ based on the $\delta N$ formalism analytically,
given by Eqs.~(\ref{pzeta_deltan_example}) and (\ref{ns_deltan_example}),
even when the multifield effects are significant. We find that in $\varphi^I _{\rm ini}$ space, 
as long as the mode exits the horizon scale
sufficiently before the turn and the resultant $n_s$ is consistent with the Planck result,
the analytic method can reproduce the numerical results very well.

After that, we have imposed observational constraints on the $\alpha$-attractor-type
double inflation model based on the Planck result. More precisely, we imposed the observational constraints
by specifying possible $\lambda$ for given $\varphi^I _{\rm ini}$ if there exist for given $M$.
The results with $M=\sqrt{3} \Mp$, $M=\Mp$ and $M=\sqrt{6} \Mp$ were summarized in the right panel in Fig.~\ref{obsconst_Msqrt3_num} and in the two panels in Fig.~\ref{obsconst_M1_Msqrt6}, respectively,
which were the main results of this paper.
 Although we have obtained the $\varphi^I_{\rm ini}$ dependent constraint on $\lambda$ 
in these plots, practically, there is no way for us to know the initial values of the fields.
Based on this viewpoint, these results shows that the observational constraint on $\lambda$ for given $M$
in this double inflation becomes weaker compared with the single field inflation
and this double inflation allow smaller $\lambda$.
This result suggests that  $\alpha$-attractor-type multiple inflation with more than two scalar fields could relax the constraints on the coupling constant $\lambda$ and the mass scale $M$ furthermore.

 Although we can obtain  these plots by numerical calculations
on the perturbations, as an alternative method with which we can intuitively understand the results,
the analytic $\delta N$ formalism played a
crucial role. Especially,  the facts that
$N_*$ in Eq.~(\ref{nexit_deltan})
does not depend on $\lambda$ and $\mathcal{P}_{\mathcal{R}}$ depends on $\lambda$ linearly
as shown in Eq.~(\ref{pzeta_deltan_example}) make the analysis simpler. 
As we can see from them, although the qualitatively they look similar,
quantitatively there are some differences depending on $M$.
With smaller $M$, the width of the part in $\varphi^I _{\rm ini}$ space 
giving $n_s$ smaller than the lower bound of  the Planck result is narrower,
while the  maximum value of $\varphi^I _{\rm ini}$
dependent enhancement factor of $\mathcal{P}_{\mathcal{R}}$ from that in the single-field model
is larger. Another evident difference is that 
there is a part in $\varphi^I _{\rm ini}$ space
giving $n_s$ larger than the upper bound of the Planck result only with $M=\Mp$.
These were shown to be explained by the excitation of the heavy field
depending on the ratio between $m$ defined by Eq.~(\ref{potential_TModel_asym_small})
and $H$ at the turn, where it is more efficient with smaller $M$.
We also discussed the regime of $M$,
where the main property of the $\alpha$-attractor-type double inflation discussed in this paper
and we found that when $M$ becomes as large as $M=6 \Mp$, the background trajectory can
no longer be regarded as the two almost straight lines connected with a turn
and the two-field model no longer possesses the good property of the 
$\alpha$-attractor-type double inflation.

As we have mentioned above, since most of the results in this paper approximated by analytic solutions are based on 
the $\alpha$-attractor-type asymptotic form of the potential like Eq.~(\ref{potential_TModel_asym_large}),
we expect that the results like sufficiently small $r$ are not restricted to the symmetric two-field T-model,
but are also applicable to other classes of $\alpha$-attractor-type double inflation. As we have shown in Appendix~\ref{sec:PNG}, we also think that
the result that the primordial non-Gaussianity is sufficiently small to be consistent with the Planck result  is 
the generic prediction of the $\alpha$-attractor-type double inflation.
On the other hand, as we have discussed in Appendix~\ref{sec:BD_TurningPhase},
since we used the information of the full potential in the symmetric two-field T-model
to estimate the excitation of the heavy field,
in principle, by investigating the detailed spectrum of the curvature perturbation around the scale
corresponding to the turn,
we can distinguish models that are all classified to  $\alpha$-attractor-type double inflation.
In this paper, as a first study of the  $\alpha$-attractor-type double inflation,
we have concentrated on the possibility that the primordial perturbations observed by CMB
are generated by this double inflation, implicitly assuming that  the scale corresponding to the mode
exits the horizon scale around the turn is within or at least near the regime observable by CMB.
Regardless of this, we can also consider that the case where the scale exits the horizon scale
around the turn is far below the CMB scales.
In this case, we do not have any stringent observational constraints on the excitation of the heavy fields
and the change in the Hubble expansion rate accompanied by the turn.
Then, the model without imposing the symmetry between the two fields considered in 
Appendix~\ref{sec:bd_asymmetriccase} can give quantitatively interesting phenomenology.
It might be interesting to consider the possibility that the primordial black holes,
which accounts for the cold dark matter of the current Universe can be produced
in the $\alpha$-attractor-type double inflation.
Finally, in this paper, we have just concentrated on the phenomenological
aspect of the  $\alpha$-attractor-type double inflation. It is worth trying to see if this class of models
is obtained out of fundamental theories. We would like to leave these topics for future works.

\begin{acknowledgements}
We would like to thank Renata Kallosh, Shinji Mukohyama, Shi Pi and Yusuke Yamada for useful discussions.
We also thank the Yukawa Institute
for Theoretical Physics, Kyoto University for the hospitality
during the workshop on Gravity and Cosmology 2018  (YITPT-17-02)
and the symposium on General Relativity  - The Next
Generation  - (YKIS2018a).
This work was supported by JSPS Grant-in-Aid for 
Scientific Research Grants 
No. 16K05362 (K. M.), No. JP17H06359 (K. M.) and No. 16K17709 (S. M.).

\end{acknowledgements}

\appendix

\section{BACKGROUND DYNAMICS IN $\alpha$-ATTRACTOR-TYPE DOUBLE INFLATION
WITH AN ASYMMETRIC POTENTIAL
\label{sec:bd_asymmetriccase}}

In the main text, we have concentrated on  the $\alpha$-attractor-type double inflation
based on the two-field T-model with a symmetric potential
satisfying $M_1 = M_2 = M$ and $\lambda_1=\lambda_2 = \lambda$. 
Here, we briefly summarize the background dynamics of a  more general  $\alpha$-attractor-type double inflation
model that is
still based on the two-field T-model, but without imposing $M_1=M_2$ nor  $\lambda_1 = \lambda_2$. 
For simplicity, we call the model considered in the main text as the symmetric model, while the one we discuss
from now on as the asymmetric model.
Since even in the asymmetric model, the potential is symmetric with  $\phi \to - \phi$ and $\chi \to -\chi$, 
we can restrict ourselves to the region $\phi \geq 0, \chi \geq 0$, without loss of generality. Furthermore, for simplicity,
we do not consider the cases $M_1 \gg \Mp$ nor $M_2 \gg \Mp$,
which makes it easy to analyze the background dynamics of double inflation, as we will see.
Then, from the similar discussion in Sec.~\ref{sec:background},  in the early stage, 
the potential is approximated by
\bea
V(\phi, \chi) &\simeq&
\frac{1}{36}\left(\lambda_1 M_1^4 + \lambda_2 M_2^4\right)
\nonumber \\
&&
-\frac{1}{9}\left(\lambda_1 M_1^4 e^{-2 \frac{\phi}{M_1}}+\lambda_2 M_2^4 e^{-2 \frac{\chi}{M_2}} \right)\,.
\label{gen_pot_app}
\ena
\begin{widetext}

\begin{figure}[htbp]
\begin{center}
\begin{tabular}{c}

\begin{minipage}{0.45\hsize}
\begin{center}
\includegraphics[width=4.5cm]{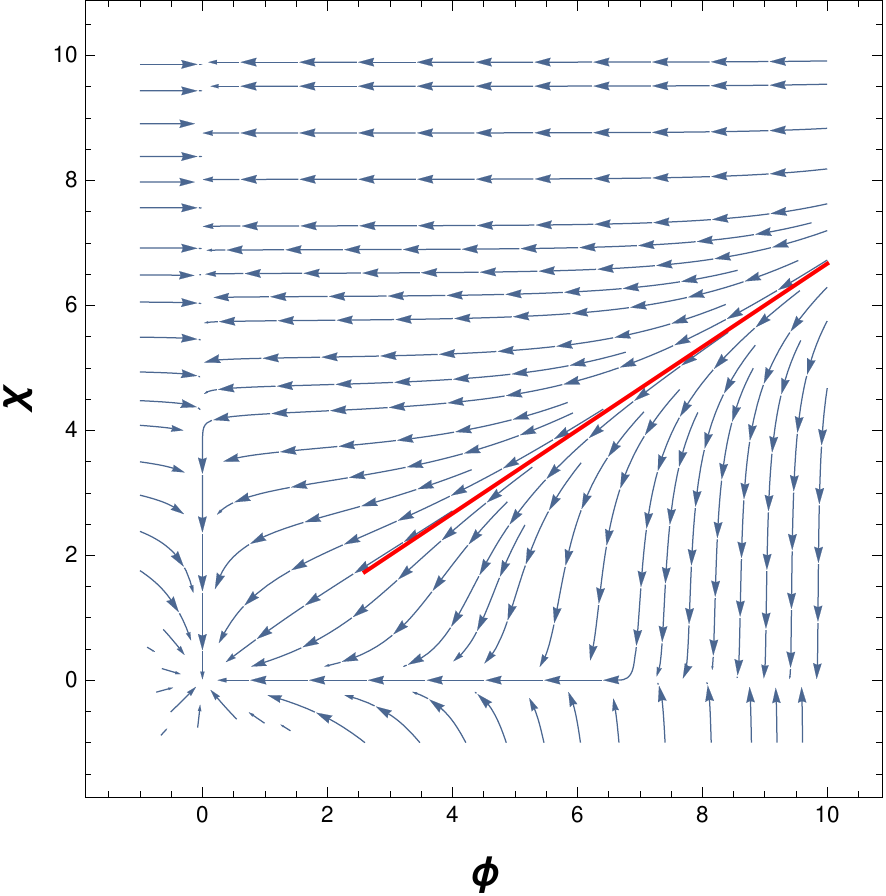}\\
\hspace{1.6cm}
\end{center}
\end{minipage}

\begin{minipage}{0.45\hsize}
\begin{center}
\includegraphics[width=4.5cm]{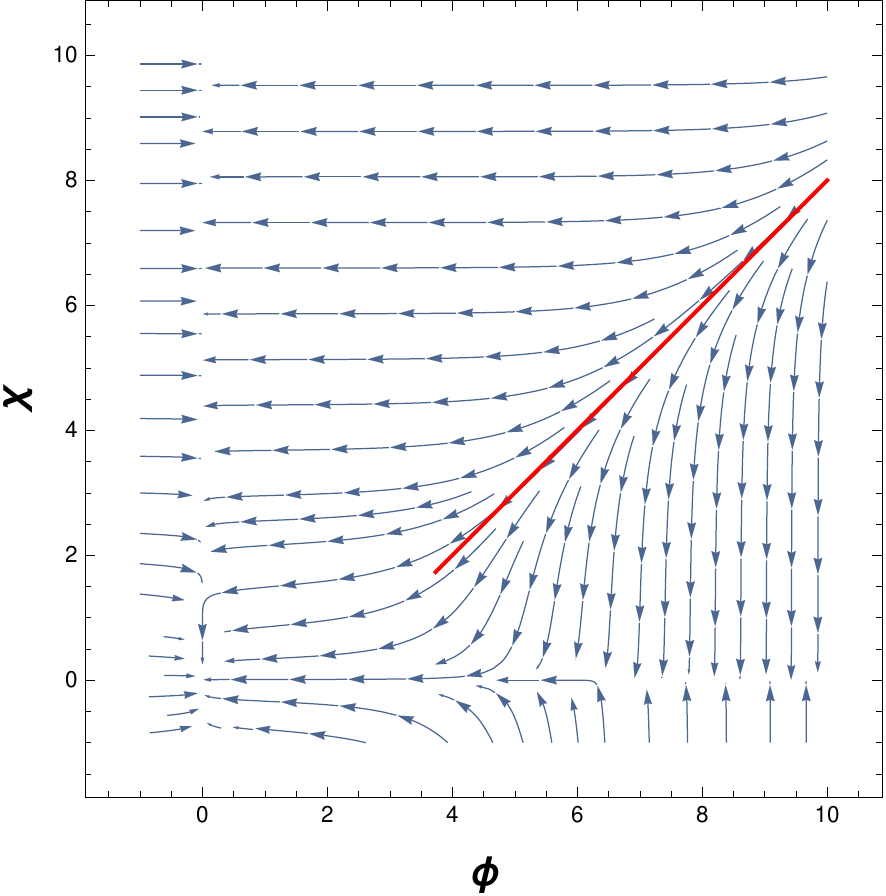}\\
\hspace{1.6cm}
\end{center}
\end{minipage}

\end{tabular}
\caption{Gradients of the inflationary potential
based on the two-field T-model potential,
which suggests the direction of the background trajectories.
The background trajectories towards $\phi =0$ and the ones towards $\chi=0$
in the early stage are separated by the red lines.
(Left) The case with $M_1 =  (3 \sqrt{3}/2 ) \Mp$, $M_2 = \sqrt{3} \Mp$, 
$\lambda_1 = 10^{-9}$ and $\lambda_2 = 2.25 \times 10^{-9}$.
The red line is given by $\chi = (2/3) \phi$. 
(Right) The case with $M_1 =M_2= \sqrt{3} \Mp$,  
$\lambda_1 = \exp [4/\sqrt{3}] \times  10^{-9}$ and $\lambda_2 = 10^{-9}$.
The red line is given by $\chi = \phi -2 \Mp$.}
\label{TModelPotential_gen}
\end{center}
\end{figure}

\end{widetext}

Actually,
by considering only the first two terms in Eq.~(\ref{gen_pot_app})
and making use of the slow-roll approximation, the Hubble expansion rate
 in the first inflationary phase in this asymmetric model  is given by 
$H/\Mp \simeq \sqrt{\lambda_1 M_1^4 + \lambda_2 M_2^4}/(6 \sqrt{3} \Mp^2)$.
About the evolution of $\varphi^I$ in the first inflationary phase, with the  slow-roll approximation, 
the equations of motion for $\varphi^I$ are given by
\bea
&&
\frac{d \phi}{d N} \simeq 8 \frac{ \lambda_1 \Mp^2 M_1^3}
{(\lambda_1 M_1^4 + \lambda_2 M_2^4)} e^{-2 \frac{\phi}{M_1}}\,,
\\
&&
\frac{d \chi}{d N} \simeq 8 \frac{\lambda_2 \Mp^2  M_2^3}{(\lambda_1 M_1^4 + \lambda_2 M_2^4)} 
e^{-2 \frac{\chi}{M_2}}\,,
\label{eom_phi_chi_slowroll_firstinflationaryphase_gen}
\ena
where $\simeq$ means that we have kept only up to the terms 
$\mathcal{O}(e^{-2 \frac{\phi}{M_1}})$ and $\mathcal{O}(e^{-2 \frac{\chi}{M_2}})$. 
Eqs.~(\ref{eom_phi_chi_slowroll_firstinflationaryphase_gen}) can be integrated as
\bea
 &&
 \phi=\frac{M_1}{2} \ln \left[\frac{16  \lambda_1  \Mp^2 M_1^2}{(\lambda_1 M_1^4 + \lambda_2 M_2^4)} 
 (N-N_{\rm ini}) + e^{2 \frac{ \phi_{\rm ini}}{M_1}}\right]\,,\\
 &&
 \chi=\frac{M_2}{2} \ln \left[\frac{16 \lambda_2 \Mp^2  M_2^2}{(\lambda_1 M_1^4 + \lambda_2 M_2^4)} 
 (N-N_{\rm ini}) + e^{2 \frac{ \chi_{\rm ini}}{M_2}} \right]\,,
 \label{phi_chi_slowroll_firstinflationaryphase_ito_N_gen}
\ena
where $N_{\rm ini}$  is again an integration constant corresponding to  $N$ at initial time. 
Notice that the above results recover the ones in the symmetric model 
with the substitution $M_1 = M_2=M$ and $\lambda_1 = \lambda_2=\lambda$, 
where the background trajectories towards $\phi=0$ and those towards $\chi=0$
in the first inflationary phase are separated by the line
$\chi_{\rm ini} = \phi_{\rm ini}$ as we can see from 
Fig.~\ref{TModelPotential2}. 
On the other hand, in this asymmetric model, 
we can show that these trajectories are separated by the line
\bea
\chi_{\rm ini} =\frac{M_2}{M_1} \phi_{\rm ini}+ 
\frac{M_2}{2} \ln \left(\frac{\lambda_2 M_2^2}{\lambda_1 M_1^2} \right)\,,
\label{boundary_bd_gen}
\ena
which means that in  $\varphi^I _{\rm ini}$ space, the slope of the line (\ref{boundary_bd_gen}) 
depends on $M_2/M_1$, while its $\chi_{\rm ini}$ intercept depends on 
$(\sqrt{\lambda_2} M_2)/(\sqrt{\lambda_1} M_1)$.

In Fig.~\ref{TModelPotential_gen}, we show the plots 
similar to Fig.~\ref{TModelPotential2} with different parameters.
In the left panel, we choose the parameters satisfying $\sqrt{\lambda_1} M_1 = \sqrt{\lambda_2} M_2$,
where the line (\ref{boundary_bd_gen}) becomes $\chi_{\rm ini} = (M_2/M_1) \phi_{\rm ini}$,
while in the right panel we choose the parameters satisfying $M_1=M_2$, where  
the line (\ref{boundary_bd_gen}) becomes $\chi_{\rm ini} = \phi_{\rm ini} +
(M_2/2) \ln [(\lambda_2 M_2^2)/(\lambda_1 M_1^2)]$.
In both cases, we can confirm that for $\phi > M_1$ and $\chi > M_2$, 
where the approximated expression of the potential (\ref{gen_pot_app}) is valid, the line (\ref{boundary_bd_gen}) separates the background trajectories
towards $\phi=0$ from the ones towards $\chi=0$ in the first inflationary phase.

If $\chi_{\rm ini}$ is larger than the value (\ref{boundary_bd_gen}) for given $\phi_{\rm ini}$, 
since this is just the generalization of discussion in Sec.~\ref{subsec:background_app},
for all the results related with the first inflationary phase below, 
we can recover the corresponding results in the symmetric model 
with the substitution $M_1 = M_2=M$ and $\lambda_1 = \lambda_2=\lambda$. 
The integration constants $N_{\rm ini}$ and $N_1$ introduced in Eq.~(\ref{def_N1})
in the symmetric model, which represent the total $e$-foldings and 
the $e$-foldings of the second inflationary phase, respectively become
\bea
&&
N_{\rm ini} = \frac{1 }{16 \Mp^2} \biggl[\frac{\lambda_2 M_2^4}{\lambda_1 M_1^2 }
+ M_2^2 e^{2 \frac{\chi_{\rm ini}}{M_2}} +M_1^2 e^{2  \frac{\phi_{\rm ini}}{M_1} }
  \biggr]\,,\\
  &&
  N_1 = \frac{M_2^2 }{16 \Mp^2} \biggl[\frac{\lambda_2 M_2^2}{\lambda_1 M_1^2 }
+ e^{2 \frac{\chi_{\rm ini}}{M_2}}  -\frac{\lambda_2 M_2^2}{\lambda_1 M_1^2}
 e^{2 \frac{\phi_{\rm ini}}{M_1}}  \biggr]\,.
\label{N_ini_N_1_gen_1}
\ena
We can also express the background trajectory originating from $\varphi^I _{\rm ini}$ 
and $\Theta$, introduced in Eq.~(\ref{def_dotsigma_Theta})
in this phase as, 
\beann
&&
\chi = \frac{M_2}{2} \ln \left[\frac{\lambda_2 M_2^2}{\lambda_1 M_1^2}e^{2 \frac{\phi}{M_1} } + e^{2 \frac{\chi_{\rm ini}}{M_2}} - \frac{\lambda_2 M_2^2}{\lambda_1 M_1^2} e^{2 \frac{\phi_{\rm ini}}{M_1}}\right]\,,\\
&&
\Theta = \tan^{-1} \left[\frac{\lambda_2 M_2^3}{\lambda_1 M_1^3}\left(\frac{\frac{16 \lambda_1 M_1^2}{(\lambda_1 M_1^4 +\lambda_2 M_2^4)}(N-N_{\rm ini}) 
+ e^{2 \frac{\phi_{\rm ini}}{M_1}}}{\frac{16 \lambda_2 M_2^2}{(\lambda_1 M_1^4 +\lambda_2 M_2^4)}(N-N_{\rm ini}) 
+ e^{2 \frac{\chi_{\rm ini}}{M_1}}}\right) \right]\,.
\label{trajectory_firstinflationaryphase_gen_1}
\enann

In the first inflationary phase, after $\phi$ becomes smaller than $M_1$ and the turning phase, 
the potential becomes of the form
\beann
V(\phi, \chi) \simeq \frac12 m_1^2 \phi^2 + \frac{\lambda_2 M_2 ^4}{36} \left[1 -4 e^{-2 \frac{\chi}{M_2}} \right]\,,
\label{potential_turning_gen}
\enann
with
\beann
m_1=\frac{\sqrt{2} \lambda_1}{6} M_1\,.
\enann
As in the symmetric model, since the duration  when $\phi$ is in the region 
$\phi > M$ is much longer than  the one in the region $\phi<M$,
we expect that the expression (\ref{phi_chi_slowroll_firstinflationaryphase_ito_N_gen}) with (\ref{N_ini_N_1_gen_1})
describing the background dynamics in the first inflationary phase is not significantly affected
by the existence of the turning phase. 
On the other hand, since we have more parameters to change 
the ratio between $m_1$ and $H$ during the turning phase in this asymmetric model,
we expect that the excitation of the heavy field discussed in Appendix~\ref{sec:BD_TurningPhase}
gives richer phenomenology in  this asymmetric model than the symmetric one discussed in the main text.

After the turning phase,
the potential giving the second inflationary phase can be  approximated by
\beann
V(\phi=0,\chi) \simeq \frac{\lambda_2 M_2^4}{36} \left[1- 4 e^{-2 \frac{\chi}{M_2}}\right]\,,
\enann
for
$\chi \gg M_2$\,.
The background dynamics for the second inflationary phase
in this asymmetric model is obtained just by repeating the discussion for the symmetric model 
in Sec.~\ref{subsec:background_app}
with the replacements $M \to M_2$ and $\lambda \to \lambda_2$. One thing to mention here is that
the Hubble expansion rate in the second inflationary phase in this asymmetric model  is given by 
$H/\Mp \simeq (\sqrt{\lambda_2} M_2^2)/(6 \sqrt{3} \Mp^2)$. Therefore, the gap of the Hubble expansion rate
defined by $H_2/H_1$, the ratio between that in the first inflationary phase and that in the second inflationary phase,
is given by $(\sqrt{\lambda_2} M_2^2)/\sqrt{\lambda_1 M_1^4 + \lambda_2 M_2^4}$,
while it is always given by $1/\sqrt{2}$ in the symmetric model. Again, this
gives richer phenomenology than the symmetric model considered in the main text.

So far, we have discussed the case with  $\chi_{\rm ini}$ greater than the value (\ref{boundary_bd_gen}) 
for  given $\phi_{\rm ini}$. On the other hand, for the opposite case with  $\chi_{\rm ini}$
smaller than the value (\ref{boundary_bd_gen}), the background trajectory in the first inflationary phase
is towards $\chi=0$, instead of $\phi=0$. In this case, the background dynamics can be discussed
by just repeating the above discussion with exchanging the roles of 
$\phi$ and $\chi$.
As we see, by considering $\alpha$-attractor-type double inflation based on the asymmetric model,  
we have more model parameters than in that based on the symmetric model.
This certainly gives richer phenomenology, like the background dynamics
in the first inflationary phase, the gap of the Hubble expansion rate
between the first and second inflationary phases, and the excitation of heavy fields in the turning phase.
However, the basic properties of $\alpha$-attractor-type double inflation
extracted by the symmetric two-field model
that there are first inflationary phase, turning phase, second inflationary phase,
and the background trajectory is composed of two almost straight lines 
in the field space remain unchanged.
Furthermore, as long as we concentrate on the constraints on model parameters and
initial field values based on the Planck result, 
such richer phenomenology obtained by this generalization is strongly constrained.
In this sense, we think that the essential properties of perturbations 
in the $\alpha$-attractor-type double inflation
is sufficiently included in the simple symmetric model.  

\section{EXCITATION OF THE HEAVY FIELD IN  $\alpha$-ATTRACTOR-TYPE
DOUBLE INFLATION
\label{sec:BD_TurningPhase}}

Here, we analyze the background dynamics in the turning phase, where
the potential is approximated by Eq.~(\ref{potential_turning}) in detail and
clarify when the oscillations in the background trajectory caused by the heavy field excitation
appear. First, we consider the motion in $\phi$ direction around $\phi=0$, 
where $V$ can be further approximated by $V \simeq \lambda M^4/36$.
Then, from the slow-roll approximation and Eq.~(\ref{bg_field_eq_orig}),
as long as $H$ is greater than $m$, with an integration constant $\phi_0$,
the evolution of $\phi$ is given by
\bea
\phi = \phi_0 \exp \left[-\frac{m^2}{3 H} t \right]\,,
\label{turn_phi_smooth}
\ena
with
\beann
 H\simeq \frac{\sqrt{\lambda} M^2}{6 \sqrt{3} \Mp}\,.
\enann
In this case, the motion towards $\chi=0$ 
becomes important as $\phi$ approaches to $0$ and $\Theta$ changes from $\pi$ to 
$(3/2) \pi$ smoothly. Although it is necessary to confirm that the slow-roll approximation
holds all the way to $\phi \sim 0$ numerically, the evolution of $\phi$
in the smooth turn with $m < H$ is described by Eq.~(\ref{turn_phi_smooth}).
The case corresponding to Fig.~\ref{phi_chi_BG}  is an example giving a smooth turn.

\begin{widetext}

\begin{figure}[htbp]
\begin{center}
\begin{tabular}{c}

\begin{minipage}{0.3\hsize}
\begin{center}
\includegraphics[width=5cm]{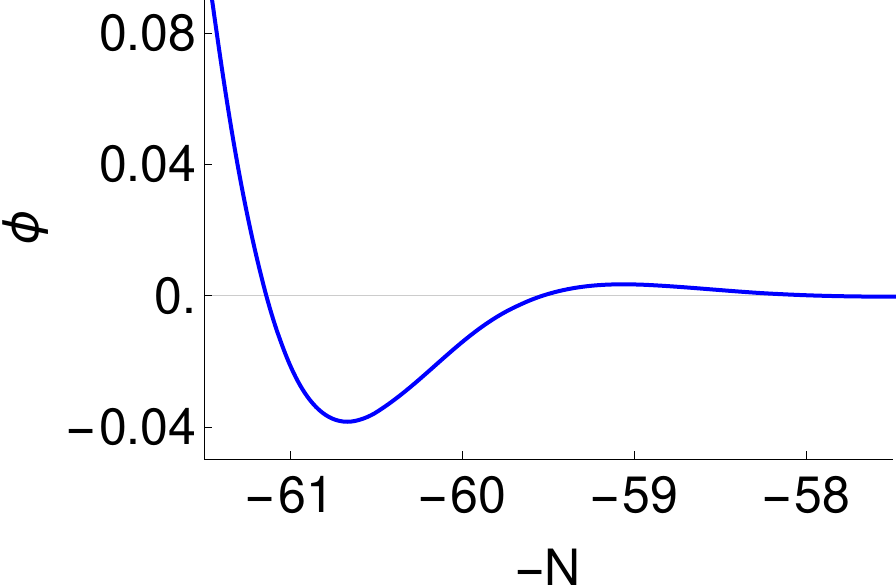}\\
\hspace{1.6cm}
\end{center}
\end{minipage}
\hspace{2cm}
\begin{minipage}{0.3\hsize}
\begin{center}
\includegraphics[width=5cm]{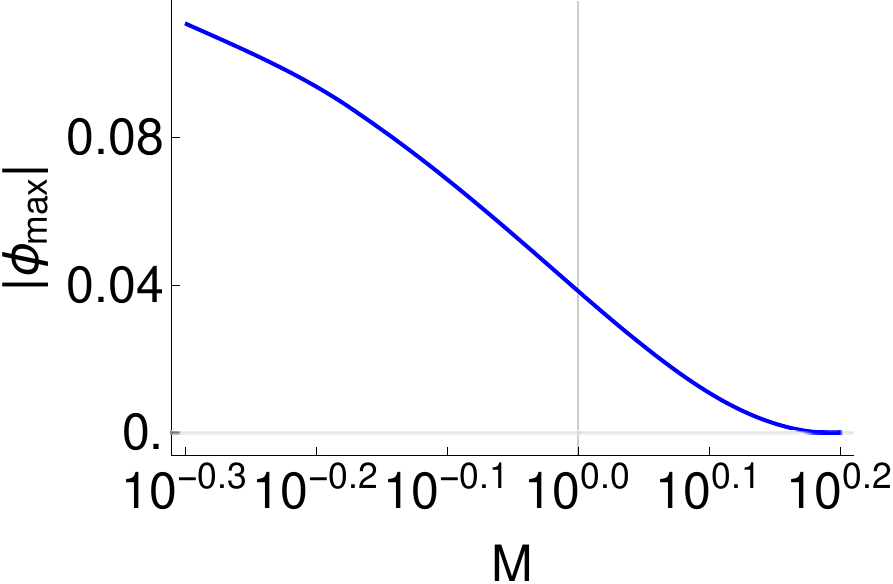}\\
\hspace{1.6cm}
\end{center}
\end{minipage}

\end{tabular}
\caption{
(Left) Time evolution of $\phi$ around the turn in terms of $-N$ with $M = \Mp$, $\lambda=6.6 \times 10^{-10}$.
Since $m^2/H^2 > 9/4$, there is excitation of the heavy field, where the maximum value of $|\phi|$ after it passes $0$
is about $0.0384$.
(Right) $M$ dependence of the efficiency of the heavy field excitation in terms of the
maximum value of $|\phi|$ after it passes $0$. We choose $\lambda$ so that $m = (\sqrt{2 \lambda} M)/6  \simeq 
6.1 \times 10^{-4} \Mp$. 
\label{heavy_field_excitation}}
\end{center}
\end{figure}

\end{widetext}

On the other hand, if $H \gg m$ does not hold,
the second derivative of $\phi$ in Eq.~(\ref{bg_field_eq_orig}) is no longer negligible.
Especially if $m^2/H^2  > 9/4 $,
with another integration constant $\theta_0$, the evolution of $\phi$ is given by
\beann
\phi=\phi_0 \exp \left[-\frac{3}{2} Ht \right] \cos[\omega t - \theta_0]\,,
\label{turn_phi_oscillation}
\enann
with
\beann
 \omega \equiv \frac12 \sqrt{4 m^2 - 9 H^2}\,,
\enann
which describes dumped oscillations. If the integration constant $\phi_0$ is large,
or equivalently the kinetic energy of $\phi$ is sufficiently large around $\phi=0$,
the background trajectory experiences oscillations during the turn,
which can be interpreted as the excitation of the heavy field.
In the examples considered in Sec.~\ref{sec:constraints} B,
with $M=\Mp$, there are oscillations of $\phi$ during the turn as shown in
the left panel in Fig.~\ref{heavy_field_excitation}.
This excitation of heavy field affects the resultant $\mathcal{P}_{\mathcal{R}}$ and $n_s$,
as discussed in Sec.~\ref{sec:constraints}B.
In the current model, the parameter dependence of the efficiency of heavy field excitation
depends on the dynamics of $\phi$ at $0 < \phi < M$ in the first inflationary phase, 
where the potential cannot be approximated
by neither Eq.~(\ref{TModelPotential_approximated}) nor Eq.~(\ref{potential_turning}).
Here, we numerically estimate
the parameter dependence of the efficiency of the heavy field excitation.
As we mentioned in Sec.~\ref{sec:constraints} B, when we change $M$,  
we also change the fiducial value of $\lambda$
that  gives $\mathcal{P}_{\mathcal{R}}$ consistent with the Planck result in the single-field model
so that $\lambda M^2$ is fixed.  From Eq.~(\ref{potential_TModel_asym_small}), we also fix 
$m = (\sqrt{2 \lambda} M)/6  \simeq 6.1 \times 10^{-4} \Mp$ 
so that it covers the fiducial values of $\lambda$ considered in Sec.~\ref{sec:constraints}.
Here, we investigate the efficiency of the heavy field excitation
in terms of the maximum value of $|\phi|$ after it passes zero by changing $M$
with $\chi_{\rm ini}$ giving the turn about $N \simeq N_* = -60$.
However, it is worth mentioning that as long as $\chi_*$ is much greater than $M$,
we expect that $\chi_{\rm ini}$ dependence is weak, as $H_* \simeq \sqrt{V_*} /(\sqrt{3} \Mp)$ is almost constant
with $\chi_* \gg M$. The right panel in Fig.~\ref{heavy_field_excitation} shows 
$M$ dependence of the efficiency of the heavy field excitation.
We see that by decreasing $M$, the heavy field excitation occurs around $M \sim 10^{0.2} \Mp$,
and the efficiency monotonously increases when $M$ decreases.
Notice that $M$ giving $m^2/H^2=9/4$ based on $H$ given by the slow-roll approximation (\ref{turn_phi_smooth})
gives $(2 \sqrt{6})/3 \Mp$, which  is very close to $10^{0.2} \Mp$.
 
\section{PRIMORDIAL NON-GAUSSIANITY IN $\alpha$-ATTRACTOR-TYPE DOUBLE INFLATION
\label{sec:PNG}}

In the main text, we imposed the observational constraints on $\lambda$ and $\varphi^I _{\rm ini}$ 
for a given $M$ in the  $\alpha$-attractor-type double inflation model  based on
the Planck result of $\mathcal{P}_{\mathcal{R}}$ and $n_s$. On the other hand, for the region in $\varphi^I _{\rm ini}$
space where the multifield effects modify the resultant $\mathcal{P}_{\mathcal{R}}$ from that 
in the single-field model, while the resultant $n_s$ is still consistent with the Planck result,
we can obtain $\mathcal{P}_{\mathcal{R}}$ and $n_s$ analytically based on
the $\delta N$ formalism with a high accuracy. Since we can easily calculate the primordial bispectrum,
which includes the leading signal of non-Gaussianity,
based on the $\delta N$ formalism, here we calculate it to see 
if additional constraints are obtained by the primordial non-Gaussianity on this model or not.

\begin{widetext}

By extending the relation between $\mathcal{R}$ and $\delta \varphi^I$ in the $\delta N$ formalism
given by Eq.~(\ref{rel_deltan}), we obtain
\bea
\mathcal{R} (t_{\rm f}, {\bf x}) \simeq \delta N  (t_{\rm f}, t_{\rm i}, {\bf x}) 
\simeq N_{,I } \delta \varphi^I (t_{\rm i}, {\bf x}) + \frac12 N_{,IJ}  \delta \varphi^I (t_{\rm i}, {\bf x})    
\delta \varphi^J (t_{\rm i}, {\bf x})\,.
\label{app_png_rel_deltan}
\ena
On the other hand, the bispectrum of the curvature perturbation 
$B_{\mathcal{R}}$ in the Fourier space
is defined by
\bea
\langle \mathcal{R}_{\bf k_1} \mathcal{R}_{\bf k_2}  \mathcal{R}_{\bf k_3} \rangle \equiv
(2 \pi)^3 \delta ({\bf k_1} + {\bf k_2} + {\bf k_3}) B_{\mathcal{R}} ({\bf k_1}, {\bf k_2}, {\bf k_3} )\,,
\label{app_png_def_bispectrum_zeta}
\ena
where the left hand side of Eq.~(\ref{app_png_def_bispectrum_zeta}) can be evaluated 
using Eq.~(\ref{app_png_rel_deltan}) as
\bea
\langle \mathcal{R}_{\bf k_1}   \mathcal{R}_{\bf k_2}    \mathcal{R}_{\bf k_3}  \rangle
=N_{,I} N_{,J} N_{,K} \langle \delta \varphi^I_{\bf k_1}  \delta \varphi^J_{\bf k_2}  \delta \varphi^K_{\bf k_3}  \rangle
+\frac12 N_{,I}   N_{,J} N_{,K L}
 \langle \delta \varphi^I_{\bf k_1}  \delta \varphi^J_{\bf k_2}  
 [ \delta \varphi^K \star \delta \varphi^L  ]_{\bf k_3}    \rangle + {\rm perms.}\,.
 \label{delta_N_upto_secondorder}
\ena
Here,  $\star$ denotes the convolution and  ``${\rm perms}$'' denotes the remaining two other permutations.
In Eq.~(\ref{delta_N_upto_secondorder}), roughly speaking, the first term in the right hand side
denotes the contribution from the intrinsic non-Gaussianity of the field perturbations,
giving nonlocal-type bispectrum, while the second term denotes the contribution from
the nonlinear dynamics of the background field on super horizon scales
giving the local-type bispectrum \cite{Lyth:2005fi}. 

\end{widetext}

Observational constraints on the primordial bispectrum are given by the nonlinear parameter 
$f_{\rm NL}$ defined by  \cite{Maldacena:2002vr}
\beann
\frac65 f_{\rm NL}= \frac{\prod_j k_j^3}{\sum_j k_j^3} \frac{B_{\mathcal{R}} }{4 \pi^4 \mathcal{P}_{\mathcal{R}}^2}\,.
\enann
Since the kinetic terms in this model
are canonical and  the former contribution is expected to be small \cite{Maldacena:2002vr},
by concentrating on the local-type primordial bispectrum, we obtain the following expression of $f_{\rm NL} ^{\rm loc}$
in the $\delta N$ formalism
\bea
f_{\rm NL} ^{\rm loc} = \frac56 \frac{N_{,I} N_{,J} N^{,IJ} }{(N_{,I} N^{,I})^2}\,.
\label{app_png_fnl_local}
\ena
Then, by just regarding $N_*$ given by Eq.~(\ref{nexit_deltan}) as $N$ 
and $\varphi^I _{*}$ given by Eq.~(\ref{phiexit_chiexit_deltan}) in Eq.~(\ref{app_png_fnl_local}),
we can obtain $f_{\rm NL} ^{\rm loc}$,
\beann
f_{\rm NL} ^{\rm loc} = \frac56 \frac{({N_*}_{,\phi_*})^2 {N_*}_{,\phi_* \phi_*} +({N_*}_{,\chi_*})^2 {N_*}_{,\chi_* \chi_*} }{(({N_*}_{,\phi_*})^2 + ({N_*}_{,\chi_*})^2)^2}\,,
\enann
where ${N_*}_{,\phi_*}$ and ${N_*}_{,\chi_*}$ are given by Eq.~(\ref{delNoverdelphi_delNoverdelchi})
and ${N_*}_{,\phi_* \phi_*}$ and ${N_*}_{,\chi_* \chi_*}$ are given by
\beann
&&
{N_*}_{,\phi_* \phi_*} = \frac{1}{4 \Mp^2} \left(\frac{8 \Mp^2}{M^2} N_{*} -\frac12 + \frac12 e^{2 \frac{\phi_{\rm ini}}{M}}
-\frac12 e^{2 \frac{\chi_{\rm ini}}{M}}\right)\,,\\
&&
{N_*}_{,\chi_* \chi_*} = \frac{1}{4 \Mp^2} \left(\frac{8 \Mp^2}{M^2} N_{*} -\frac12 + \frac12 e^{2 \frac{\chi_{\rm ini}}{M}}
-\frac12 e^{2 \frac{\phi_{\rm ini}}{M}}\right)\,.
\label{deldelNoverdelphidelphi_deldelNoverdeldelchi}
\enann

We find that for the three cases with $M=\Mp$, $M=\sqrt{3} \Mp$ and $M=\sqrt{6} \Mp$
we considered in Sec.~\ref{sec:constraints}, in the part in $\varphi^I _{\rm ini}$ space,  
$f_{\rm NL} ^{\rm loc}$ does not depends on 
the parameter $M$ or the initial values $\varphi^I_{\rm ini}$ so much and takes the value 
$0.013 < f_{\rm NL} ^{\rm loc} < 0.016$, which is consistent with the Planck result \cite{Ade:2015ava}.
This result shows that the primordial non-Gaussianity cannot give additional constraints
on the $\alpha$-attractor-type double inflation
other than the ones considered in the main text.


~\\


\begin{thebibliography}{99}

\bibitem{Guth:1982ec}
  A.~H.~Guth and S.~Y.~Pi,
  Phys.\ Rev.\ Lett.\  {\bf 49} (1982) 1110.
  doi:10.1103/PhysRevLett.49.1110


\bibitem{Hawking:1982cz}
  S.~W.~Hawking,
  Phys.\ Lett.\  {\bf 115B} (1982) 295.
  doi:10.1016/0370-2693(82)90373-2

\bibitem{Starobinsky:1982ee}
  A.~A.~Starobinsky,
  Phys.\ Lett.\  {\bf 117B} (1982) 175.
  doi:10.1016/0370-2693(82)90541-X

\bibitem{Bardeen:1983qw}
  J.~M.~Bardeen, P.~J.~Steinhardt and M.~S.~Turner,
  Phys.\ Rev.\ D {\bf 28} (1983) 679.
  doi:10.1103/PhysRevD.28.679


\bibitem{Kodama:1985bj}
  H.~Kodama and M.~Sasaki,
  Prog.\ Theor.\ Phys.\ Suppl.\  {\bf 78} (1984) 1.
  doi:10.1143/PTPS.78.1

\bibitem{Mukhanov:1990me}
  V.~F.~Mukhanov, H.~A.~Feldman and R.~H.~Brandenberger,
  Phys.\ Rept.\  {\bf 215} (1992) 203.
  doi:10.1016/0370-1573(92)90044-Z
  
  
  
\bibitem{Ade:2015xua}
  P.~A.~R.~Ade {\it et al.} [Planck Collaboration],
  Astron.\ Astrophys.\  {\bf 594} (2016) A13
  doi:10.1051/0004-6361/201525830
  [arXiv:1502.01589 [astro-ph.CO]].

\bibitem{Ade:2015ava}
  P.~A.~R.~Ade {\it et al.} [Planck Collaboration],
  Astron.\ Astrophys.\  {\bf 594} (2016) A17
  doi:10.1051/0004-6361/201525836
  [arXiv:1502.01592 [astro-ph.CO]].


\bibitem{Baumann:2014nda}
  D.~Baumann and L.~McAllister,
  arXiv:1404.2601 [hep-th].


\bibitem{Kallosh:2013hoa} 
  R.~Kallosh and A.~Linde,
  JCAP {\bf 1307}, 002 (2013)
  doi:10.1088/1475-7516/2013/07/002
  [arXiv:1306.5220 [hep-th]].
  
  
  
\bibitem{Kallosh:2013xya}
  R.~Kallosh and A.~Linde,
  JCAP {\bf 1306} (2013) 028
  doi:10.1088/1475-7516/2013/06/028
  [arXiv:1306.3214 [hep-th]].

\bibitem{Ferrara:2013rsa}
  S.~Ferrara, R.~Kallosh, A.~Linde and M.~Porrati,
  Phys.\ Rev.\ D {\bf 88} (2013) no.8,  085038
  doi:10.1103/PhysRevD.88.085038
  [arXiv:1307.7696 [hep-th]].
  
\bibitem{Kallosh:2013yoa}
  R.~Kallosh, A.~Linde and D.~Roest,
  JHEP {\bf 1311} (2013) 198
  doi:10.1007/JHEP11(2013)198
  [arXiv:1311.0472 [hep-th]].
  
  
  
\bibitem{Galante:2014ifa}
  M.~Galante, R.~Kallosh, A.~Linde and D.~Roest,
  Phys.\ Rev.\ Lett.\  {\bf 114} (2015) no.14,  141302
  doi:10.1103/PhysRevLett.114.141302
  [arXiv:1412.3797 [hep-th]].
  
 
\bibitem{Carrasco:2015uma}
  J.~J.~M.~Carrasco, R.~Kallosh, A.~Linde and D.~Roest,
  Phys.\ Rev.\ D {\bf 92} (2015) no.4,  041301
  doi:10.1103/PhysRevD.92.041301
  [arXiv:1504.05557 [hep-th]].
  
\bibitem{Carrasco:2015rva}
  J.~J.~M.~Carrasco, R.~Kallosh and A.~Linde,
  Phys.\ Rev.\ D {\bf 92} (2015) no.6,  063519
  doi:10.1103/PhysRevD.92.063519
  [arXiv:1506.00936 [hep-th]].
  
\bibitem{Cremonini:2010ua}
  S.~Cremonini, Z.~Lalak and K.~Turzynski,
  JCAP {\bf 1103} (2011) 016
  doi:10.1088/1475-7516/2011/03/016
  [arXiv:1010.3021 [hep-th]].
 
  
\bibitem{Renaux-Petel:2015mga} 
  S.~Renaux-Petel and K.~Turzynski,
  Phys.\ Rev.\ Lett.\  {\bf 117}, no. 14, 141301 (2016)
  doi:10.1103/PhysRevLett.117.141301
  [arXiv:1510.01281 [astro-ph.CO]].

 
\bibitem{Achucarro:2016fby} 
  A.~Achucarro, V.~Atal, C.~Germani and G.~A.~Palma,
  JCAP {\bf 1702}, no. 02, 013 (2017)
  doi:10.1088/1475-7516/2017/02/013
  [arXiv:1607.08609 [astro-ph.CO]].
    
     
      
\bibitem{Brown:2017osf}
  A.~R.~Brown,
  arXiv:1705.03023 [hep-th].

\bibitem{Mizuno:2017idt}
  S.~Mizuno and S.~Mukohyama,
  Phys.\ Rev.\ D {\bf 96} (2017) no.10,  103533
  doi:10.1103/PhysRevD.96.103533
  [arXiv:1707.05125 [hep-th]].
        
         
\bibitem{Garcia-Saenz:2018ifx}
  S.~Garcia-Saenz, S.~Renaux-Petel and J.~Ronayne,
  arXiv:1804.11279 [astro-ph.CO].
           
            
\bibitem{Ferrara:2016fwe}
  S.~Ferrara and R.~Kallosh,
  Phys.\ Rev.\ D {\bf 94} (2016) no.12,  126015
  doi:10.1103/PhysRevD.94.126015
  [arXiv:1610.04163 [hep-th]].
  
\bibitem{Kallosh:2017ced}
  R.~Kallosh, A.~Linde, T.~Wrase and Y.~Yamada,
  JHEP {\bf 1704} (2017) 144
  doi:10.1007/JHEP04(2017)144
  [arXiv:1704.04829 [hep-th]].
  
\bibitem{Gordon:2000hv}
  C.~Gordon, D.~Wands, B.~A.~Bassett and R.~Maartens,
  Phys.\ Rev.\ D {\bf 63} (2001) 023506
  doi:10.1103/PhysRevD.63.023506
  [astro-ph/0009131].

  
\bibitem{GrootNibbelink:2001qt}
  S.~Groot Nibbelink and B.~J.~W.~van Tent,
  Class.\ Quant.\ Grav.\  {\bf 19} (2002) 613
  doi:10.1088/0264-9381/19/4/302
  [hep-ph/0107272].
  
\bibitem{Polarski:1992dq} 
  D.~Polarski and A.~A.~Starobinsky,
  Nucl.\ Phys.\ B {\bf 385}, 623 (1992).
  doi:10.1016/0550-3213(92)90062-G
  
  
  
\bibitem{Polarski:1994rz}
  D.~Polarski and A.~A.~Starobinsky,
  Phys.\ Rev.\ D {\bf 50} (1994) 6123
  doi:10.1103/PhysRevD.50.6123
  [astro-ph/9404061].
  
  
  
\bibitem{Langlois:1999dw} 
  D.~Langlois,
  Phys.\ Rev.\ D {\bf 59}, 123512 (1999)
  doi:10.1103/PhysRevD.59.123512
  [astro-ph/9906080].
  
\bibitem{Tsujikawa:2002qx}
  S.~Tsujikawa, D.~Parkinson and B.~A.~Bassett,
  Phys.\ Rev.\ D {\bf 67} (2003) 083516
  doi:10.1103/PhysRevD.67.083516
  [astro-ph/0210322].
  
\bibitem{Vernizzi:2006ve}
  F.~Vernizzi and D.~Wands,
  JCAP {\bf 0605} (2006) 019
  doi:10.1088/1475-7516/2006/05/019
  [astro-ph/0603799].
  
\bibitem{Byrnes:2006fr}
  C.~T.~Byrnes and D.~Wands,
  Phys.\ Rev.\ D {\bf 74} (2006) 043529
  doi:10.1103/PhysRevD.74.043529
  [astro-ph/0605679].
  
  
\bibitem{yamada:2018pc} 
Y.~Yamada, private communication (2018).
   
    
     
      
\bibitem{Kallosh:2013daa} 
  R.~Kallosh and A.~Linde,
  JCAP {\bf 1312}, 006 (2013)
  doi:10.1088/1475-7516/2013/12/006
  [arXiv:1309.2015 [hep-th]].
  
\bibitem{Kallosh:2017wku}
  R.~Kallosh, A.~Linde, D.~Roest, A.~Westphal and Y.~Yamada,
  JHEP {\bf 1802} (2018) 117
  doi:10.1007/JHEP02(2018)117
  [arXiv:1707.05830 [hep-th]].
    
\bibitem{Achucarro:2017ing}
  A.~Achucarro, R.~Kallosh, A.~Linde, D.~G.~Wang and Y.~Welling,
  JCAP {\bf 1804} (2018) no.04,  028
  doi:10.1088/1475-7516/2018/04/028
  [arXiv:1711.09478 [hep-th]].
  
\bibitem{Krajewski:2018moi} 
  T.~Krajewski, K.~Turzynski and M.~Wieczorek,
  arXiv:1801.01786 [astro-ph.CO].
  
\bibitem{Yamada:2018nsk}
  Y.~Yamada,
  JHEP {\bf 1804} (2018) 006
  doi:10.1007/JHEP04(2018)006
  [arXiv:1802.04848 [hep-th]].
  
\bibitem{Linde:2018hmx}
  A.~Linde, D.~G.~Wang, Y.~Welling, Y.~Yamada and A.~Achucarro,
  arXiv:1803.09911 [hep-th].
    
\bibitem{Christodoulidis:2018qdw}
  P.~Christodoulidis, D.~Roest and E.~I.~Sfakianakis,
  arXiv:1803.09841 [hep-th].
  
\bibitem{Dias:2018pgj}
  M.~Dias, J.~Frazer, A.~Retolaza, M.~Scalisi and A.~Westphal,
  arXiv:1805.02659 [hep-th].

\bibitem{Iarygina:2018kee}
  O.~Iarygina, E.~I.~Sfakianakis, D.~G.~Wang and A.~Achucarro,
  arXiv:1810.02804 [astro-ph.CO].




\bibitem{Starobinsky:1980te}
  A.~A.~Starobinsky,
  Phys.\ Lett.\  {\bf 91B} (1980) 99.
  doi:10.1016/0370-2693(80)90670-X
  
  
\bibitem{Ellis:2014gxa}
  J.~Ellis, M.~A.~G.~Garcia, D.~V.~Nanopoulos and K.~A.~Olive,
  JCAP {\bf 1408} (2014) 044
  doi:10.1088/1475-7516/2014/08/044
  [arXiv:1405.0271 [hep-ph]].
  
\bibitem{Abe:2014vca}
  H.~Abe and H.~Otsuka,
  JCAP {\bf 1411} (2014) no.11,  027
  doi:10.1088/1475-7516/2014/11/027
  [arXiv:1405.6520 [hep-th]].
  
\bibitem{Ellis:2014opa}
  J.~Ellis, M.~A.~G.~Garcia, D.~V.~Nanopoulos and K.~A.~Olive,
  JCAP {\bf 1501} (2015) 010
  doi:10.1088/1475-7516/2015/01/010
  [arXiv:1409.8197 [hep-ph]].
  
  
\bibitem{vandeBruck:2015xpa}
  C.~van de Bruck and L.~E.~Paduraru,
  Phys.\ Rev.\ D {\bf 92} (2015) 083513
  doi:10.1103/PhysRevD.92.083513
  [arXiv:1505.01727 [hep-th]].
  
\bibitem{Kaneda:2015jma}
  S.~Kaneda and S.~V.~Ketov,
  Eur.\ Phys.\ J.\ C {\bf 76} (2016) no.1,  26
  doi:10.1140/epjc/s10052-016-3888-0
  [arXiv:1510.03524 [hep-th]].
  
\bibitem{Wang:2017fuy}
  Y.~C.~Wang and T.~Wang,
  Phys.\ Rev.\ D {\bf 96} (2017) no.12,  123506
  doi:10.1103/PhysRevD.96.123506
  [arXiv:1701.06636 [gr-qc]].
  
\bibitem{Mori:2017caa}
  T.~Mori, K.~Kohri and J.~White,
  JCAP {\bf 1710} (2017) no.10,  044
  doi:10.1088/1475-7516/2017/10/044
  [arXiv:1705.05638 [astro-ph.CO]].
  
\bibitem{Pi:2017gih}
  S.~Pi, Y.~l.~Zhang, Q.~G.~Huang and M.~Sasaki,
  JCAP {\bf 1805} (2018) no.05,  042
  doi:10.1088/1475-7516/2018/05/042
  [arXiv:1712.09896 [astro-ph.CO]].
  
\bibitem{He:2018gyf}
  M.~He, A.~A.~Starobinsky and J.~Yokoyama,
  JCAP {\bf 1805} (2018) no.05,  064
  doi:10.1088/1475-7516/2018/05/064
  [arXiv:1804.00409 [astro-ph.CO]].

  
 
\bibitem{Linde:1983gd}
  A.~D.~Linde,
  Phys.\ Lett.\  {\bf 129B} (1983) 177.
  doi:10.1016/0370-2693(83)90837-7
 
  





  
\bibitem{Sasaki:1986hm} 
  M.~Sasaki,
  Prog.\ Theor.\ Phys.\  {\bf 76}, 1036 (1986).
  doi:10.1143/PTP.76.1036
  
\bibitem{Mukhanov:1988jd} 
  V.~F.~Mukhanov,
  Sov.\ Phys.\ JETP {\bf 67}, 1297 (1988)
  [Zh.\ Eksp.\ Teor.\ Fiz.\  {\bf 94N7}, 1 (1988)].

\bibitem{Bunch:1978yq}
  T.~S.~Bunch and P.~C.~W.~Davies,
  Proc.\ Roy.\ Soc.\ Lond.\ A {\bf 360} (1978) 117.
  doi:10.1098/rspa.1978.0060
  
\bibitem{Lyth:1984gv}
  D.~H.~Lyth,
  Phys.\ Rev.\ D {\bf 31} (1985) 1792.
  doi:10.1103/PhysRevD.31.1792
  
\bibitem{Bardeen:1980kt} 
  J.~M.~Bardeen,
  Phys.\ Rev.\ D {\bf 22}, 1882 (1980).
  doi:10.1103/PhysRevD.22.1882
  
 

\bibitem{Starobinsky:1979ty} 
  A.~A.~Starobinsky,
  JETP Lett.\  {\bf 30}, 682 (1979)
  [Pisma Zh.\ Eksp.\ Teor.\ Fiz.\  {\bf 30}, 719 (1979)].

\bibitem{Starobinsky:1986fxa} 
  A.~A.~Starobinsky,
  JETP Lett.\  {\bf 42}, 152 (1985)
  [Pisma Zh.\ Eksp.\ Teor.\ Fiz.\  {\bf 42}, 124 (1985)].
  
  
\bibitem{Sasaki:1995aw}
  M.~Sasaki and E.~D.~Stewart,
  Prog.\ Theor.\ Phys.\  {\bf 95} (1996) 71
  doi:10.1143/PTP.95.71
  [astro-ph/9507001].
  
\bibitem{Sasaki:1998ug}
  M.~Sasaki and T.~Tanaka,
  Prog.\ Theor.\ Phys.\  {\bf 99} (1998) 763
  doi:10.1143/PTP.99.763
  [gr-qc/9801017].
  
\bibitem{Salopek:1990jq}
  D.~S.~Salopek and J.~R.~Bond,
  Phys.\ Rev.\ D {\bf 42} (1990) 3936.
  doi:10.1103/PhysRevD.42.3936
  
\bibitem{Wands:2000dp}
  D.~Wands, K.~A.~Malik, D.~H.~Lyth and A.~R.~Liddle,
  Phys.\ Rev.\ D {\bf 62} (2000) 043527
  doi:10.1103/PhysRevD.62.043527
  [astro-ph/0003278].
  
\bibitem{Lyth:2005fi}
  D.~H.~Lyth and Y.~Rodriguez,
  Phys.\ Rev.\ Lett.\  {\bf 95} (2005) 121302
  doi:10.1103/PhysRevLett.95.121302
  [astro-ph/0504045].
  
  
\bibitem{Tolley:2009fg}
  A.~J.~Tolley and M.~Wyman,
  Phys.\ Rev.\ D {\bf 81} (2010) 043502
  doi:10.1103/PhysRevD.81.043502
  [arXiv:0910.1853 [hep-th]].
  
  

  
  
  
\bibitem{Achucarro:2010da}
  A.~Achucarro, J.~O.~Gong, S.~Hardeman, G.~A.~Palma and S.~P.~Patil,
  JCAP {\bf 1101} (2011) 030
  doi:10.1088/1475-7516/2011/01/030
  [arXiv:1010.3693 [hep-ph]].
  
 
  
\bibitem{Shiu:2011qw}
  G.~Shiu and J.~Xu,
  Phys.\ Rev.\ D {\bf 84} (2011) 103509
  doi:10.1103/PhysRevD.84.103509
  [arXiv:1108.0981 [hep-th]].
  
\bibitem{Pi:2012gf}
  S.~Pi and M.~Sasaki,
  JCAP {\bf 1210} (2012) 051
  doi:10.1088/1475-7516/2012/10/051
  [arXiv:1205.0161 [hep-th]].
  
  
\bibitem{Gao:2012uq}
  X.~Gao, D.~Langlois and S.~Mizuno,
  JCAP {\bf 1210} (2012) 040
  doi:10.1088/1475-7516/2012/10/040
  [arXiv:1205.5275 [hep-th]].
  
  
\bibitem{Noumi:2012vr}
  T.~Noumi, M.~Yamaguchi and D.~Yokoyama,
  JHEP {\bf 1306} (2013) 051
  doi:10.1007/JHEP06(2013)051
  [arXiv:1211.1624 [hep-th]].
   
      
         
            
               
\bibitem{Maldacena:2002vr}
  J.~M.~Maldacena,
  JHEP {\bf 0305} (2003) 013
  doi:10.1088/1126-6708/2003/05/013
  [astro-ph/0210603].




\end{thebibliography}
\end{document}